\documentclass[prd,nofootinbib,floats,superscriptaddress,eqsecnum,tightenlines,preprintnumbers,11pt]{revtex4}

\usepackage{hyperref}
\usepackage{graphicx}
\usepackage{amsmath,amssymb,amsfonts,amsthm,latexsym}
\usepackage{marginnote}
\usepackage{xcolor}
\usepackage{physics}
\usepackage{nicefrac}

\usepackage{amsmath}%
\usepackage{amsfonts}%
\usepackage{amssymb}%
\usepackage{graphicx}

\usepackage{mathrsfs}
\usepackage{xspace}
\usepackage{mathtools, amssymb}

\usepackage{tikz}
\tikzset{every picture/.style={line width=0.75pt}} 

\def\be{\begin{equation}}
\def\ee{\end{equation}}
\def\beq{\begin{equation}}
\def\eeq{\end{equation}}
\newcommand{\bea}{\begin{eqnarray}}
\newcommand{\eea}{\end{eqnarray}}
\def\bi{\begin{itemize}}
\def\ei{\end{itemize}}
\def\ba{\begin{array}}
\def\ea{\end{array}}
\def\bfig{\begin{figure}}
\def\efig{\end{figure}}

\newcommand{\ac}{a}

\newcommand{\overbar}[1]{\mkern 1.5mu\overline{\mkern-1.5mu#1\mkern-1.5mu}\mkern 1.5mu}

\newcommand{\hodge}{{\star}}

\newcommand{\s}{s}
\newcommand{\ssb}{\bar{s}}
\newcommand{\x}{x}
\newcommand{\y}{\bar{x}}
\newcommand{\z}{\varphi}
\newcommand{\zb}{\bar{\varphi}}

\newcommand{\p}{p}
\newcommand{\pp}{\bar{p}}

\newcommand{\po}{o}
\newcommand{\poo}{\bar{o}}

\newcommand{\LL}{J}

\newcommand{\RNum}[1]{\uppercase\expandafter{\romannumeral #1\relax}}

\begin{document}
\title{Degenerate higher-order Maxwell-Einstein theories}

\author{Aimeric Coll\'{e}aux}
\affiliation{Institute of Theoretical Physics, Faculty of Mathematics and  Physics, Charles University,  V Hole\v{s}ovi\v{c}k\'{a}ch  2,  Prague  180  00,  Czech  Republic}
\author{Karim Noui}
\affiliation{Laboratoire de Physique des deux Infinis IJCLab, CNRS/IN2P3, Universit\'e Paris-Saclay, France}

\date{\today}

\begin{abstract}
We classify higher-order Maxwell-Einstein theories linear in the curvature tensor and quadratic in the derivatives of the electromagnetic field strength whose kinetic matrices are degenerate. This provides a generalisation of quadratic degenerate higher-order scalar-tensor theories for a U(1) gauge field. After establishing a classification of the independent Lagrangians, we obtain all the theories with at most third order field equations involving only second order derivatives of the metric, thus generalising Horndeski's quadratic theory for a gauge field. Some of these are shown to be conformally invariant. We then classify degenerate non-minimally coupled interactions, obtaining all conformally invariant ones. Finally, we investigate the effect of U(1)-preserving disformal transformations on these degenerate Lagrangians.  The ``mimetic" singular transformations are obtained and new ghost-free degenerate theories are generated.
\end{abstract}

\maketitle


\section{Introduction}

The classification of four-dimensional metric gravitational theories which modify General Relativity (GR) at small or large scales is an important step towards a better understanding and possible resolution of its inconsistencies or shortcomings, such as singularities and perturbative non-renormalisability or dark matter and dark energy. Probably because all the presently-known fundamental fields have at most second order field equations, the first unicity theorems have mostly restricted to this conservative assumption, in addition to diffeomorphism invariance. In particular, Lovelock famously showed that the unique purely metric theory of this kind is GR and found the unique theory with third order equations of motion \cite{Lovelock1969TheUO}, while Horndeski obtained the most general scalar-tensor theory satisfying these properties \cite{Horndeski1974SecondorderSF}, which has since been generalised to two scalar fields\footnote{Although the most general second order field equations for a four-dimensional bi-scalar-tensor theories was obtained in \cite{Ohashi:2015fma}, the complete form of the Lagrangian is still unknown, see \cite{Horndeski:2024hee}} \cite{Ohashi:2015fma}.

Although allowing for higher-order field equations typically introduces new degrees of freedom, known as Ostrogradski ghosts, which make the Hamiltonian unbounded from below and usually break unitarity at the quantum level, it was shown in \cite{Langlois:2015cwa} that ghost-free higher-order scalar-tensor theories can be obtained provided that their kinetic matrix be degenerate, what followed from the discovery of the so-called Beyond-Horndeski \cite{Zumalacarregui:2013pma, Gleyzes:2014dya} and resulted in the more general degenerate higher-order scalar-tensor theories (DHOST) \cite{Langlois:2015cwa, Langlois:2015skt}, which have been classified up to cubic order in the second derivatives of the scalar field \cite{BenAchour:2016fzp}, see also \cite{Kobayashi:2019hrl}. All these theories, as well as the khronometric \cite{Blas:2009yd, Blas:2010hb} and the spatially covariant gravity with two tensorial degrees of freedom \cite{Gao:2019twq} are part of the more general U-DHOST, which are degenerate in the unitary gauge, i.e. when the scalar field is chosen as time-coordinate \cite{DeFelice:2018ewo}. 
Many generalisations to higher orders \cite{Babichev:2024eoh} or to other field content have been undertaken, such as generalised Proca \cite{Heisenberg:2014rta, Allys:2015sht, Heisenberg:2016eld}.

From a phenomenological point of view, these scalar field models have been used in particular to model an inflationary epoch in the very early universe \cite{Kobayashi:2010cm}, to mimic the effect of dark matter at cosmological scales using e.g. Mimetic Gravity \cite{Chamseddine:2013kea, Sebastiani:2016ras} and to provide an alternative explanation to the accelerated expansion of the universe as that given by the cosmological constant \cite{Armendariz-Picon:2000ulo, Deffayet:2010qz, Langlois:2018dxi}. Their simplicity also make them particularly adapted to include additional symmetries such as conformal \cite{Fernandes:2021dsb, Fernandes:2022zrq, Babichev:2024krm} and disformal ones \cite{Domenech:2023ryc}, or as a laboratory to obtain exact solutions, in particular black holes \cite{Bakopoulos:2023fmv, Lecoeur:2024kwe} and non-singular exotic compact objects \cite{Barrientos:2022avi}.

\medskip

However, it is important to mention that Ostrogradski ghosts are not always problematic, neither at the classical \cite{Pagani:1987ue, Deffayet:2023wdg, Heredia:2024wbu, Smilga:2020elp, ErrastiDiez:2024hfq, Deffayet:2025aa} nor at the quantum level \cite{Donoghue:2021eto, Platania:2020knd, Platania:2022gtt}, and that theories propagating such ghosts can widely differ in the type of instabilities (if any \cite{Deffayet:2021nnt}) they generate, as for example Mimetic gravity which is a conformal invariant DHOST \cite{Chaichian:2014qba, Langlois:2018jdg}. Other interpretations have also been advocated, see e.g. \cite{Babichev:2024uro, Stoilov:1997ek}. Furthermore, the constraints implied by the degeneracy of a theory can sometimes remove healthy degrees of freedom from the spectrum, instead of the problematic ones, as it happens for instance in Class \RNum{3} of DHOST \cite{BenAchour:2016cay}, see also \cite{Crisostomi:2016czh}, or do not remove sufficiently many ghost degrees of freedom \cite{Colleaux:2024ndy}.

Thus, degeneracy is by itself neither necessary nor sufficient to ensure both stability and ghost-freeness, but it has been a fruitful avenue to obtain theories enjoying both of these properties. In particular, classifying degenerate theories might enable to assess if renormalisable gravitational theories which are stable and ghost-free (unlike, say, Stelle gravity \cite{Stelle:1976gc}) exist, what would be quite remarkable. Ghost-free higher-order theories can also be considered as low-energy effective field theories, although ghosts are by construction harmless in this setting, whose degeneracy (which is a type of fine-tuning) would be protected by some mechanism against quantum corrections \cite{Santoni:2018rrx} (such as weakly broken Galilean invariance in scalar-tensor theories \cite{Pirtskhalava:2015nla}).

From a different perspective, the classification of degenerate theories can be viewed as an invariant way to partition the space of theories in terms of the number and type of degrees of freedom generated by the higher-order interactions, as illustrated by the fact that degeneracy classes in scalar-tensor theories are invariant under (invertible) disformal transformation \cite{BenAchour:2016cay, BenAchour:2024hbg}. Moreover, such classification necessitates to write the action in terms of power series of covariant derivatives with generic (possibly non-perturbative) coupling functions, making it in some sense a complementary approach to form-factors and non-local theories\footnote{More precisely, denoting $n$ the number of explicit covariant derivatives and $m$ the number of ``field strengths" (e.g. $d \phi$, $dA$ or the Riemann tensor, depending on the field content), the form-factor approach usually restricts to theories with $n <  \infty$ and $m\to \infty$, while the more standard approach we follow considers $n \to \infty$ and $m  <  \infty$.} \cite{Modesto:2011kw, Kolar:2023gqi}.

\medskip

In this paper, we generalise many aspects of the classifications of four-dimensional degenerate scalar-tensor theories when the scalar field is replaced by a U(1) gauge field, such as the electromagnetic (EM) field. There is a gap in the literature regarding these higher-order Maxwell-Einstein theories (HOME) where derivatives of the Faraday field strength tensor and its non-minimal couplings with the Riemann curvature are taken into account. Indeed, despite Horndeski's unique non-minimally coupled theory (NMC) with second order field equations \cite{Horndeski:1976gi, Buchdahl:1979wi} and the complementary no-go result for vector Galileons on flat spacetime \cite{Deffayet:2013tca} (see also \cite{Heisenberg:2018acv, Mironov:2024idn, Mironov:2024umy} for new scalar-vector-tensor theories with second order field equations), we know now that ghost-free theories can be obtained from the more general condition of degeneracy.

While HOME can be encountered in various context such as quantum effective actions \cite{Drummond:1979pp, Barvinsky:1985an, Daniels:1993yi, Daniels:1995yw, Shore:1995fz, Shore:2002gn, Gusynin:1998bt, Bittar:2024xuc}, (asymptotically safe) quantum gravity \cite{Knorr:2024yiu}, astrophysical phenomenology \cite{Prasanna:2003ix, Chen:2015cpa, Chen:2016hil, Jana:2021lqe, Chen:2023wna, Carballo-Rubio:2025zwz}, Bopp-Podolsky electrodynamics \cite{Podolsky:1942zz, Gratus:2015bea}, or to construct duality invariant theories \cite{Cano:2021tfs, Cano:2021hje} and regular black holes \cite{Cano:2020ezi}, there is, as far as we know, only a few results on degenerate higher-order Maxwell-Einstein theories (DHOME). Furthermore, their study has mostly restricted so far to the use of U(1)-preserving disformal transformation \cite{Goulart:2013laa, DeFelice:2019hxb}. In particular, these transformations have been used as a solution-generating technique in \cite{Minamitsuji:2020jvf, Minamitsuji:2021rtw}, while \cite{Gumrukcuoglu:2019ebp} has investigated the U(1) conformal transformation of GR and \cite{Gorji:2018aa, Hammer:2020aa, Gorji:2020aa} some singular mimetic transformations of this type. Using a disformal metric given by Maxwell stress-energy tensor, a detailed study of solutions and singularities has been given in \cite{Goulart:2020wkq, Bittencourt:2023ikm}. Interestingly, it was recently shown that the optical metrics obtained from flat-space non-linear electrodynamics\footnote{Concerning non-linear electrodynamics, see e.g. \cite{Bandos:2020jsw, Sorokin:2021tge} and the interesting recent results of \cite{Gies:2024xzy}} are able to mimic rotating black holes \cite{Goulart:2024ldm, Goulart:2024ugj}. Similarly, the study of two-form field theories, which is to HOME what generalised Proca is to HOST, has mostly restricted to the massless case, known as Kalb-Ramon gauge fields, although some partial results have been obtained in \cite{Heisenberg:2019akx}.

In a recent series of papers, we started a systematic study of HOME and two-form theories, restricting to quadratic order in the first derivative of the field strength and to linearity in the Riemann tensor, which is the natural setting to generalise quadratic Horndeski and DHOST theories for a gauge field, as well as their generalised Proca counterpart when considering a non-integrable two-form instead of the field strength. In \cite{Colleaux:2023cqu}, we obtained a 21 dimensional basis of independent U(1) Lagrangians and 34 dimensional basis for the kinetic Lagrangians of two-form fields\footnote{Notice that explicit basis for the independent \textit{invariants} in two and three dimensions and for both the U(1) and two-form field cases are also provided in appendix (A)3 of \cite{Colleaux:2023cqu}.}. This enabled us to investigate the fate of Ostrogradski ghosts in flat space-time \cite{Colleaux:2024ndy}, resulting in new no-go results for ghost-free higher-order Maxwell theories. Although there remains in principle some gaps in our Hamiltonian analysis, this strongly indicates that gravity is a necessary ingredient for ghost-freeness in this kind of model. This is also illustrated by the DHOME theory studied in \cite{Gumrukcuoglu:2019ebp} whose ghost-freeness is explicitly lost in flat space. 
 
 \medskip

The main result of this paper is to obtain (quadratic) DHOME with at most third order field equations. Within this subset, while in quadratic HOST, assuming that the field equations are second order in the derivatives of the metric implies the same for the scalar and thus uniquely yields quadratic Horndeski theory \cite{Langlois:2015cwa} (see also \cite{Horndeski:2016bku}), we show that the unique HOME theory involving at most second derivatives of the metric involves in general third order derivatives of the gauge field, which is why it has remained unseen by previous investigations. This theory is parametrised by a couple of free functions, each depending on one of the EM invariants, while GR and Horndeski's NMC are obtained from a combination of these two theories with respectively constant and quadratic functions. The linear ones give a pair of theories which are totally degenerate\footnote{Remark that in scalar-tensor theories, totally degenerate theories, i.e. those with vanishing kinetic matrices, exist only in the unitary gauge \cite{DeFelice:2018ewo}.}, conformal invariant and U(1)-preserving disformal invariant. In the non-integrable case, we obtain all quadratic theories whose field equations are second order in the metric and first order in the two-form field.

The remaining results of this paper are first to classify degenerate non-minimally coupled theories linear in the curvature, resulting in a vast bestiary containing in particular a theory which is both fully degenerate in the electromagnetic sector and conformal invariant. Finally, we investigate the effect of U(1)-preserving disformal transformations on these degenerate Lagrangians.  The analogue of mimetic gravity for a gauge field is obtained from the classification of (non-invertible) U(1) transformations and new ghost-free DHOME theories are generated taking GR and Horndeski's NMC as seeds.

Finally, let us emphasise that, apart from these lasts, the degenerate theories we have obtained are not necessarily ghost-free but a detailed constraint analysis to characterise their degrees of freedom goes beyond the scope of this work.

 \medskip

The paper is organised as follows. In Sec.\ref{HOME}, we present the action \ref{HOME1} and introduce from the onset four-dimensional projectors, built from the Faraday tensor and metric field, which drastically simplify all the subsequent analysis \ref{Sec.EMProj}. In particular, we give a new classification of quadratic HOME Lagrangians, which are shown to split into two sets related by a complex transformation \ref{Classification}. We then perform a 3+1 decomposition of the action adapted to these variables and discuss the conditions under which the kinetic matrix is degenerate \ref{SEC3+1}. In Sec.\ref{Sec.DHOME}, we obtain all the theories  with third order field equations discussed above \ref{Trivially degenerate theories}, as well as the classification of degenerate non-minimally coupled theories \ref{NMCDeg}. Finally, Sec.\ref{SecDisfTransf} is devoted to the study of U(1)-preserving disformal transformations, in particular the mimetic transformations \ref{Sec.Mimetic}, as well as the U(1) frames of General Relativity and Horndeski's unique non-minimal coupling, yielding new ghost-free DHOME \ref{Sec.GhostFree}.

\section{Higher-order Maxwell-Einstein theories}\label{HOME}

In this section, we  give an alternative analysis to that given in \cite{Colleaux:2023cqu} for the classification, in four dimensions, of the invariants built from arbitrary powers of the field strength $F$ which are quadratic in $\nabla F$ or linear in the Riemann curvature. 

\subsection{Action and field equations}\label{HOME1}
We look for the most general action of the form\footnote{This is the straightforward analogue of quadratic higher-order scalar-tensor theories and generalised Proca, given by $ A^{\mu\nu\rho\sigma} R_{\mu\nu\rho\sigma}+B^{\mu\nu,\rho\sigma}\nabla_\mu v_\nu \nabla_\rho v_\sigma$, where in the former case $v=d\phi$ and $\{A, B\}$ are the most general tensors built from the metric, $v$ and $\phi$, while in the latter $v$ is non-integrable and $\{A, B\}$ only depend on $v$ and $g$.}
\begin{eqnarray}
I[g_{\mu\nu}, F_{\mu\nu}] =  \int d^4 x \sqrt{-g}  \left( \mathscr{A}^{\mu\nu\rho\sigma} R_{\mu\nu\rho\sigma}+\mathscr{B}^{\gamma\mu\nu,\delta\rho\sigma} \nabla_\gamma F_{\mu\nu}\nabla_\delta F_{\rho\sigma} \right)\, ,\label{Action}
\end{eqnarray} 
where  $\mathscr{A}$ and $\mathscr{B}$ are tensors constructed from the two-form $F$ and the metric. For later convenience we define the  linear piece in the Riemann tensor as $\cal R$  and $\cal L$ for the quadratic term in $\nabla F$, i.e.
\begin{eqnarray}
\mathcal{R} \; = \;  \mathscr{A}^{\mu\nu\rho\sigma} R_{\mu\nu\rho\sigma} \, , \qquad
\mathcal{L} \; = \;  \mathscr{B}^{\gamma\mu\nu,\delta\rho\sigma} \nabla_\gamma F_{\mu\nu}\nabla_\delta F_{\rho\sigma}   \, . 
\end{eqnarray} 
Although it is always possible to add a potential for the two-form field $V\left(F_{\mu\nu} \right)$, which contains the cosmological constant, or in the case of a gauge field, to consider the Maxwell action and its first order modifications, so-called non-linear electrodynamics, these terms do not influence the possible degeneracy of the kinetic matrix of \eqref{Action}, so we will omit such terms in this paper. However, it is important to notice that adding these to the action would modify the constraint structure and in some cases the number of degrees of freedoms.

\medskip

We will investigate both cases where $F$ is respectly an arbitrary two-form field or the curvature associated with a U(1) potential, $F_{\mu\nu} = \partial_{[\mu} A_{\nu]}$\footnote{In this paper, all the (anti)symmetrisations are taken without normalisation, for example, $X_{[ab]}=X_{ab}-X_{ba}$.}, so that $\nabla_{[\mu}F_{\rho\sigma]}=0$. In that case, the action can be reformulated as follows
\begin{eqnarray}
I_\lambda[g_{\mu\nu}, F_{\mu\nu},\lambda^{\mu\nu\sigma}] = I[g_{\mu\nu}, F_{\mu\nu}] +  \int d^4 x \sqrt{-g}\,  \lambda^{\mu\nu\sigma} \, \nabla_{[\mu} F_{\nu\sigma]} \,,  \label{ActionU1}
\end{eqnarray}
where $\lambda$ is a three-form Lagrange multiplier. It is trivially equivalent to the usual formulation in terms of a gauge potential due to the Bianchi identity of the Riemann tensor $R_{[\mu\nu\rho]\sigma}=0$. Any Bianchi identity of $F$ in the action can then be reabsorbed into a redefinition of the Lagrange multiplier. 

\medskip

Moreover, it follows from the diffeomorphism invariance of the theory that the field equations of the two-form theory satisfy the following off-shell identity\footnote{This identity is satisfied for any diffeomorphism invariant action involving solely a metric and a two-form field. It is the analogue of the off-shell identity for scalar-tensor theories, $\nabla_\sigma \frac{\delta I}{\delta g_{\mu\sigma}}  = \frac{1}{2}  \nabla^\mu \phi \frac{\delta I}{\delta \phi}$, as it can be seen for example in the equation (2.4) of \cite{Ohashi:2015fma}.}
\begin{eqnarray}
g_{\mu\nu} \nabla_\sigma \frac{\delta I}{\delta g_{\nu\sigma}}  = \left( \frac{1}{4}  \nabla_{[\mu} F_{\nu \sigma]} + F_{\mu\nu} \nabla_\sigma \right) \frac{\delta I}{\delta F_{\nu\sigma}}  \, .
\end{eqnarray}
When the U(1) constraint in \eqref{ActionU1} is taken into account, this equation is modified by terms which vanish when the field equations of $\lambda^{\mu\nu\rho}$ are satisfied (i.e. when the Bianchi identity is imposed). Thus, we obtain in this case the off-shell (for $g$ and $A$) identity,
\begin{eqnarray}
g_{\mu\nu} \nabla_\sigma \frac{\delta I_\lambda}{\delta g_{\nu\sigma}} =\frac{1}{2} F_{\mu\nu} \frac{\delta I_\lambda}{\delta A_{\nu}} \, .
\end{eqnarray} 
Therefore, when $\text{det}\left(F_{\mu\nu}\right) \neq 0$, what we will assume without loss of generality\footnote{Indeed, we restrict in this paper to off-shell calculations for which the fields are always assumed algebraically general.} throughout this paper, the field equations for the gauge potential are redundant with the equations for the metric, just like in the case of scalar-tensor theories.  As for the metric field equations, they are generically of the form 
\begin{eqnarray}
 \frac{1}{\sqrt{-g}}\frac{\delta I}{\delta g^{\alpha\beta}} =   \mathscr{N}_{\alpha\beta}^{\mu\nu\rho\sigma} R_{\mu\nu\rho\sigma} + \mathscr{S}_{\alpha\beta}^{\gamma\mu\nu,\delta\rho\sigma} \nabla_\gamma F_{\mu\nu}\nabla_\delta F_{\rho\sigma}  + \mathscr{T}_{\alpha\beta}^{\mu\nu\rho\sigma} \nabla_{(\mu} \nabla_{\nu)} F_{\rho\sigma} =0 \, ,
\end{eqnarray}
where $\mathscr{N}$, $\mathscr{S}$ and $\mathscr{T}$ are tensors which are linear in $\mathscr{A}$ and $\mathscr{B}$ as well as their derivatives with respect to $g$ and $F$. As they are quite complicated and not particularly useful for our analysis, we omit to write  their explicit expressions.

\medskip

In the following, we present an alternative classification of the Lagrangians \eqref{Action} compared to the one given in our recent paper \cite{Colleaux:2023cqu}.
The purpose is to simplify as much as possible the structure of the action, its 3+1 decomposition, as well as its transformation properties under U(1)-preserving disformal transformations, by choosing four-dimensional variables from the onset, similar to those used to derive the non-perturbative one-loop effective action of QED in the presence of slowly-varying electromagnetic backgrounds \cite{Gusynin:1998bt} (see also the more recent \cite{Goulart:2013laa}).

\subsection{Electromagnetic projectors}\label{Sec.EMProj}

Let us first show that in four dimensions, any two-form field $F_{\mu\nu}$ together with the metric induce a covariant $2+2$ splitting of the space-time,  characterised by a pair of orthogonal two-dimensional projectors. Using these to classify tensorial quantities and Lagrangians presents many advantages due to their orthonormality properties, which in our particular case also encode the four-dimensionality of spacetime and enable to treat dimensionally dependent identities (DDIs) \cite{Edgar:2001vv} in a trivial way. In particular, these projectors are by definition invariant under global scale transformation of the two-form (or gauge field), so that only the scalar invariants and explicit covariant derivatives carry physical dimensions in this setting\footnote{This is the straightforward analogue of the covariant $3+1$ splitting induced by a scalar field, see e.g. \cite{Langlois:2021aa}. Indeed, given a scalar field $\phi$ with non-vanishing kinetic term, one can construct respectively the one and three-dimensional projectors $u_\mu = \partial_\mu \phi / \sqrt{ |X |}$ and $\gamma_{\mu\nu} = g_{\mu\nu} - \epsilon u_\mu u_\nu$, where $X=\partial_\gamma \phi \partial^\gamma \phi \neq 0$ and $\epsilon = \text{sign}\left(X\right)$. They are both by construction invariant under the global scale transformations $\phi \to \lambda \phi$, so that the physical dimension of $\phi$ only appears in a Lagrangian through $X$ (and through undifferentiated appearances of $\phi$, e.g. in a potential).}.

\subsubsection{Construction of the projectors}
Given a two-form field $F_{\mu\nu}$ in a four-dimensional space-time, two scalar invariants can be constructed as follows,
\begin{eqnarray}
\mathcal{F}= -\frac{1}{4} F^{\mu\nu}F_{\mu\nu} \,,\;\;\;\;\; \mathcal{G}= \frac{1}{4}\hodge F^{\mu\nu} F_{\mu\nu}  \,, \label{EMinv}
\end{eqnarray}
where $\hodge$ is the Hodge dual defined by 
\begin{eqnarray}
\hodge F_{\mu\nu} = \frac{1}{2} \varepsilon_{\mu\nu\rho\sigma}F^{\rho\sigma} \, .
\end{eqnarray} 
However, we will see that it is more convenient for our purpose to introduce two alternative scalars $\x$ and $\y$, which we will commonly denote as $z_a$ for $a \in \{+,-\}$ ($z^+ =\x$ and $z^-=\y$), defined through
\begin{eqnarray}
\mathcal{F} =\frac{1}{2} \left(\x^2 - \y^2\right)  \,,\;\;\;\;\; \mathcal{G}= - \x \y\,, \label{EMinvxy}
\end{eqnarray}
and corresponding to the real eigenvalues of the field strength $F$, see \cite{Batalin1971PHOTONGF}. We will always assume algebraically general configurations of the electromagnetic field\footnote{This is the analogue of the non-vanishing of the scalar field kinetic term mentioned in the previous footnote, see also the footnote 9.}, so that $\x \neq 0$ and $\y \neq 0$. 

 In order to classify the tensors $\mathscr{A}$ and $\mathscr{B}$ appearing in the action \eqref{Action}, which are arbitrary functions of the metric and field strength tensors, we proceed as follows. We start by showing (in appendix \ref{PropProj}) that the Cayley-Hamilton identity in four-dimensions implies that the basis of independent matrices constructed from $g$ and $F$ is also four-dimensional and can be chosen as $\{g, F,F_2,\hodge F \}$, where 
 \begin{eqnarray}
 F_2^{\mu\nu} \; = (F^2)^{\mu\nu}\; = \; F^{\mu\sigma}F_{\sigma}{}^\nu \, . 
 \end{eqnarray}
We can construct an alternative parametrisation based on symmetric and antisymmetric space-time projectors, similar to the decomposition of the field strength into bivectors \cite{Santos:2021ypa,CordeirodosSantos:2016eaj}.

Starting with symmetric projectors $s$, i.e. satisfying $s^2 = s$, we can show using \eqref{CH} that besides the metric field, there are only two solutions given by
\begin{eqnarray}
\p^{\mu\nu} =\left( \frac{\y^2}{\x^2+\y^2} \right)g^{\mu\nu} +\left(  \frac{1}{\x^2+\y^2} \right)F_2^{\mu\nu}   \,,\;\;\;\;\;  \pp^{\mu\nu} =\left( \frac{\x^2}{\x^2+\y^2} \right)g^{\mu\nu} -\left(   \frac{1}{\x^2+\y^2} \right) F_2^{\mu\nu} \,. \label{proj}
\end{eqnarray}
Denoting these two by $\p^a$, for $a=1,2$, it can be shown that they satisfy the relations
\begin{eqnarray}
\p^a_{\mu\alpha}\, \p_{b}^{\alpha\nu}=\delta^a_b \,  \p^{a\nu}_\mu \,,\;\;\;\;\; \p^a_{\mu\nu} g^{\mu\nu}=2 \,,  \label{projdef}
\end{eqnarray}
meaning that they are orthogonal to each other and project any tensors on two-dimensional surfaces, that we denote $\mathscr{M}$ and $\mathscr{S}$ for $\p$ and $\pp$ respectively.  Interestingly, as shown in appendix \ref{App:tetrads}, $\mathscr{S}$ is a space-like surface for any (algebraically general) configuration of the electromagnetic field while $\mathscr{M}$ is a space-time\footnote{This is different from the scalar-tensor case for which the (space-like or time-like) characters of the surfaces defined by the projectors depend on the sign of the kinetic term of the scalar, which is given by the dynamics of the theory.}.  Inverting \eqref{proj} leads to
\begin{eqnarray}
g_{\mu\nu} = \p_{\mu\nu}+\pp_{\mu\nu} \,,\;\;\;\;\; F_2^{\mu\nu} = \x^2 \p^{\mu\nu}- \y^2 \pp^{\mu\nu} \,. \label{gF2}
\end{eqnarray}

Similarly we define the antisymmetric projectors as square-roots of these projectors, i.e. as solutions of the equations 
\begin{eqnarray}
\po_{\mu\alpha} \po^{\alpha}{}_\nu = \p_{\mu\nu} \,,\;\;\;\;\; \poo_{\mu\alpha} \poo^{\alpha}{}_\nu =- \pp_{\mu\nu} \,. \label{asprojdef}
\end{eqnarray}
This yields up to unimportant overall signs,
\begin{eqnarray}
\po_{\mu\nu} =\left( \frac{\x}{\x^2+\y^2} \right)F_{\mu\nu} +\left(   \frac{\y}{\x^2+\y^2} \right)\hodge F_{\mu\nu}  \,,\;\;\;  \poo_{\mu\nu} =\left( \frac{\y}{\x^2+\y^2} \right)F_{\mu\nu} -\left(  \frac{\x}{\x^2+\y^2} \right)\hodge F_{\mu\nu}, \label{asproj}
\end{eqnarray}
or alternatively
\begin{eqnarray}
F_{\mu\nu} = \x \po_{\mu\nu} + \y \poo_{\mu\nu} \,,\;\;\;\;\; \hodge F_{\mu\nu} =\y \po_{\mu\nu} - \x \poo_{\mu\nu} \,.
\end{eqnarray}
Notice that they are orthogonal to each other, related by duality and can be used to construct the Levi-Civita tensor as follows 
\begin{eqnarray}
\hodge \po = - \poo \,,\;\;\;\;\; \hodge \poo =  \po  \,,\;\;\;\;\; \varepsilon_{\mu\nu\rho\sigma}= \frac{1}{4} \po_{[\mu\nu} \poo_{\rho\sigma]} \,. \label{duality}
\end{eqnarray}
In the sequel, we will use the following notation to denote  the four projectors
\begin{eqnarray}
q_I \in  \{\p, \po, \pp, \poo\} \,,\label{quartet}
\end{eqnarray}
where $I \in \{0,1,\bar{0} , \bar{1}\}$, as well as the following notation for the indices: $i,j \in \{0,1\}$ and $\bar{i},\bar{j} \in \{\bar{0},\bar{1}\}$.

\subsubsection{Complex transformation}

The previous decomposition into projectors naturally leads to consider a ``complex'' transformation  or a ``rotation'' which interchanges the two-dimensional hypersurfaces $\mathscr{S}$ and $\mathscr{M}$ and will play an important role in the classification of HOME theories and degenerate ones. The transformation acts on the projectors and on the scalars $\x,\y$ according to, 
\begin{eqnarray}
\p \leftrightarrow  \pp \,,\;\;\;\; \x \leftrightarrow \y  \,,\;\;\;\; \po \to i \poo  \,,\;\;\;\; \poo \to i \po \,. \label{bartransf}
\end{eqnarray}
In order for the transformation to be an involution, we assume that it extends to  any complex number as follows,
\begin{eqnarray}
\lambda \in \mathbb C \, \to \, \bar{\lambda} \, , 
\end{eqnarray}
where now the bar denotes the usual complex conjugation.
Hence, this transformation extends to any product of sums of these projectors (exactly as an algebra morphism), and we will denote it  by an overall ``bar"\footnote{Thus, we have for instance $\overbar{\left(\po\right)} = i \poo$ and $\overbar{\left(\poo\right)} = i \po$, so that considering functionals of these objects $F\left(\po, \poo, \dots \right)$, we have $\bar{F}=F\left(i \poo , i \po, \dots \right)$.  In the following, all quantities with a bar should be understood as such functionals, unless their specific transformation laws are given.}. Notice that the properties  \eqref{projdef} and \eqref{asprojdef}  of the projectors, as well as  the decomposition of the metric field in terms of  $\p$ and $\pp$ given by the first equation of \eqref{gF2}, are left invariant by this transformation.

When applied to the Levi-Civita tensor, the  field strength  $F$ and its dual $\hodge F$, we obtain
\begin{eqnarray}
\varepsilon \to - \varepsilon \,,\qquad F \to i F  \,,\qquad \hodge F \to -i \hodge F \label{bartransfF} \, .
\end{eqnarray} 
As a consequence, the transformation of any invariant which vanishes due to the Bianchi identity is itself vanishing. We remark that it acts on the Levi-Civita tensor as a parity transformation.

\subsection{Classification of higher-order electromagnetic invariants}\label{Classification}

Using the previous decompositions of $F$ and $g$, it is clear that the tensors $\mathscr{A}$ and $\mathscr{B}$ appearing in the action \eqref{Action} can be decomposed schematically as follows,
\begin{eqnarray}
\mathscr{A} = \sum_{I,J} \alpha_{IJ}\left(\x,\y\right) q_I q_J \,,\;\;\;\;\;  \mathscr{B} = \sum_{I,J,K} \beta_{IJK}\left(\x,\y\right) q_I q_J q_K \,. \label{invariantAB}
\end{eqnarray}
with the notations \eqref{quartet} for the projectors.

\subsubsection{Riemann invariants}
Let us start by classifying the tensors $\mathscr{A}$, i.e. the terms which are linear in the Riemann  tensor.
Taking into account the Bianchi identity $R_{[\mu\nu\rho]\sigma}=0$ and using the tetrad representation of the projectors \eqref{ProjTetrad} to trivialise the treatment of DDIs, 
it is straightforward to show that we recover the four independent Riemann terms, as in \cite{Colleaux:2023cqu}, which can be chosen as follows,
\begin{eqnarray}
\begin{split}
\mathcal{R}_0&= \p^{\mu\nu} \pp^{\rho\sigma} R_{\mu\rho\nu\sigma}\,, \\
\mathcal{R}_1 &=  \po^{\mu\nu} \poo^{\rho\sigma} R_{\mu\rho\nu\sigma} \,,\\
\mathcal{R}_2 &= \p^{\mu\nu} \p^{\rho\sigma}   R_{\mu\rho\nu\sigma} \propto   \po^{\mu\nu} \po^{\rho\sigma} R_{\mu\rho\nu\sigma} \,,\\
\mathcal{R}_3 &= \pp^{\mu\nu} \pp^{\rho\sigma}   R_{\mu\rho\nu\sigma}  \propto  \poo^{\mu\nu} \poo^{\rho\sigma}  R_{\mu\rho\nu\sigma}  \,.
\end{split}\label{RiemBasis}
\end{eqnarray}
We used the notation  $\propto$ for a proportionality up to DDIs. Notice that under a rotation \eqref{bartransf}, $\mathcal{R}_0$ is invariant and $\mathcal{R}_1 \to -\mathcal{R}_1$, while $\mathcal{R}_3=\bar{\mathcal{R}}_2$.

\medskip

Interestingly, one can show that two of these invariants are related to Weyl scalars due to the relations,
\begin{eqnarray}
 \po^{\mu\nu} \poo^{\rho\sigma} C_{\mu\rho\nu\sigma} = \mathcal{R}_1 \,,\;\;\;   \p^{\mu\nu} \pp^{\rho\sigma} C_{\mu\rho\nu\sigma} = \frac{1}{3} \left( \mathcal{R}_0 - \mathcal{R}_2 - \mathcal{R}_3\right)\,, \label{RiemWeyl}
\end{eqnarray}
where  the four-dimensional Weyl tensor is given by,
\begin{eqnarray}
C_{\mu\nu}{}^{\rho\sigma}= R_{\mu\nu}{}^{\rho\sigma} - \frac{1}{2} \delta_{[\mu}^{[\rho}R_{\nu]}^{\sigma]} + \frac{1}{6} \delta_{[\mu}^{\rho}\delta_{\nu]}^{\sigma} R \,, \label{WeylDef}
\end{eqnarray}
while the remaining two can be related to the Ricci scalar and Horndeski's non-minimal coupling as follows 
\begin{eqnarray}
R=  2 \mathcal{R}_0+\mathcal{R}_2+\mathcal{R}_3   \,,\;\;\;\;\; \hodge F^{\mu\nu} \hodge F^{\rho\sigma}  R_{\mu\nu\rho\sigma}= 2 \left( \x^2 \mathcal{R}_3 -2  \x \y \mathcal{R}_1 -\y^2 \mathcal{R}_2 \right) \,. \label{RicciNPCProj}
\end{eqnarray}

\subsubsection{Quadratic invariants}\label{SecQuadInv}

Now, we focus on invariants quadratic in derivatives of the field strength, i.e. we classify the tensors $\mathscr{B}$. As shown in appendix \ref{Classification of higher-order Lagrangian}, it is sufficient to consider terms quadratic in the derivatives $\nabla o$ instead of the derivatives of the four projectors $\nabla q_I$. However, it appears to be more convenient to parametrise these $ \mathscr{B}$ terms in terms of  $\{ \nabla \x , \nabla \y , \nabla \po, \nabla \poo \}$, because it enables to divide the set of invariants into two  classes related by the ``rotation" transformation introduced in section \ref{Sec.EMProj}. Hence, we want to classify quadratic invariants which are linear combinations of the following  terms,
\begin{eqnarray}
 q_I^{\mu\nu} \nabla_\mu z_a \nabla_\nu z_b  \,,\;\;\;\;\; 
\left( q_I q_J \right)^{\mu\nu\rho\sigma} \nabla_\mu \po_{\nu\rho} \nabla_\sigma z_a \,,\;\;\;\;\; 
 \left( q_I q_J q_K \right)^{\mu\nu\rho\sigma\alpha\beta} \nabla_\mu \po_{\nu\rho} \nabla_\sigma \po_{\alpha\beta} \,,
\end{eqnarray}
where the parenthesis notation refers to all the possible configurations of the indices. We recall that $z_a \in \{\x,\y\}$ with $a\in \{+,-\}$. These terms clearly fall into three different classes.

 The first class of invariants is $(4\times2)$-dimensional and we consider a basis defined by the following invariants 
\begin{eqnarray}
\mathcal{Q}_a =  \pp^{\mu\nu}\nabla_\mu z_a \nabla_\nu z_a \,, \qquad
\mathcal{Q}_{\bar{i}}= q_{\bar{i}}^{\mu\nu}  \nabla_\mu \x \nabla_\nu \y \,,  \label{FirstClassInv}
\end{eqnarray}
together with their rotations $\bar{\mathcal{Q}}_a$ and $\bar{\mathcal{Q}}_{\bar{i}}$. They correspond to the elements of the class $\tilde{\mathcal{B}}$ in the classification \cite{Colleaux:2023cqu}. The second class is found to be $(8\times2)$-dimensional and using \eqref{DprojEq}, we obtain a basis defined by the invariants,
\begin{eqnarray}
\mathcal{Q}^a_{i\bar{j}} = q^{\mu\nu}_i q^{\rho\sigma}_{\bar{j}} \nabla_\mu \po_{\nu\rho} \nabla_\sigma z_a \,, \label{SecondClassInv}
\end{eqnarray}
together with their rotations $\bar{\mathcal{Q}}^a_{i\bar{j}}$. They correspond to the irreducible subset of the class $\hat{\mathcal{B}}$ appearing in \cite{Colleaux:2023cqu}.  Finally, using again \eqref{DprojEq}, we can show that the invariants of the third class can be decomposed into the following set of $12\times2$ terms, 
\begin{eqnarray}
\begin{split}
\mathcal{Q}^{I}_1 &=   s_I^{\mu\nu\sigma\alpha\rho\beta}  \nabla_\mu \po_{\nu\rho} \nabla_\sigma \po_{\alpha\beta} \,,\\
\mathcal{Q}^{I}_2 &=   s_I^{\mu\alpha\nu\sigma\rho\beta}  \nabla_\mu \po_{\nu\rho} \nabla_\sigma \po_{\alpha\beta} \,,\\
\mathcal{Q}^{I}_3 &=   t_I^{\mu\sigma\nu\alpha\rho\beta}  \nabla_\mu \po_{\nu\rho} \nabla_\sigma \po_{\alpha\beta} \,,\\
\end{split} \label{ThirdClassInv}
\end{eqnarray}
and their rotations, where $s_I$ and $t_I$ belong to,
\begin{eqnarray}
\begin{split} 
s_I& \in \{ \p \otimes \p\otimes \pp, \po \otimes \p \otimes \pp,\po\otimes\po\otimes\pp,\p\otimes\po\otimes\poo \}\,,\\
t_I & \in \{ \p\otimes\p\otimes\pp, \po\otimes\p\otimes\poo,\po\otimes\po\otimes\pp,\p\otimes\po\otimes\poo\}\,,
\end{split}
\end{eqnarray}
with $I \in \{0,1,\bar{0} , \bar{1}\}$. The invariants  $\mathcal{Q}^I_1$, $\mathcal{Q}^I_2$ and $\mathcal{Q}^I_3$ (together with their rotations) correspond respectively to the invariants in the classes $\mathcal{B}$, $\mathcal{A}$ and $\tilde{\mathcal{A}}$ in the classification \cite{Colleaux:2023cqu}. We show in appendix \eqref{6DDIs} that there are $6\times2$ left-over DDIs relating the invariants of this class.

\medskip

To summarise, a basis for the quadratic invariants in the derivatives of the two-form $F$ is $(18\times2)$-dimensional and obtained from the following classes,
\begin{eqnarray}
\begin{split}
&\text{First Class :} \;\;\;\;\;\;\;  \{  \mathcal{Q}_a ,\mathcal{Q}_{\bar{i}} \}\,, \;\;\,\;\;\;\;\;\; \{   \bar{\mathcal{Q}}_a ,\bar{\mathcal{Q}}_{\bar{i}} \}
\\
&\text{Second Class :} \quad  \{\mathcal{Q}^a_{i\bar{j}} \} \, , \;\;\;\;\;\;\;\;\;\;\;\,\,\,\, \{\bar{\mathcal{Q}}^a_{i\bar{j}} \}
\\
&\text{Third Class :}  \;\;\quad   \{\mathcal{Q}^{I}_1,\mathcal{Q}^{I}_2 , \mathcal{Q}^{I}_3   \} \, ,\;\;  \{ \bar{\mathcal{Q}}^{I}_1,\bar{\mathcal{Q}}^{I}_2 , \bar{\mathcal{Q}}^{I}_3   \}  \,,
\end{split}
\end{eqnarray}
where $6\times2$ DDIs \eqref{6DDIs} have to be taken into account in the third class.

It is interesting to notice that these invariants can be divided into two families depending on whether they involve an odd or an even total number of projectors $\po$ and $\poo$, or equivalently an odd and even number of field strength $F$.  In particular, there are only $7 \times 2$ odd terms given by 
\begin{eqnarray}
\mathcal{Q}_{\bar{1}}\,, \;\;\;\;  \mathcal{Q}^+_{0\bar{0}} \,, \;\;\;\;  \mathcal{Q}^-_{1\bar{1}} \,, \;\;\;\; \mathcal{Q}^+_{1\bar{1}} \,, \;\;\;\; \mathcal{Q}^-_{0\bar{0}} \,, \;\;\;\; \mathcal{Q}^{1}_1 \,, \;\;\;\; \mathcal{Q}^{1}_2 \,, \label{oddterms}
\end{eqnarray}
together with their conjugates. As the latter are purely imaginary, it is intended in the following that their corresponding couplings are also purely imaginary in order to have a real Lagrangian density. 

\subsubsection{Basis of invariants}
We have now all the ingredients to complete the construction of a basis for the invariants \eqref{invariantAB}.  However, it is quite difficult to know a priori which choice of representatives would be more adapted to our purpose\footnote{For a concrete illustration of this, we refer the interested reader to appendix \ref{AppLstar}.}. For this reason, we write the  two terms of the action \eqref{Action}  as the following linear combination of $(4+8+6+6)\times 2$ invariants\footnote{They correspond to $4\times 2$ terms in the first class, $8\times 2$ terms in the second class, $6\times 2$ independent terms in the third class together with the $6\times 2$ DDIs.}, 
\begin{eqnarray}
\mathcal{L} = \mathscr{L}+\bar{\mathscr{L}} \,,\;\;\;\;\;\; \mathscr{L}= \sum_a \beta_a \mathcal{Q}_a + \sum_{\bar{i}} \beta_{\bar{i}} \, \mathcal{Q}_{\bar{i}} + \sum_{a,i,{\bar{j}}} \beta^a_{i{\bar{j}}} \, \mathcal{Q}^a_{i\bar{j}} + \sum_{n,I} \beta_n^I  \, \mathcal{Q}^{I}_n  \,, \label{Lag48}
\end{eqnarray} 
 plus a linear combination of the 4 curvature invariants,
\begin{eqnarray}
\mathcal{R} = \sum_I \alpha_I \,  \mathcal{R}_I \,.
\end{eqnarray}
In the expression of $\mathscr{L}$ \eqref{Lag48}, the last sum runs over  $n\in \{1,2,3\}$. Let us also precise that the couplings are functions $\beta\left( \x ,\y\right)$ and $\alpha_I\left( \x ,\y\right)$.

Finally, in order to obtain a complete classification of the independent higher-order invariants, we need to consider both the U(1) symmetry via the Bianchi identity and the possible boundary terms. All this has been analysed in the  appendix \ref{Classification of higher-order Lagrangian}, where we have recovered that the basis for quadratic-higher-order Maxwell-Einstein theories is $21-$dimensional which corresponds to $4$ Riemann terms, $7\times 2$ even quadratic invariants and $3$ odd quadratic invariants.

\subsection{3+1 decomposition of the action}
\label{SEC3+1}

In this section, we perform a (manifestly covariant) 3+1 decomposition of the action. We start by decomposing the building blocs appearing in the higher-order invariants in a way which is adapted to the transformation \eqref{bartransf}. Introducing the necessary boundary terms, we then decompose the action into kinetic and potential parts. Finally, we focus on the kinetic Lagrangian and write the degeneracy equations, i.e. the conditions for which the kinetic matrix  possesses null eigenvalues.

\subsubsection{Decomposition of the projectors}

In order to make the 3+1 decomposition, we introduce  a normalised time-like (i.e. $g^{\mu\nu} n_\mu n_\nu = -1$) and surface-forming vector field $n$ in term of which we can construct the projector $h_{\mu\nu}$ to the spatial hypersurfaces $\Sigma$ orthogonal to $n$ as follows
\begin{eqnarray}
\label{hmetric}
h_{\mu\nu} \; = \; g_{\mu\nu} + n_\mu n_\nu \, .
\end{eqnarray}
It is convenient to introduce the acceleration of the normal and the extrinsic curvature of the hypersurfaces $\Sigma$ which are respectively defined by 
\begin{eqnarray}
\ac_\mu =- n^\sigma \nabla_\sigma n_\mu \, , \qquad K_{\mu\nu} = h_\mu^\rho h_\nu^\sigma \nabla_\rho n_\sigma \, ,
\end{eqnarray}
with $K_{[\mu\nu]}=0$. We can also compute the curvature tensor of $\Sigma$ and the covariant derivative compatible with $h$ as follows,
\begin{eqnarray}
R^{(3)}_{\mu\nu\rho\sigma} = h_\mu^\gamma \, h_\nu^\delta \, h_\rho^\alpha \, h_\sigma^\beta \, R_{\gamma\delta\alpha\beta} - K_{\mu [\rho} K_{\sigma] \nu}  \,,\;\;\;\;\;\; D_\mu s_\nu =h_{\mu}^\rho h_\nu^\sigma \nabla_\rho s_\sigma \, ,
\end{eqnarray}
where $s$ is any spatial vector fields. The expression of the covariant derivative can be generalised in a straightforward way to higher rank tensors.

To preserve the structure of the Lagrangian \eqref{Lag48}, it is convenient to introduce the electric field $e$ and the magnetic field $b$ 
associated with the antisymmetric projector $\po$ as follows,
\begin{eqnarray}
\po_{\mu\nu} = \psi \,  \varepsilon_{\mu\nu\sigma} b^\sigma- \bar{\psi}  \, n_{[\mu} e_{\nu]} \,, \label{Projo}
\end{eqnarray}
where $\varepsilon_{\rho\sigma\mu} = n^\nu \varepsilon_{\nu\rho\sigma\mu}$ is the 3-dimensional Levi-Civita tensor, and $(\psi,\bar{\psi})$ is a couple of scalar fields. The two vector fields $e$ and $b$ are orthonormal and space-like, their relation with the electric and magnetic fields associated with $F$ is given by \eqref{EeBb}  and many other properties can be found in appendix \ref{App:tetrads}. 

In order to set up an orthonormal spatial triad, we introduce the Poynting vector $r$ associated to $e$ and $b$. Its definition, and some of its properties  
(its relations with the spatial metric and Levi-Civita tensors) are given by, 
\begin{eqnarray}
r_\sigma  =\varepsilon_{\mu \sigma \nu} b^\mu e^\nu   \,,\;\;\;\;\;  \varepsilon_{\mu\nu\sigma} = r_{[\mu} e_\nu b_{\sigma]}    \,,\;\;\;\;\;  h_{\mu\nu} = e_\mu e_\nu + b_\mu b_\nu +r_\mu r_\nu \,.
\end{eqnarray}
The two scalars $\psi$ and $\bar{\psi}$ are not independent and satisfy the relation,    
\begin{eqnarray}
 \bar{\psi}^2 - \psi^2 = 1 \, , \label{psi}
\end{eqnarray} 
what can be seen from the two-dimensional nature of $\po$ and its relation to the two-dimensional projector $\p$. Details can be found in appendix \ref{App:tetrads}. 

Finally, the quantities we have just introduced transform under the complex conjugation as
\begin{eqnarray}
n \to n \,, \;\;\;\;\;\; e \leftrightarrow b \,,  \;\;\;\;\;\; r \to r \, , \;\;\;\;\;\; \psi \to - i \bar{\psi} \,, \;\;\;\;\;\; \bar{\psi} \to - i \psi  \,.
\end{eqnarray}

\subsubsection{Elements of the kinetic matrix}
As we have constructed a triad, we can decompose the 6 extrinsic curvature components in the corresponding basis and obtain 
 the extrinsic curvature scalars $K_A$ for $A\in\{1,\dots, 6\}$, 
\begin{eqnarray}
\begin{split}
K_1 &= e^\mu e^\nu K_{\mu\nu} \,,\;\;\;  K_2 = e^\mu b^\nu K_{\mu\nu}   \,,\;\;\;  K_3 = b^\mu r^\nu K_{\mu\nu}  \,,\\
 K_4 &= b^\mu b^\nu K_{\mu\nu} \,,\;\;\;  K_5 = r^\mu r^\nu K_{\mu\nu} \,,\;\;\;  K_6 = e^\mu r^\nu K_{\mu\nu} \, . \label{veloK}
\end{split}
\end{eqnarray}
Considering such projections on the triad instead of the tensor itself is useful to avoid considering DDIs. Finally, it will be useful to compute the action of the complex conjugation on these scalars which is given by,
\begin{eqnarray}
 \bar{K}_1 = K_4 \,,  \;\;\;\;\;\;\;\;\;\;  \bar{K}_3 = K_6 \,,  \;\;\;\;\;\;\;\;\;\; \bar{K}_2 =  K_2 \, ,  \;\;\;\;\;\;\;\;\;\; \bar{K}_5 =  K_5 \, .
\end{eqnarray}

We can proceed in a similar way with the ``velocities" of the strength field $\nabla F$ (or equivalently the accelerations of the gauge potential $A_\mu$) which can be fully parametrised in terms of the derivatives of the projector $\po$ and  the two invariants $\{\x,\y\}$. Hence,  using the relation \eqref{psi} together with the orthonormality of the frame $\{e,b,r\}$, we can construct the six independent components of these velocities $\nabla F$, denoted $\LL_A$ with $A\in\{1,\dots, 6\}$ and defined by\footnote{When expressed in terms of the modified velocities $\LL_5 - \bar{\psi} r^\mu a_\mu$, $\LL_3 + \bar{\psi} \psi^{-1} b^\mu a_\mu$ and  $\LL_6 + \psi \bar{\psi}^{-1} e^\mu a_\mu$, while the others are unchanged, the 3+1 decomposition of $\nabla \po$ does not depend on the acceleration of the normal anymore, which means that in the ADM formalism, $\nabla \po$ does not depend on the spatial derivatives of the lapse function.}
\begin{eqnarray}
\begin{split}
\LL_1 = n^\mu \nabla_\mu \x \,, \;\;\;\;\;\; \LL_2 &=e^\nu  n^\mu \nabla_\mu b_\nu \,, \;\;\;\;\;\, \LL_3 = r^\nu  n^\mu \nabla_\mu b_\nu  \,, \\
\LL_4 = n^\mu \nabla_\mu \y  \,, \;\;\;\;\;\;  \LL_5 &=n^\mu \nabla_\mu \psi \,,  \;\;\;\;\;\;\;\;\;\;  \LL_6 =r^\nu  n^\mu \nabla_\mu e_\nu \, .
 \end{split}
\end{eqnarray} 
The complex conjugation  acts on these components as follows,
\begin{eqnarray}
 \bar{J}_1= J_4  \,,   \;\;\;\;\;\;\;\;\;\; \bar{J}_3=J_6  \, ,  \;\;\;\;\;\;\;\;\;\; \bar{J}_2=- J_2  \, ,\;\;\;\;\;\;\;\;\;\; \bar{J}_5= - i \frac{\psi}{\bar{\psi}} J_5 \, .
\end{eqnarray}

So far, we have assumed that $F$ is an arbitrary two-form. When the U(1) symmetry is taken into account, we get 4 scalar Bianchi identities which are obtained by projection of the usual tensorial Bianchi identity in the space-time basis formed by the triad and the normal $n$. Three out of these four identities are given by,
\begin{eqnarray}
\begin{split}
 \x \psi \left( K_3  - \LL_3  \right)+ \y \bar{\psi} \left( \bar{K}_3 - \bar{\LL}_3  \right)\approx & \; 0 \, ,\\
 \x \LL_5+\psi \LL_1 + \x \psi \left( K_1 + K_5    \right) -   \y \bar{\psi} \left( K_2 + \LL_2 \right) \approx &\; 0 \, ,
\end{split}\label{Bianchiscalar}
\end{eqnarray}
together with the conjugate of the last one. For simplicity, we are using the symbol $\approx$ which means an equality modulo spatial derivatives. This reduces the number of independent accelerations of the gauge field from 6 to 3. In terms of the velocities of the electric and magnetic fields associated with the Faraday tensor $F$, this means that the velocities of the magnetic field can be expressed in terms of spatial derivatives, as it is well known. Hence, we can choose  3 independent components  for the acceleration of the gauge field and, for convenience, we consider the set $\{ \LL_a , \LL \}$ with
\begin{eqnarray}
\LL= \x \bar{\psi} \bar{\LL}_3 - \y \psi \LL_3 \,,\;\;\;\;\;\;\LL_a = n^\mu \nabla_\mu z_a \,, \label{veloJ}
\end{eqnarray}
the former being chosen such that $\bar{\LL}\propto \LL$. The last Bianchi identity is similar to the vanishing of the divergence of the magnetic field and does not involve any components of the acceleration of the gauge field, thus it is irrelevant for our purposes.

\subsubsection{Total action and boundary terms}

We turn to the 3+1 decomposition of the action  \eqref{Action}. As it is usually the case with metric gravity, this requires to add boundary terms to compensate the second derivatives of the metric appearing in the curvature invariants\footnote{In order to simplify the calculations, it is convenient to introduce  the identity in the space of tensors with the symmetries of the Riemann tensor, denoted  $I^{\mu\nu\rho\sigma}_{\xi\zeta\gamma\delta}$ and defined explicitly by 
 \begin{eqnarray}
 I^{\mu\nu\rho\sigma}_{\xi\zeta\gamma\delta}=\frac{1}{8} \left( \delta^{\mu}_{[\xi} \delta^\nu_{\zeta]} \delta^\rho_{[\gamma} \delta^{\sigma}_{\delta]}+\delta^{\mu}_{[\gamma} \delta^\nu_{\delta]} \delta^\rho_{[\xi} \delta^{\sigma}_{\zeta]} \right) \, ,
 \end{eqnarray}
 so that  
\begin{eqnarray}
\mathscr{A}^{\mu\nu\rho\sigma} = I^{\mu\nu\rho\sigma}_{\xi\zeta\gamma\delta} \left( \alpha_0 \left( \x,\y\right) \p^{\xi\gamma} \pp^{\zeta\delta} + \alpha_1\left( \x,\y\right) \po^{\xi\gamma} \po^{\zeta\delta} + \alpha_2\left( \x,\y\right) \p^{\xi\gamma} \p^{\zeta\delta} + \alpha_3\left( \x,\y\right) \pp^{\xi\gamma} \pp^{\zeta\delta} \right) \, . \label{mathscrA}
\end{eqnarray}},
\begin{eqnarray}
\mathcal{R} = \sum_{I=0}^3 \alpha_I \mathcal{R}_I =  \mathscr{A}^{\mu\nu\rho\sigma} R_{\mu\nu\rho\sigma} \, ,
\end{eqnarray}
where $\mathcal{R}_I$ are given by \eqref{RiemBasis}.  Using the Gauss-Codazzi decomposition of the Riemann tensor, we obtain the following boundary terms, 
\begin{eqnarray}
I_{\rm{bound}}[n_{\mu}, g_{\mu\nu}, F_{\mu\nu}]= \int d^4 x \sqrt{-g}  \,\nabla_\gamma  \left( n^\gamma K^{\mu\nu} \mathscr{A}_{\mu\nu}    +    K^{\mu\nu} \mathscr{A}^{\gamma}{}_{\mu\nu} + \mathscr{A}^{\mu\gamma}  a_\mu \right)\, ,
\end{eqnarray}
where we have introduced the  spatial tensors
 \begin{eqnarray}
 \mathscr{A}_{\sigma\nu\rho} =  n^\mu  h_\sigma^\beta h_{(\nu}^\gamma  h_{\rho)}^\alpha \,  \mathscr{A}_{\mu\gamma\alpha\beta} \,,\;\;\;\;\mathscr{A}_{\mu\nu} =n^\rho n^\sigma \mathscr{A}_{\mu\rho\nu\sigma}  \, .
 \end{eqnarray}
 As the first divergence enables to put the action into the required kinetic form, it might yield the correct boundary integral required to ensure a well-defined Dirichlet variational principle for the metric field, once Stokes theorem is used and once the normal to $\Sigma$ is extended to that of the complete boundary. The effect of the second one is to cancel the spatial divergence of the extrinsic curvature in the action, enabling to express, if possible, the velocities in terms of momenta. Although it is vanishing for General Relativity, it has to be taken into account for higher order theories. Finally, the last divergence enables to render the action polynomial in the acceleration of the normal.  
 
 As a consequence, we can now consider the total action
 \begin{eqnarray}
I_{\rm{tot}} = I+ I_{\rm{bound}} = I_{\rm{kin}} + I_{\rm{lin}} + I_{\rm{pot}} \, ,
\end{eqnarray}
 which decomposes as a sum of a kinetic Lagrangian $I_{\rm{kin}} $ (quadratic in the velocities), a term linear in the velocities $ I_{\rm{lin}}$ and a potential term $ I_{\rm{pot}}$. Using the components of the velocities we have constructed in the previous subsection, \eqref{veloK} for the extrinsic curvature and  \eqref{veloJ} for the acceleration of the gauge field,
 we immediately show that the kinetic Lagrangian is of the form,
\begin{eqnarray}
I_{\text{kin}}= \int d^4 x \sqrt{-g} \left(\mathscr{C}^{AB} K_{A} K_{B} +2 \mathscr{D}^{Ai} K_{A} \LL_i+\mathscr{E}^{ij} \LL_i \LL_j \right), \label{LKin}
\end{eqnarray} 
where $A \in \{1,...,6\}$ while  $J_i \in \{J_1,...,J_6\}$ if we assume $F$ to be a non-integrable two-form and $J_i \in \{J,J_a\}$ if we assume it to be the field strength of a gauge field. Notice that $\mathscr{E}$  depends only on the couplings $\beta$, i.e. on the invariants quadratic in the derivative of $F$. The linear term takes the form,
\begin{eqnarray}
I_{\rm{lin}} = \int d^4 x \sqrt{-g} \left( \mathscr{F}^{A} K_{A} +\mathscr{G}^{i} \LL_i  \right), \label{IPotLin} 
\end{eqnarray} 
where $\mathscr{G}$ only depends on $\beta$. Finally, the potential term $I_{\rm{pot}}$, which does not involve any velocities,  is irrelevant for our analysis.

\subsubsection{Degeneracy equations}\label{SecDegEq}

We are now ready to study the degeneracy conditions.  A first and obvious necessary condition for these HOME theories to be free of Ostrogradski ghosts is that the kinetic matrix $\mathbb{M}$, whose block form is given by,
\begin{eqnarray}
\mathbb{M}=\left( \begin{matrix}
\mathscr{E}  & \mathscr{D} \\
\mathscr{D}^\intercal  & \mathscr{C} 
\end{matrix}
\right)\,,
\end{eqnarray}
admits a non-trivial kernel. When $F$ is an arbitrary two-form, the dimension of $\mathscr{E} $ is 6 and the dimension of $\mathbb{M}$ is 12 while these two matrices have dimensions 3 and 6 respectively for U(1) theories.

Furthermore, following the analysis of scalar-tensor theories \cite{Langlois:2015cwa}, a natural assumption is to require the associated null eigenvectors to be in the electromagnetic sector so that the degeneracy conditions do not eliminate gravitational degrees of freedom. This implies that the 6-dimensional matrix $\mathscr{C} $ is invertible\footnote{Although having $\mathscr{C}$ invertible ensures that no gravitational degrees of freedom are lost, its non-invertibility does not necessary imply less gravitational degrees of freedom.}. Hence, $\mathbb{M}$ could possess up to $6$ or up to $3$ null eigenvectors depending on whether we are considering two-form or U(1) theories. If, in addition,  we assume  the degeneracy to be maximal (i.e. the kernel of $\mathbb{M}$ has dimension $6$ or $3$),  the kinetic Lagrangian can be factorised as follows,
\begin{eqnarray}
I_{\text{kin}}= \int d^4 x \sqrt{-g} \, \mathscr{C}^{AB}  \left( K_A + \Theta_A{}^i \LL_i \right)\left( K_B + \Theta_{B}{}^j \LL_j \right) =  \int d^4 x \sqrt{-g} \, \mathscr{C}^{AB} \mathcal{K}_A \mathcal{K}_B \,,\label{Fact}
\end{eqnarray}
for some functions $ \Theta_A{}^i $. Expanding the square and comparing with \eqref{LKin}, we obtain the conditions, 
\begin{eqnarray}
\mathscr{E}^{ij}= \mathscr{C}^{AB}  \Theta_{A}{}^i \Theta_{B}{}^j \,,\;\;\;\;\;\;\;\;\; \mathscr{D}^{Ai}= \mathscr{C}^{AB} \Theta_{B}{}^i \, .
\end{eqnarray}
If $\mathscr{C}$ is invertible, the second equation can be solved $\Theta_{A}{}^i=  \left(\mathscr{C}^{-1}\right)_{AB} \mathscr{D}^{Bi}$ and we obtain the following sufficient condition for a maximal degeneracy, 
\begin{eqnarray}
\mathscr{E}^{ij} = \mathscr{D}^{Bi} \left(\mathscr{C}^{-1}\right)_{AB} \mathscr{D}^{Aj} \, . \label{SuffMaxDeg}
\end{eqnarray}
Those are the equations we would like to solve in principle, as it has been done in the context of DHOST theories.

However, contrary to DHOST, the maximal  degeneracy of the kinetic matrix in the electromagnetic sector is not sufficient to guarantee the absence of Ostrogradski ghosts if the theory admits higher-order field equations. For instance, we obtained in \cite{Colleaux:2024ndy} theories which are purely linear in the acceleration of the gauge field and propagate ghost modes, even though their kinetic matrix has vanishing rank. Moreover, the maximal degeneracy is not even a necessary condition because we could find  theories with less primary constraints but with a sufficient number of secondary, or higher order constraints to eliminate the eventual ghosts. As a consequence, at this stage, the maximal degeneracy does not enable us to conclude anything about the number and the nature of degrees of freedom propagating in the theory.

\medskip

Let us discuss more specifically the case of DHOME theories with U(1) gauge invariance. In that case, the maximal degeneracy in the accelerations of the gauge field implies the existence of 3 primary constraints, which would be second class in general. If so, and if the theories are consistent, the Dirac-Bergmann algorithm would then produce at least one secondary constraint (as there should be an even number of second class constraints)\footnote{In this case, we would expect the two constraints which fix the two acceleration components $J_a$ to form a second class system  while the constraint which fix the remaining component $J$ would produce a secondary one.}, but this would not be sufficient to get rid of the $3\times 2$ variables associated with the acceleration of the gauge field in the phase space. Therefore, we would expect that ghost-freeness would be obtained  if each primary constraint generate a secondary one. Such a scenario seems possible  given that $J_a$ can be viewed as the velocities of two independent scalar fields (analogous to the kinetic term of a scalar field), and the associated primary constraints, by analogy with the case of DHOST, might each produce a secondary constraint, while the consistency of the theory would impose one more secondary second class constraint associated to $J$. 
Some of these constraints may also be first class. In fact, there exist many possibilities\footnote{For instance, it is also possible that even though the 3 primary constraints enable (by construction) to suppress 3 out of the 6 phase space variables in the electromagnetic sector, the additional constraints, if they exist, could relate only gravitational variables, in which case solving them would not eliminate  Ostrograski ghosts, but some healthy degrees of freedom.} and only a detailed Hamiltonian analysis could enable us to count the number of degrees of freedom and to determine their nature. However, such an analysis  is beyond the scope of this work, so we do not discuss further on these aspects.

\section{Degenerate higher-order Maxwell-Einstein theories}\label{Sec.DHOME}

In this section we construct DHOME theories with at most third order field equations\footnote{This means that $\mathscr{E} =0$. In \cite{Colleaux:2024ndy}, we have classified all the flat space theories with this property and dubbed these ``quasi-linear".}. In particular, we first obtain all the theories of this class whose field equations contain second order derivatives of the metric field. While these assumptions uniquely yield quadratic Horndeski theory for a scalar-field, we show that this is not the case for a gauge field. Thus, the corresponding theories can be seen as the generalisation of quadratic Horndeski gravity to a gauge field.

Then, following the discussion of \ref{SecDegEq}, we adopt an agnostic viewpoint on the type of degeneracy we are looking for and classify as exhaustively as possible the non-minimally coupled DHOME (i.e. keeping only the gravitational part of the action $\mathcal{R}$). Two classes of theories which are fully degenerate in the electromagnetic sector are obtained, one subclass of which being conformally invariant.

\subsection{Theories with second and third order field equations} \label{Trivially degenerate theories}

Let us start by investigating the conditions implied on the coupling functions of the theory when the latter is ``trivially degenerate", in the sense that instead of considering quadratic factorisations \eqref{Fact}, we impose the linear conditions 
\begin{eqnarray}
\mathscr{E} =0 \,, \qquad \mathscr{D}=0  \,,
\end{eqnarray}
meaning, by covariance, that the field equations must be at most first order for the two-form field, or at most third order for a gauge field, while being at most second order for the metric field in both cases. 

\subsubsection{Totally degenerate theories}

We first obtained all the totally degenerate theories, in the sense that
\begin{eqnarray}
\mathscr{C}=\mathscr{D}=\mathscr{E}=0\,,
\end{eqnarray}
 which are not trivial or boundary terms. These theories might be useful when discussing the constraint structure and the number of degrees of freedom of the degenerate theories we obtain in the following sections, as by definition they leave the degeneracy intact while modifying the terms linear in the velocities, i.e. the momenta.

Interestingly, there is no such theory in higher-order quadratic scalar-tensor theory, unless the unitary gauge $n_\mu \propto \partial_\mu \phi$ is chosen (see (2.13) of \cite{DeFelice:2018ewo}). However, we find for U(1) gauge field that there is a unique class of totally degenerate odd theory given by 
\begin{eqnarray}
\boxed{\mathscr{L}_{\text{o}} = \beta_{\bar{1}}\left( \x, \y\right) \mathcal{Q}_{\bar{1}}= \beta_{\bar{1}}\left( \x, \y\right) \poo^{\mu\nu}  \nabla_\mu \x \nabla_\nu \y } \label{Ltdo}
\end{eqnarray}
which, in terms of the field strength $F$, corresponds to the quadratic invariant
\begin{eqnarray}
\hodge F^{\mu\nu}  \nabla_\mu \mathcal{F} \nabla_\nu \mathcal{G}\,. \label{F17}
\end{eqnarray}
Indeed, its 3+1 decomposition reads 
\begin{eqnarray}
\mathscr{L}_{\text{o}} =  \beta_{\bar{1}} \Big( \bar{\psi} (n.\nabla) \x (b.D)\y  + \psi  (b.D)\x (r.D)\y  - \left( \x \leftrightarrow \y\right) \Big)\,,
\end{eqnarray}
where the first term is linear in the velocities (accelerations) while the second is the potential. Notice that in the U(1) case, the conjugated theory $\bar{\mathscr{L}}_{\text{o}}$ is equivalent to the former, while for two-form fields, both theories are independent and yield first order field equations for the two-form field.

There is another specificity of the U(1) gauge field compared to the two-form, given that only in the former case, there is a unique couple of conjugated totally degenerate even theories given by
\begin{eqnarray}
\boxed{\mathscr{L}_{\text{e}} =  \y \left(  \mathcal{R}_2 +  \mathcal{Q}_3^{\bar{0}}  \right) +    \x \mathcal{Q}_3^{1}  + \mathcal{Q}^{+}_{0\bar{1}} - \bar{\mathcal{Q}}^{-}_{0\bar{1}} } \label{Ltde}
\end{eqnarray}
together with its conjugate $\bar{\mathscr{L}}_{\text{e}}$. Performing a 3+1 decomposition, projecting all the spatial derivatives on the triad $\{e,b,r\}$ in order to avoid dimensionally dependent identities and using the spatial scalar Bianchi identity, it can be shown that the terms linear in the velocities (and the potential), defined by \eqref{IPotLin}, are non-vanishing for these theories
\begin{eqnarray}
\mathscr{F} \neq 0 \,,\qquad \mathscr{G} \neq 0 \,.
\end{eqnarray}
Although covariant boundary terms also share this property, given that we took into account all of these, the theory is not topological\footnote{Remark that the theory could still in principle be a purely U(1) boundary term of the form $\nabla_\mu \left(A_\nu  \mathcal{H}^{\mu\nu}  \right)= F_{\mu\nu}  \mathcal{H}^{\mu\nu}$, with $\mathcal{H}$ an antisymmetric divergenceless tensor involving an odd total number of $o$ and $\bar{o}$, a piece linear in the Riemann tensor and a piece quadratic in $\nabla F$. However, it can be shown that there is unique antisymmetric odd tensor linear in the Riemann tensor whose divergence is second order and whose contraction with $F$ is linear in $\mathcal{R}_2$ and $\mathcal{R}_3$, which is given by $\delta \hodge F^{\alpha\beta} \hodge F^{\rho\sigma}R_{\alpha\beta\rho\sigma} / \delta F_{\mu\nu}$. When contracted with the field strength, it gives Horndeski's non-minimal coupling rather than the Riemann terms of $\mathscr{L}_{\text{e}}$ and its conjugate.}.

Using the tools of Sec.\ref{SecDisfTransf}, it can be shown that the even theory \eqref{Ltde} and its conjugate are invariant under conformal transformations
\begin{eqnarray}
g_{\mu\nu} \longrightarrow  e^\sigma g_{\mu\nu} \,, \label{TTDEConf}
\end{eqnarray}
for $\sigma$ a scalar field. Furthermore, they are also invariant under a generic U(1) disformal transformation given by \eqref{DisfMetric}, meaning under
\begin{eqnarray}
g_{\mu\nu} \longrightarrow \z \left(\x, \y\right) \p_{\mu\nu} + \zb \left(\x, \y\right) \pp_{\mu\nu}\,,\label{TTDEDisf}
\end{eqnarray}
for any positive functions $\z$ and $\zb$ of the EM invariants.

\subsubsection{Field equations involving only second derivatives of the metric}\label{SecondOrderg}

Let us now relax the previous assumption by allowing a quadratic piece in the extrinsic curvature, i.e. we impose $\mathscr{D}=\mathscr{E}=0$ and $\mathscr{C}\neq0$. Interestingly, within the context of quadratic higher-order scalar-tensor theories, the unique solution for these constraints is Horndeski's quadratic theory with second order field equations. Thus, we are looking for the analogue of that theory when the scalar field is replaced by a U(1) gauge field. The unique two conjugated classes of theories solutions of these equations are even and given by 
\begin{eqnarray}
\boxed{\mathscr{L}_{\hodge}\left[ \upsilon\right] = \upsilon(\y) \left( \mathcal{R}_2 + 2 \mathcal{Q}_3^{\bar{0}} \right) +  \upsilon_{\y}\left(\y\right) \left(  \x \mathcal{Q}_3^{1} - \y \mathcal{Q}_3^{\bar{0}} + \mathcal{Q}^{+}_{0\bar{1}} - \bar{\mathcal{Q}}^{-}_{0\bar{1}}\right) } \label{L*}
\end{eqnarray}
together with $\bar{\mathscr{L}}_{\hodge}\left[\bar{\upsilon}\right]$, where the subscript $\y$ means partial derivative w.r.t. that variable. Recall that by definition
\begin{eqnarray}
\begin{split}
\mathcal{Q}_3^{1}&= o^{\mu\sigma} \bar{o}^{\rho\beta} p^{\nu\alpha} \nabla_\mu o_{\nu\rho} \nabla_{\sigma} o_{\alpha\beta} \,,\\
\mathcal{Q}_3^{\bar{0}}&= o^{\mu\sigma} o^{\nu\alpha} \bar{p}^{\rho\beta} \nabla_\mu o_{\nu\rho} \nabla_{\sigma} o_{\alpha\beta} \,, \\
\mathcal{Q}^{+}_{0\bar{1}} - \bar{\mathcal{Q}}^{-}_{0\bar{1}} &= \bar{o}^{\mu[\nu} p^{\rho]\sigma} \nabla_{\rho} \x \nabla_{\nu} o_{\mu\sigma}\,,
\end{split}
\end{eqnarray}
so that the quadratic pieces are ``quasi-linear", using the terminology of \cite{Colleaux:2024ndy}, meaning that both derivatives are antisymmetrised. As we see, these theories depend on an arbitrary function of a single kinetic invariant. These two properties are shared with the second order scalar-tensor theories of Horndeski.  In an early attempt to classify DHOME using the basis of Lagrangians of \cite{Colleaux:2023cqu}, we derived an alternative expression for this theory in terms of the field strength tensor which is reported in appendix \ref{AppLstar}. Although it would be interesting to find a simpler formulation in terms of totally antisymmetrised quantities (which might be more easily generalised to higher orders), we have been unable to find one so far, except for the ``Galileons-looking" expression \eqref{Galileonlike}. In this spirit, we provide a dictionary between projectors and field strength in appendix \ref{App:dictio}.

Notice that when considering a two-form field, the equations $\mathscr{D}=\mathscr{E}=0$ (and $\mathscr{C}\neq 0$) are solved when these functions are constant, so that we obtain a pair of theories with first order field equations for the two-form field,
\begin{eqnarray}
 \mathscr{L}_{\text{F}}=  \mathcal{R}_2 + 2 \mathcal{Q}_3^{\bar{0}}  = -\frac{1}{2}  \po^{\mu\nu} \po^{\rho\sigma} \left( R_{\mu\nu\rho\sigma} -4 \pp^{\gamma\delta} \nabla_\mu \po_{\rho\gamma} \nabla_\nu \po_{\sigma\delta} \right)  \label{Ltf}
\end{eqnarray}
and its conjugate, where we have used DDIs and the Bianchi identity for the Riemann tensor to obtain the equation. Furthermore, as we will see later, the sum of the two for $\upsilon=\bar{\upsilon}=1$ is equivalent to General Relativity up to DDIs and a boundary term. Thus, there remains one new theory, say \eqref{Ltf}.

Back to the U(1) case, the kinetic pieces of these theories are given by 
\begin{eqnarray}
\int d^4 x \sqrt{-g} \, \mathscr{C}_{\hodge}^{AB} K_{A} K_{B} =\int d^4 x \sqrt{-g}  \left( \upsilon  -\y \upsilon_{\y} \right) \left(   K_6^2 -K_1 K_5- \bar{\psi}^2 \left( K_4 K_1 - K_2^2 \right)  \right) 
\end{eqnarray}
and similarly for the conjugated theory. Thus, in order for the associated kinetic matrix to be degenerate only in the EM directions, it is necessary to consider a generic sum of these theories, because $\mathscr{L}_{\hodge}$ lacks $K_3$ while $\bar{\mathscr{L}}_{\hodge}$ lacks $K_6$. Furthermore, we recover the totally degenerate even theories \eqref{Ltde} for linear functions
\begin{eqnarray}
\upsilon= \y  \,,\;\;\;\; \bar{\upsilon}=\x \,. \label{TDEfunctions}
\end{eqnarray}
Given the disformal \eqref{TTDEDisf} and conformal \eqref{TTDEConf} invariances of these theories, they can be expected to be generated from non-invertible ``mimetic" disformal transformations applied to some seed HOME theory, as it is the case in higher-order scalar-tensor theories, see e.g. \cite{Domenech:2023ryc}. Using the tools of section \ref{SecDisfTransf} and in particular the classification of these singular transformations appearing in \ref{Sec.Mimetic}, it is straightforward to show that the corresponding seed theory has quadratic functions and is given by 
\begin{eqnarray}
\mathscr{L}_{\text{seed}}=\frac{1}{2} \left(  \alpha \mathscr{L}_{\hodge}\left[\y_0 +  \frac{\y^2}{\y_0}  \right]  + \bar{\alpha} \bar{\mathscr{L}}_{\hodge}\left[ \x_0 +  \frac{\x^2}{\x_0} \right]  \right) \,,
\end{eqnarray}
where $\alpha$, $\x_0$ and their conjugates are constants. Indeed, applying the following singular disformal transformation
\begin{eqnarray}
\tilde{g}_{\mu\nu} = \frac{\x}{\x_0} \p_{\mu\nu} +\frac{\y}{\y_0}  \pp_{\mu\nu}
\end{eqnarray}
on the seed, we obtain up to boundary terms, dimensional and Bianchi identities,
\begin{eqnarray}
\sqrt{-\tilde{g}}\, \tilde{\mathscr{L}}_{\text{seed}}= \sqrt{-g} \left( \alpha \mathscr{L}_{\text{e}} + \bar{\alpha} \bar{\mathscr{L}}_{\text{e}} \right) \,, \label{MimeticTDE}
\end{eqnarray}
what confirms the fact that the totally degenerate even theories are mimetic.

\subsubsection{Theories with second order equations of motion}\label{SecondOrder}

Given that all second order theories must belong to this class, notice that General Relativity and Horndeski's non-minimal coupling are obtained summing both theories with the following choice of functions
\begin{eqnarray}
\upsilon_{\text{H}}= 1 -\frac{\gamma \y^2}{2} \,,\;\;\;\;\; \bar{\upsilon}_{\text{H}}= 1 + \frac{\gamma \x^2}{2}\,,
\end{eqnarray}
up to dimensional identities and even boundary terms defined by \eqref{BTEven}, where the subscript ``H" refers to Horndeski's theory \cite{Horndeski:1976gi}. Indeed 
\begin{eqnarray}
\mathcal{L}_{\text{H}}
= \mathscr{L}_{\hodge}\left[ \upsilon_{\text{H}} \right]  + \bar{\mathscr{L}}_{\hodge}\left[ \bar{\upsilon}_{\text{H}} \right]  - \mathcal{B}_0\left[2\right] - \gamma \mathcal{B}_1 \left[\mathcal{G} \right]
=R + \frac{\gamma}{4} \hodge F^{\mu\nu} \hodge F^{\rho\sigma}  R_{\mu\nu\rho\sigma}\,, \label{GRNMC}
\end{eqnarray}
using \eqref{RicciNPCProj}. Furthermore, it is straightforward to show using the classification of trivial terms (boundary, Bianchi and DDIs) that these are the only two theories of this class which can be written purely in terms of Riemann invariants. Given that $\mathscr{G}=0$ for curvature terms, we recover the well-known fact that both yield second order field equations as their total action does not depend on second time derivatives and the theories are invariant under 4D-diffeomorphisms, so their 3+1 action should be possible to express in terms of first order spatial derivatives as well.

\subsection{Degenerate non-minimally coupled theories}\label{NMCDeg}

Let us now restrict our investigation to non-minimally coupled theories, meaning $\mathscr{B}=0$ in the action \eqref{Action}. As we are going to see, this restriction already enables to uncover a rich structure of degenerate theories, with notably a class of conformally invariant theories which are fully degenerate in the EM sector (meaning $\text{rank}(\mathbb{M})=\text{rank}(\mathscr{C})=6$). Although we could further restrict to this type of degeneracy, which we also encountered with the ``Horndeski-like" theories obtained in \ref{SecondOrderg} and \ref{SecondOrder}, it is (a priori) neither necessary nor sufficient to ensure the absence of Ostrogradski ghosts (see the discussion at the end of \ref{SecDegEq}). Similarly, having a degenerate gravitational sector  (i.e. $\text{rank}(\mathscr{C})<6$)  is not necessarily problematic, as this happens also in conformal gravity, for which the trace of the rescaled extrinsic curvature is absent (see \cite{Nikolic:2017hfi} and (133) of \cite{Kiefer:2017nmo}) while the theory still contains two gravitational wave polarizations. Therefore, we will not put any restriction on $\text{rank}(\mathscr{C})$ and $\text{rank}(\mathbb{M})$.

Finally, notice that the main reason for setting $\mathscr{B}= 0$ is that degenerate theories with quadratic derivatives of the field strength exist in very large number and are therefore quite complicated to fully classify. At the end of this section, we will give an example of degenerate theories satisfying both $\mathscr{B}\neq 0$ and $\text{rank}(\mathbb{M})=\text{rank}(\mathscr{C})=6$ and involving up to ten free coupling functions.

\subsubsection{Classification}

 The action \eqref{Action} reduces for non-minimally coupled theories to
\begin{eqnarray}
 I = \int d^4 x \, \sqrt{-g}\, \mathcal{R} =   \int d^4 x \, \sqrt{-g}\,   \mathscr{A}^{\mu\nu\rho\sigma} R_{\mu\nu\rho\sigma} =  \int d^4 x \, \sqrt{-g}\, \sum_{I=0}^3 \alpha_I \, \mathcal{R}_I \,,\label{NMCAction}
\end{eqnarray}
where $\mathscr{A}$ is given by \eqref{mathscrA}. These models have by construction a vanishing flat-space limit and satisfy the relation $\mathscr{E}=\mathscr{G}=0$, while the expressions of the building blocks, $\mathscr{C}$ and $\mathscr{D}$, of the kinetic matrix $\mathbb{M}$ are given in appendix \ref{AppNMC}. In particular, it can be shown that the determinant of the latter factorises 
\begin{eqnarray}
\text{det}\left( \mathbb{M} \right) = m_0 m_1 \,,\label{FacM}
\end{eqnarray}
where $m_0$ is given in appendix \ref{AppNMC}. Therefore, degenerate theories can be divided into two families 
\begin{eqnarray}
\begin{split}
&\text{Family}\; 1 :  m_0=0 \,,\\
&\text{Family} \; 2 :  m_0\neq 0  \;\; \&  \;\, m_1= 0 \,.
\end{split} \label{Family12}
\end{eqnarray}
In the following, we fully classify the first family and give some examples for the second one, which we were not able to classify fully.  Interestingly, as shown in appendix \ref{AppNMC}, the defining condition \eqref{Family12} of the first family is algebraic in the coupling functions $\alpha_I$. Solving it for any $\psi$, we obtain four different classes :
\begin{eqnarray}
\begin{split}
&\text{Class 0 (C$_0$) :}\;\;\;\;\;\;\;\; \alpha_2 = \frac{1}{2} \left( \alpha_0 + \frac{\bar{x}}{\x} \alpha_1\right) \,,\;\;\;\;\, \alpha_3 = \frac{1}{2} \left( \alpha_0 - \frac{\x}{\y} \alpha_1 \right)\\
&\text{Class 1 (C$_1$) :}\;\;\;\;\;\;\;\; \alpha_2 =  \alpha_0 \,,\;\;\;\;\;\;\;\;\;\;\;\;\;\;\;\;\;\;\;\;\;\;\;\; \alpha_3 = \alpha_0  \\
&\text{Class 2 (C$_2$) :}\;\;\;\;\;\;\;\; \alpha_1 = - \frac{\y}{\x}\alpha_0  \,,\;\;\;\;\;\;\;\;\;\;\;\;\;\;\;\;\;\; \alpha_3 = \alpha_0  \\
&\text{Class 3 (C$_3$) :} \;\;\;\;\;\;\;\; \alpha_1 =   \frac{\x}{\y}\alpha_0 \,,\;\;\;\;\;\;\;\;\;\;\;\;\;\;\;\;\;\;\;\;\; \alpha_2 = \alpha_0   \,,
\end{split}\label{FourClasses}
\end{eqnarray}
what, without further fine-tuning of the coupling functions\footnote{More precisely, we have $\text{rank}\left(\mathbb{M}\right)=8$ for any coupling functions (fine-tuned or not), unless the specific fine-tunings obtained in the following sections are imposed.}, implies
\begin{eqnarray}
\text{rank}\left(\mathbb{M}\right)=8 \,.
\end{eqnarray} 
It is noticeable that Horndeski's non minimal coupling and General Relativity are obtained from the Class 0. Moreover, the numbering $\{0,...,3\}$ of the classes is chosen so that (setting $\alpha_0=0$)
\begin{eqnarray}
\alpha_1 \left( \x,\y\right) \mathcal{R}_1 \in \text{C}_1 \,,\;\;\;\; \alpha_2 \left( \x,\y\right) \mathcal{R}_2 \in \text{C}_2  \,,\;\;\;\; \alpha_3 \left( \x,\y\right) \mathcal{R}_3 \in \text{C}_3 \,,
\end{eqnarray}
meaning that these elementary Lagrangians are degenerate for arbitrary functions $\alpha_1$, $\alpha_2$ and $\alpha_3$ of the two electromagnetic invariants  (while $\mathcal{R}_0$ is not).

Remark also that Class 2 and 3 are related by conjugation, assuming that $\alpha_I$ transforms under a conjugation as $\mathcal{R}_I$, meaning $\alpha_1 \to \alpha_1$, $\alpha_2 \to - \alpha_2$, $\alpha_3 \leftrightarrow \alpha_4$. This should not be viewed as a restriction on their specific dependence on $(\x,\y)$. Therefore, it is sufficient to restrict to say Class $3$, and to apply a rotation on the resulting degenerate theories to complete the classification. Furthermore, the different intersections of these four classes are such that
\begin{eqnarray}
C_0 \cap C_1 = \emptyset \,,\;\;\;\; C_2 \cap C_3 = \emptyset \,,\;\;\;\; C_0 \cap C_3 \neq \emptyset \,,\;\;\;\; C_1 \cap C_3 \neq \emptyset 
\end{eqnarray}
and similarly for the conjugates. Thus, we start by considering the classification of degenerate theories in Class $3$, including those which also belong to Classes $0$ and $1$, and will ignore these afterwards.

\subsubsection{Gravitationally degenerate theories}

 The degenerate Lagrangian associated with the Class $3(=\bar{2})$ is given by 
\begin{eqnarray}
\boxed{\mathcal{R}_{\text{C}_3}= \alpha_0 \left(\x,\y\right) \left(  \mathcal{R}_0 +  \mathcal{R}_2 + \frac{\x}{\y}  \mathcal{R}_1 \right)  + \alpha_3 \left(\x,\y\right)  \mathcal{R}_3}
\end{eqnarray}
and remark that $\mathcal{R}_0 +  \mathcal{R}_2 =- G_{\mu\nu} \pp^{\mu\nu}$. Notice also that $\text{det} \left(\mathscr{C} \right)= 0$ for this class, so that the gravitational sector is degenerate. Although not ideal, it does not automatically imply the presence of Ostrogradski ghost, or that gravitational waves have less than two polarizations, as explained above. There are three options to make the kernel two-dimensional and all the degenerate theories of Class $3$ are summarized in Table \ref{table:1}.

\paragraph{Theories C$_3$\RNum{1} : }

 The first case, denoted \RNum{1}$_1$, corresponds to the intersection between Class $3$ and Class $0$ and can be obtained imposing
\begin{eqnarray}
\alpha_3\left(\x,\y\right) = \frac{\y^2 - \x^2}{2 \y^2}\, \alpha_0\left(\x,\y\right)\,,
\end{eqnarray}
what also makes the kernel of $\mathscr{C}$ two-dimensional. In this case, the theory can be rewritten as 
\begin{eqnarray}
\mathcal{R}_{\text{C}_3\text{\RNum{1}}_1}= \frac{1}{2} \alpha_0\left(\x,\y\right) \left(  R - \frac{1}{2 \y^2} \hodge F^{\mu\nu} \hodge F^{\rho\sigma}  R_{\mu\nu\rho\sigma} \right)\,.
\end{eqnarray}
 Setting further 
\begin{eqnarray}
\alpha_{0\x}=0 
\end{eqnarray}
yields $\text{dim}\left(\text{ker}\left( \mathbb{M}\right) \right) = 3$. We denote the corresponding theory \RNum{1}$_2$. Finally, it is possible to completely fix the dependence of the coupling functions as
\begin{eqnarray}
\alpha_{0}= \y 
\end{eqnarray}
and obtain a theory, denoted \RNum{1}$_3$, with 4 null eigenvalues, while the kernel of $\mathscr{C}$ is still two-dimensional.

\paragraph{Theories C$_3$\RNum{2} :}

The second possibility, denoted \RNum{2}$_1$, is to keep $\alpha_3$ arbitrary while setting 
\begin{eqnarray}
\alpha_{0}\left(\x,\y\right)= \y c \,,
\end{eqnarray}
where $c$ is a constant, in which case the kernel of $\mathbb{M}$ is two-dimensional. Setting further $c=0$ yields a theory \RNum{2}$_2$  with a three-dimensional kernel which is given by one of our elementary Lagrangian 
\begin{eqnarray}
\boxed{\mathcal{R}_{\text{C$_3$\RNum{2}$_2$}} = \alpha_3 \left(\x,\y\right)  \mathcal{R}_3 \propto  \alpha_3 \left(\x,\y\right)\left( \x^4 R - 2 \x^2 F_2^{\mu\nu} R_{\mu\nu} + F_2^{\mu\nu}F_2^{\rho\sigma} R_{\mu\rho\nu\sigma} \right)}
\end{eqnarray}
Notice that the kernel of $\mathscr{C}$ is still one-dimensional.

\paragraph{Theories C$_3$\RNum{3} :}

Finally, the last possibility to obtain a two-dimensional kernel for $\mathbb{M}$ is to set 
\begin{eqnarray}
\alpha_{0\x}=0 \,,\;\;\;\; \alpha_{3}=\x c\left(\y\right) \,,
\end{eqnarray}
where $c$ is a function and $\text{dim}\left(\text{ker}\left( \mathscr{C}\right) \right) = 1$ still holds. The corresponding theory is denoted \RNum{3}.

\begin{table}[h!]
\centering
\begin{tabular}{|c| c| c| c|} 
 \hline
Class & free functions & dim(ker($\mathbb{M}$)) & dim(ker($\mathscr{C}$)) \\ 
 \hline
  \hline
  C$_3$ & $\alpha_0\left(\x,\y\right)$,$\alpha_3\left(\x,\y\right)$  &  1 & 1 \\ 
  \hline
  \hline
\RNum{1}$_1$ & $\alpha_0\left(\x,\y\right)$ & 2 & 2 \\ 
 \hline
\RNum{1}$_2$ & $\alpha_0\left(\y\right)$ & 3 & 2 \\
 \hline
\RNum{1}$_3$ & none & 4 & 2 \\
 \hline
 \hline
\RNum{2}$_1$ & $\alpha_0\left(\y\right)=c\, \y$, $\alpha_3\left(\x,\y\right)$ & 2 & 1 \\
 \hline
\RNum{2}$_2$ & $\alpha_3\left(\x,\y\right)$ & 3 & 1 \\
 \hline
 \hline
\RNum{3} & $\alpha_0\left(\y\right)$, $\alpha_3\left(\x,\y\right)=\x c\left( \y \right)$ & 2 & 1 \\ \hline
\end{tabular}
\caption{Class 3 of degenerate theories : $\{ \text{\RNum{1}, \RNum{2}, \RNum{3}} \} \subset \text{C}_3, \;\; \text{\RNum{1}}_3 \subset \text{\RNum{1}}_2 \subset \text{\RNum{1}}_1, \;\;\text{\RNum{2}}_2 \subset \text{\RNum{2}}_1$.}
\label{table:1}
\end{table}

\subsubsection{Partially degenerate embeddings of GR and Horndeski's non-minimal coupling}

We continue with the Class 0, excluding its intersection with the Classes  $2$ and $3$ that we just studied. The Lagrangian is given by 
\begin{eqnarray}
\boxed{\mathcal{R}_{\text{C}_0}=\frac{1}{2} \alpha_0\left(\x,\y\right)  R - \frac{1}{4 \x\y}  \alpha_1\left(\x,\y\right) \hodge F^{\mu\nu} \hodge F^{\rho\sigma}  R_{\mu\nu\rho\sigma}} \label{RC0}
\end{eqnarray}
and in this case $\text{det} \left(\mathscr{C} \right)\neq 0$, so that the gravitational sector is generically not degenerate.  The only left-over possibility to make the kinetic matrix more degenerate is to impose $m_1=0$, where $m_1$ appears in \eqref{FacM}. This yields the theories C$_0$\RNum{1}$_1$ which are such that the coupling functions satisfy the condition
\begin{eqnarray}
\alpha_{0\x} \, a_{1 \y} - \alpha_{0\y} \, a_{1 \x} = 0 \, ,
\label{F1C0I1}
\end{eqnarray}
where $a_1=\alpha_1/(\x\y)$. Such a condition is equivalent to requiring that the Jacobian of the following ``coordinate'' transformation
\begin{eqnarray}
\x \longrightarrow \alpha_0 (\x,\y) \, , \qquad
\y \longrightarrow a_1(\x,\y) 
\end{eqnarray}
vanishes. Hence, the pair of coupling functions $(\alpha_0,a_1)$ is a solution of \eqref{F1C0I1} if there exists a real-valued function $f$ such that
$f(\alpha_0,a_1)=0$, or locally a function $g$ such that either $\alpha_0=g(a_1)$ or $a_1=g(\alpha_0)$.

Requiring one more null eigenvalue is equivalent to requiring that the previous transformation is completely degenerate, meaning that $\alpha_0$ and $a_1$ are two constants. In this case, the theory reduces to General Relativity $+$ Horndeski's non minimal coupling, given by \eqref{GRNMC}.

In all these cases, $\text{rank}(\mathscr{C})=6$ so that the degeneracy is localised in the EM sector and only C$_0$\RNum{1}$_2$ is fully degenerate there. This is summarised in Table \ref{table:2}.
\begin{table}[h!]
\centering
\begin{tabular}{|c| c| c| c|} 
 \hline
Class & free functions & dim(ker($\mathbb{M}$)) & dim(ker($\mathscr{C}$)) \\ 
 \hline
  \hline
  C$_0$ & $\alpha_0\left(\x,\y\right)$,$\alpha_1\left(\x,\y\right)$  &  1 & 0 \\ 
  \hline
  \hline
\RNum{1}$_1$ & \eqref{F1C0I1} & 2 & 0 \\ 
 \hline
\RNum{1}$_2$ & $G, \gamma$ & 3 & 0 \\
 \hline
\end{tabular}
\caption{Class 0 of degenerate theories : $\text{\RNum{1}} \subset \text{C}_0, \;\;  \text{\RNum{1}}_2 \subset \text{\RNum{1}}_1.$ $\{G, \gamma\}$ are coupling constants.}
\label{table:2}
\end{table}

\subsubsection{Fully degenerate \& conformally invariant theories}\label{Sec.FullyDegConf}

The last class, summarised in Table \ref{table:3} and given by the Lagrangian
\begin{eqnarray}
\boxed{\mathcal{R}_{\text{C}_1}=\alpha_0 \left(\x,\y\right) \left( \mathcal{R}_0 + \mathcal{R}_2 +\mathcal{R}_3 \right) + \alpha_1\left(\x,\y\right)  \mathcal{R}_1 }
\end{eqnarray}
is the only one which yields fully degenerate theories in the electromagnetic sector, meaning that $\text{rank}\left(\mathbb{M}\right)=\text{rank}\left(\mathscr{C}\right)=6$, other than General Relativity $+$ Horndeski's NMC, as we are going to see. However, it is rather interesting to notice that although the gravitational sector is degenerate when
\begin{eqnarray}
\det\left(\mathscr{C}\right) = -  \frac{4 \psi^2 \bar{\psi}^2}{\left( \x^2 + \y^2 \right)^4}  \left( 2 \x \y \alpha_0 + \left( \x^2 - \y^2 \right) \alpha_1  \right)^2  c^2 =0 \,,
\end{eqnarray}
where 
\begin{eqnarray}
c=   \left(  \left( \x^2 - \y^2 \right)\alpha_0\alpha_1  + \x\y \left( \alpha_0^2 - \alpha_1^2 \right) \right)\,,
\end{eqnarray}
this does not affect the dimension of the kernel of $\mathbb{M}$. Thus, the rank of $\mathscr{C}$ should not be physically relevant in this class, as it does not seem to be related to the number of constraints.

The kinetic matrix has two vanishing eigenvalues when the following relations between the coupling functions are satisfied : 
\begin{eqnarray}
\alpha_0 \alpha_{1\x} c= \alpha_0 \left( \x \alpha_0 - \y \alpha_1 \right) \left( \alpha_0^2 + \alpha_1^2 \right) + \alpha_{0\x} \left( \alpha_0^3 \left( \y^2 - \x^2 \right) + \x\y \alpha_1 \left( 3 \alpha_0^2 + \alpha_1^2 \right) \right) \label{F1C1I}
\end{eqnarray}
together with its conjugate, and the corresponding theories are denoted $\mathcal{R}_{\text{C}_1\text{\RNum{1}}}$. There are only two possibilities to obtain 3 vanishing eigenvalues and these cases are summarised in Table \ref{table:3}.
\\

 The first class of theories which are fully degenerate in the EM sector are simply given by one of our elementary Lagrangian, 
\begin{eqnarray}
\boxed{ \mathcal{R}_{\text{C}_1\text{\RNum{2}}} = \alpha_1 \left(\x , \y\right) \mathcal{R}_1  = \alpha_1 \left(\mathcal{F},\mathcal{G}\right) \Big( 2 \mathcal{F} \hodge F^{\mu\nu} F^{\rho\sigma} + \mathcal{G} \left( F^{\mu\nu} F^{\rho\sigma} - \hodge F^{\mu\nu} \hodge F^{\rho\sigma}  \right)\Big) C_{\mu\rho\nu\sigma} } \label{R1theory}
\end{eqnarray}
where we use the same symbol $\alpha_1$ by abuse of notation\footnote{This is to avoid a proliferation of symbols, the relation between both functions being clear from \eqref{EMinvxy}.}, and where $C$ is the Weyl tensor defined by \eqref{WeylDef}, see the classification of Riemann invariants \eqref{RiemBasis} and the relation \eqref{RiemWeyl}. Thus, it constitutes a class of degenerate theories parametrized by an arbitrary function of the two electromagnetic invariants, whose kinetic term factorises as follows
\begin{eqnarray}
 \int d^4 x \sqrt{-g} \left( \mathscr{C}^{AB} K_{A} K_{B} +2 \mathscr{D}^{Ai} K_{A} \LL_i  \right)=  \int d^4 x \sqrt{-g}\,   C^{AB} \mathcal{K}_{A} \mathcal{K}_{B}  \,,\label{FactoR1}
\end{eqnarray}
where $C$ and $\mathcal{K}\left( K, J_a, J \right)$ are given in appendix \ref{AppNMC}. Besides this complete degeneracy in the accelerations of the gauge field, another theoretical appeal of these interactions lies in the fact that a  subclass of \eqref{R1theory} is conformally invariant. Indeed, using the transformation properties of the electromagnetic invariants and projectors (see \eqref{DisfTransfProj}), it is straightforward to show that for 
\begin{eqnarray}
\alpha_1 \left(\x, \y \right) =\x \alpha_1 \left( \frac{\y}{\x}\right) \,, \label{ConfInvCF}
\end{eqnarray}
the theory is invariant under conformal transformation
\begin{eqnarray}
g_{\mu\nu} \longrightarrow  e^\sigma g_{\mu\nu}\,,
\end{eqnarray}
for $\sigma$ a scalar field. In terms of the field strength, it means that the following theory
\begin{eqnarray}
\boxed{ \mathcal{C}_1 = \left(\mathcal{F}^2+\mathcal{G}^2 \right)^{-\frac{3}{4}}  \alpha_1 \left( \frac{\mathcal{F}}{\mathcal{G}}\right)   \Big( 2 \mathcal{F} \hodge F^{\mu\nu} F^{\rho\sigma} + \mathcal{G} \left( F^{\mu\nu} F^{\rho\sigma} - \hodge F^{\mu\nu} \hodge F^{\rho\sigma}  \right)\Big) C_{\mu\rho\nu\sigma}} \label{R1theoryConf}
\end{eqnarray}
is both fully degenerate in the electromagnetic sector and conformal invariant\footnote{Of course, the argument of the function $\alpha_1$ can be any conformally invariant quantity such as $\frac{\mathcal{F}}{\sqrt{\mathcal{F}^2+\mathcal{G}^2}}$.}. 
\\

The second class of fully degenerate theories in the EM sector is more restricted and given by the pair 
\begin{eqnarray}
\boxed{
\mathcal{R}_{\text{C}_1\text{\RNum{3}}_\pm} = \gamma \left(\mathcal{R}_0+\mathcal{R}_2+\mathcal{R}_3 \right) - \frac{1}{2 \gamma \delta} \left( \x \y \pm \sqrt{\left( \x^2 - 2 \gamma^2 \delta \right) \left( \y^2 + 2 \gamma^2 \delta \right)}  \right)\mathcal{R}_1 } \label{Squareroot}
\end{eqnarray}
where $\gamma$ and $\delta$ are coupling constants transforming under a rotation as $\gamma \to \gamma$ and $\delta \to - \delta$. As shown in appendix \ref{AppNMC}, its kinetic term can also be written in a factorised form, similar to \eqref{FactoR1}.
\\

Let us conclude this section emphasising first that by construction none of these theories can be summed with GR or Horndeski's NMC without spoiling the degeneracy, although allowing quadratic terms in the derivative of the field strength might enable to maintain it. Secondly, given that $\mathscr{G}=0$ for curvature terms, it is clear that while expressing the extrinsic curvature in terms of the effective one makes the kinetic action purely quadratic in the latter, it also introduces the acceleration of the gauge field at the linear level as a by-product.

\begin{table}[h!]
\centering
\begin{tabular}{|c| c| c| c|} 
 \hline
Class & free functions & dim(ker($\mathbb{M}$)) & dim(ker($\mathscr{C}$)) \\ 
 \hline
  \hline
  C$_1$ & $\alpha_0\left(\x,\y\right)$,$\alpha_1\left(\x,\y\right)$  &  1 & 0 \\ 
  \hline
  \hline
\RNum{1} & \eqref{F1C1I} & 2 & 0 \\ 
 \hline
\RNum{2} & $\alpha_1\left(\x,\y\right)$ & 3 & 0 \\
 \hline
 \RNum{3}$_\pm$ & $\gamma, \delta$ & 3 & 0 \\
 \hline
\end{tabular}
\caption{Class 1 of degenerate theories : $\text{\RNum{1}} \subset \text{C}_1, \;\;  \text{\RNum{2}} \subset \text{\RNum{1}}, \;\; \text{\RNum{3}}_\pm \subset \text{\RNum{1}}.$ $\{\gamma,\delta\}$ are coupling constants.}
\label{table:3}
\end{table}
\subsubsection{Family 2 : Degenerate conformal invariant theories}

The second family of degenerate theories, which satisfies $m_0\neq0$ and $m_1=0$, is more complicated to fully classify because, contrary to Family 1, it does not seem to imply necessary algebraic relations between the coupling functions. We will focus here on an example which might present some interest\footnote{While, it is relatively simple to obtain particular classes of solutions, we do not see a clear interest in most of the corresponding theories.}.

Let us consider the following ansatz
\begin{eqnarray*}
\alpha\left(\x,\y\right) \left( \mathcal{R}_0 - \mathcal{R}_2 - \mathcal{R}_3\right) = 3\, \alpha\left(\x,\y\right) \p^{\mu\nu} \pp^{\rho\sigma} C_{\mu\rho\nu\sigma}\,,
\end{eqnarray*}
which is the second Weyl scalar built from the action \eqref{NMCAction}. As we saw in the previous section, the first Weyl scalar $\mathcal{R}_1$ yields a class of degenerate theories, so one might expect the same here. Indeed, this ansatz yields
\begin{eqnarray}
m_1 = 3^4 \psi^4 \bar{\psi}^4 \left( \psi^2 + \bar{\psi}^2 \right)^2  \alpha^4 \left(\alpha - \x \alpha_{\x} - \y \alpha_{\y} \right)^2 \,.
\end{eqnarray}
Therefore, degeneracy implies the coupling function $\alpha\left(\x,\y\right) = \x \alpha \left( \y / \x\right)$ encountered previously (see \eqref{ConfInvCF}), which yields conformal invariance, meaning that the following theory 
\begin{eqnarray}
\boxed{ \mathcal{C}_2 =  \left(\mathcal{F}^2+\mathcal{G}^2 \right)^{-\frac{3}{4}}  \alpha \left( \frac{\mathcal{F}}{\mathcal{G}}\right) F_2^{\mu\nu}F_2^{\rho\sigma} C_{\mu\rho\nu\sigma}} \label{ConfTh2}
\end{eqnarray}
is both conformally invariant and degenerate with $\text{det} \left(\mathscr{C} \right)\neq 0$. However, contrary to the theory \eqref{R1theoryConf}, the kinetic matrix of $\mathcal{C}_2$ has only one null eigenvalue and we cannot fix $\alpha$ further to obtain more null eigenvalues.

\subsubsection{Beyond non-minimal couplings : a ten-functions family of DHOME theories}

As explained previously, the main reason to restrict to non-minimally coupled Lagrangians is because of the very large number of degenerate theories existing in the generic case where $\mathscr{B}\neq 0$, as we will illustrate here.

Interestingly, it can be shown that the factorisation \eqref{FacM}, division into two families \eqref{Family12} and subsequently into four classes \eqref{FourClasses} are all preserved in the most general case. Thus, although the specific relations between couplings are modified, the structure stays the same. However, even by restricting to the generalisation of the Family 1, which we have been able to fully classify above, and to fully degenerate theories in the electromagnetic sector, meaning $\text{rank}(\mathbb{M})=\text{rank}(\mathscr{C})=6$, a complete classification remains complicated.

As a concrete example, it can be shown that taking into account dimensional and Bianchi identities, as well as boundary terms (see appendix \ref{Classification of higher-order Lagrangian}), the unique theories belonging to the generalisation of Family 1 Class 1, satisfying $\alpha_2=\alpha_3=0$ and full degeneracy are given by the following ten-functions family 
\begin{eqnarray}
\boxed{\mathcal{L}_{\text{quadratic}}= \mathscr{L}_{\text{quadratic}} +\bar{\mathscr{L}}_{\text{quadratic}} }
\label{BeyondNMC}
\end{eqnarray}
where
\begin{eqnarray}
\boxed{ \mathscr{L}_{\text{quadratic}}=\beta \left( - \x \mathcal{Q}^{\bar{0}}_1 + \mathcal{Q}^+_{1\bar{0}}  \right) + \beta^0_1 \mathcal{Q}^{0}_1 + \beta^{\bar{1}}_2  \mathcal{Q}^{\bar{1}}_2   + \beta^+_{0\bar{0}}  \mathcal{Q}^+_{0\bar{0}} + \beta^-_{0\bar{1}}  \mathcal{Q}^-_{0\bar{1}}} 
\end{eqnarray}
for any functions $\{\beta,\beta^0_1,\beta^{\bar{1}}_2,\beta^+_{0\bar{0}},\beta^-_{0\bar{1}}  \}$, and their conjugates, of the two electromagnetic invariants $\x$ and $\y$. Indeed, the null eigenvectors of its kinetic matrix are given in appendix \ref{AppNMC}.

Some comments are in order. Firstly, the presence of the odd terms $\mathcal{Q}^+_{0\bar{0}}$ and its conjugate implies that in order to have a real Lagrangian, the function $\bar{\beta}^+_{0\bar{0}}$ should be imaginary, as discussed in section \ref{SecQuadInv} (see \eqref{oddterms}). Secondly, given the characterisation of this theory, it must be a generalisation of the theory \eqref{R1theory}. Indeed, it can be shown that up to Bianchi identities and a boundary term, this fully degenerate non-minimal coupling is obtained setting
\begin{eqnarray}
\beta= \frac{\alpha_1- \x \alpha_{1\x}}{\y} \,,\;\;\;\; \beta^0_1= - \y \alpha_{1\x} \,,\;\;\;\; \beta^{\bar{1}}_2 = \alpha_1 \,,\;\;\;\; \beta^+_{0\bar{0}} =0   \,,\;\;\;\; \beta^-_{0\bar{1}} = \alpha_{1\y} \,,
\end{eqnarray}
and similarly for the conjugates, assuming as before that under a rotation $\alpha_1 \to -  \alpha_1$.

Therefore, given the very large number of fully degenerate theories with $\mathscr{B}\neq 0$ and the subsequent difficulty to obtain a classification, it seems better to understand first if these DHOME can be ghost-free and stable. If possible, this might also yield additional physical constraints sufficient to complete the classification.

\section{Disformal transformations}\label{SecDisfTransf}

In this section, we study the action of U(1)-preserving disformal transformations on HOME theories.  When invertible, these transformations should yield classically equivalent theories in the (electro) vacuum, thus, although we do not have a general proof, the nature of their degrees of freedom should also be preserved, in the sense that if a theory propagates $n$ gravitational and $m$ vectorial degrees of freedom, it should also be the case for the disformed theory. Hence, it should also be possible to classify DHOME theories in terms of disformal classes, which makes the study of disformal transformations interesting. 

\medskip

We first define (U(1)-preserving) disformal transformations of the metric, reformulate them in terms of the projectors introduced previously and establish the conditions for these transformations to be invertible. We identify the class of singular (i.e. non-invertible) transformations from which we construct mimetic-like HOME theories, similar to those introduced in scalar-tensor theories \cite{Chamseddine:2013kea, Sebastiani:2016ras, Langlois:2018jdg}, which generalise the U(1) singular transformations used in \cite{Gorji:2018aa, Hammer:2020aa, Gorji:2020aa}.

Then, we  apply disformal transformations to the most general ghost-free U(1) theory with second order field equations, consisting of the Einstein-Hilbert action and Horndeski's non-minimal coupling. This results into new ghost-free HOME theories when the transformation is invertible, or to a mimetic-like HOME theories when the transformation is singular.

 Notice that  conformal U(1)-preserving transformations of General Relativity have been studied in \cite{Gumrukcuoglu:2019ebp}, while its disformal transformations  have been investigated in \cite{Goulart:2013laa, Goulart:2020wkq, Minamitsuji:2020jvf, Minamitsuji:2021rtw, Bittencourt:2023ikm}.

\subsection{Metric field and projectors}

We parametrise any disformal transformation of the metric in a block diagonal form as follows,
\begin{eqnarray}
\tilde{g}_{\mu\nu} = \z \left(\x, \y\right) \p_{\mu\nu} + \zb \left(\x, \y\right) \pp_{\mu\nu}\,, \label{DisfMetric}
\end{eqnarray}
where $\z$ and $\zb$ are two arbitrary functions of the EM invariants. We define the ``disformality"  by 
\begin{eqnarray}
\Delta = \z - \zb \,,
\end{eqnarray}
so that $\Delta=0$ corresponds to conformal transformations. The Jacobian of this transformation and the invertibility conditions on $\z$ and $\zb$ are obtained in the next section. Notice that when parametrised in this way, the disformal transformation is invariant under the complex transformation \eqref{bartransf}, provided that the functions $\z$ and $\zb$ transform as
\begin{eqnarray}
\z \leftrightarrow \zb \,,
\end{eqnarray}
what we will assume in the following.  Using the orthogonality relations between $\p$ and $\pp$, we show that the inverse disformal metric is also block diagonal and given by\footnote{Remark that all indices are lowered and raised by the metric $g_{\mu\nu}$ and its inverse.}
\begin{eqnarray}
\left( \tilde{g}^{-1} \right)_{\mu\nu} = \z^{-1}\left(\x, \y\right) \p_{\mu\nu} + \zb^{-1}\left(\x, \y\right) \pp_{\mu\nu}\,,
\end{eqnarray}
which is defined when $\z \neq0$ and $\zb\neq 0$, what we assume from now on. Furthermore, the volume element in the disformal frame is given by 
\begin{eqnarray}
\sqrt{- \tilde{g}}= \sqrt{-g} \, |\z \zb |   \,,
\end{eqnarray}
which is real and positive for any functions $\z$ and $\zb$, contrary to the case of  disformal transformations in the context of scalar-tensor theories (see  \cite{Alinea:2020sei} for example). 

The electromagnetic invariants and the projectors transform under disformal transformations as follows,
\begin{eqnarray}
\begin{split}
\x &\longrightarrow \s \x \z^{-1}  \,,\;\;\;\;\; \p_{\mu\nu} \longrightarrow \z \p_{\mu\nu}\,,\;\;\;\;\;  \po_{\mu\nu} \longrightarrow \s \z \po_{\mu\nu} \,, \\
\y &\longrightarrow \ssb \y \zb^{-1} \,,\;\;\;\;\; \pp_{\mu\nu}  \longrightarrow \zb \pp_{\mu\nu}\,,\;\;\;\;\;  \poo_{\mu\nu} \longrightarrow \ssb \zb \poo_{\mu\nu}   \,,  \label{DisfTransfProj}
\end{split}
\end{eqnarray}
where the two signs $\s$ and $\ssb$  satisfy $\s \ssb = \text{sign}\left(  \z \zb\right)$ (which can be seen from the definition of the invariant $\mathcal{G}$)\footnote{There is a sign ambiguity in the definition of the disformal transformations on the projector $\po$ and $\poo$.}.

\medskip

Before studying their invertibility conditions, we discuss an interesting effect of a disformal transformation on the metric.  For that purpose, we consider a surface-forming unit time-like vector $n$ for the metric $g_{\mu\nu}$, i.e. $n_\mu \propto \partial_\mu \phi$ for some scalar field $\phi$ and $g^{\mu\nu} n_\mu n_\nu=-1$. We assume that the normal $\tilde n$ associated to the disformed metric $\tilde g_{\mu\nu}$ is such that $\tilde{n}\propto n$, it is  time-like and of unit norm as well (for the disformed metric $\tilde{g}_{\mu\nu}$). These hypotheses yield the transformed normal 
\begin{eqnarray}
n_\mu \longrightarrow \, \tilde{n}_\mu \; = \;  \pm  \, \sqrt{\frac{\z \zb}{\zb \bar{\psi}^2 - \z \psi^2}}  \, n_\mu  \, . \label{DisfTransfNormal}
\end{eqnarray}
We can choose the plus sign to ensure that if $n$ is future (or past) directed, so is $\tilde{n}$. Recall that $\psi$ and its conjugate $\bar{\psi}$ are scalar fields, introduced in \eqref{Projo}, which are related to the magnetic field of $F$ \eqref{psiB} and satisfy $ \bar{\psi}^2 -\psi^2 = 1$. Assuming that the disformed normal be real for generic configurations of the electromagnetic field (i.e. $\forall \psi \in \mathbb{R}$) yields the conditions $\z>0$ and ${\z}/{\zb}<1$ which lead to two distinct classes of transformations\footnote{This is to be contrasted with  disformal transformations in scalar-tensor theories defined by $\tilde{g}_{\mu\nu} = A  g_{\mu\nu} + B \partial_\mu \phi \partial_\nu \phi$ where $A,B$ and $\phi$ are scalar fields. In that case,  the same requirements (together with $\sqrt{- \tilde{g}}>0$) yield the class of transformations given by $A+ B X \geq 0$ and $A>0$, for any configurations of 
$\phi$ and $X= g^{\mu\nu}\partial_\mu \phi \partial_\nu \phi$. The reader is invited to see \cite{Bettoni:2013diz,Bekenstein:1992pj} for more details.},
\begin{eqnarray}
0 < \z  \leq \zb  \;\;\;\;\;\;\;\;\;  \text{or} \;\;\;\;\;\;\;\;\;   \zb < 0 < \z \, \label{CdtNormalDisf} \, .
\end{eqnarray}
The first one, the ``conformal component", is connected to conformal transformations. An example of invertible (see next section) transformation which belongs  to the second class, the ``non-conformal component", has been studied in \cite{Goulart:2020wkq}. Interestingly, it is given by a disformal metric proportional to Maxwell's stress-energy tensor as recalled in appendix \ref{AppDisf} where more details are presented. In particular, we obtain there the disformal transformation of the scalar fields $\{ \psi, \bar{\psi} \}$ and that of the spatial triad $\{e_\mu, b_\mu, r_\mu \}$ appearing in the 3+1 decomposition of $\po$, defined by \eqref{Projo}. It is shown that, unlike the conformal component, this class of transformations implies that the dyad $\{ b_\mu, r_\mu \}$ becomes imaginary (while remaining space-like).

\medskip

Remark that the explicit expressions for the disformal transformations of the Christoffel symbols and the Riemann curvature, given by
\begin{eqnarray}
\delta \Gamma^\sigma _{\mu\nu}{}=  \tilde{\Gamma}^\sigma _{\mu\nu}  -  \Gamma_{\mu\nu}^\sigma \,,\;\;\;\;\;\;\;\;\;\;\;\; \delta R_{\mu\nu\rho}{}^\sigma = \tilde{R}_{\mu\nu\rho}{}^\sigma - R_{\mu\nu\rho}{}^\sigma \, ,
\end{eqnarray}
can be found  in appendix \ref{AppDisf}.

\subsection{Jacobian and ``mimetic" singular transformations}\label{Sec.Mimetic}

In order for the disformed theory to possess the same degrees of freedom (number and nature) as the seed theory, the Jacobian of the transformation should be invertible.
After a short calculation using some results of appendix \ref{App:dictio}, one shows that the Jacobian is given by
\begin{eqnarray}
\frac{\partial \tilde{g}_{\mu\nu}}{\partial g_{\rho\sigma}} = \frac{\y^2 \z + \x^2 \zb}{2 \left(\x^2+\y^2 \right)} \p^{(\rho}_{(\mu} \pp^{\sigma)}_{\nu)} + \frac{\x \y \Delta}{2 \left(\x^2+\y^2 \right)} \po^{(\rho}{}_{(\mu} \poo^{\sigma)}{}_{\nu)} + \frac{1}{2}\left( \Upsilon^{\rho\sigma}_{\mu\nu} + \bar{\Upsilon}^{\rho\sigma}_{\mu\nu} \right)\,,
\end{eqnarray}
where we have introduced the notation
\begin{eqnarray}
\Upsilon^{\rho\sigma}_{\mu\nu}=\frac{\x \z_{\x}}{2}  \po^\rho{}_{(\mu} \po^\sigma{}_{\nu)}  + \left(  \z -\frac{ \x \z_{\x}}{2} \right) \p^\rho_{(\mu} \p^\sigma_{\nu)}  - \y \z_{\y} \p_{\mu\nu} \pp^{\rho\sigma} \, .
\end{eqnarray}
The eigentensors $\xi_{\mu\nu}$ of the Jacobian, defined by the relation,
\begin{eqnarray}
 \frac{\partial \tilde{g}_{\mu\nu}}{\partial g_{\rho\sigma}} \,\xi _{\rho\sigma}  = \Lambda  \, \xi_{\mu\nu} \, ,
\end{eqnarray}
can be parametrised in terms of the electromagnetic tetrad $e^I_\mu$ defined in appendix \ref{App:tetrads}, as follows,
\begin{eqnarray}
\xi_{\mu\nu} = \sum_{I\leq J} \zeta_{IJ} \, e^I_{(\mu} e^J_{\nu)} \, .
\end{eqnarray}
Using the expression of the projectors in terms of $e^I_\mu$, we obtain the 10 eigenvalues $\Lambda$ together with their associated eigentensors.
We show that the eigenvalues can be divided into three types : there are two pairs of identical and complex eigenvalues, two other pairs of degenerate eigenvalues given by $\Lambda = \z$ and $\Lambda=\zb$, which cannot vanish (otherwise the inverse metric $\tilde{g}^{-1}$ would not exist) and finally the last two eigenvalues that we denote $\Lambda_\pm$ are solutions of
\begin{eqnarray}
\Lambda^2 - \left( \z + \zb - \x \z_{\x} - \y \zb_{\y} \right) \Lambda +\left( \zb - \y \zb_{\y} \right)   \left( \z - \x \z_{\x} \right) - \x \y \z_{\y}\zb_{\x}   =0 \, .
\end{eqnarray}
These are the only eigenvalues of the Jacobian which can vanish, and their respective eigentensors are  fully covariant and given by,
\begin{eqnarray}
\xi^{\pm}_{\mu\nu} =\left( \Lambda_\pm - \zb + \y \left( \zb_{\y} - \z_{\y} \right) \right) p_{\mu\nu} +\left( \Lambda_\pm - \z + \x \left( \z_{\x} - \zb_{\x} \right) \right)  \pp_{\mu\nu} \, .
\end{eqnarray}
Therefore any transformation satisfying\footnote{This equation can be reformulated as follows
\begin{eqnarray}
\epsilon_{\x} \overline{\epsilon}_{\y} - \epsilon_{\y} \overline{\epsilon}_{\x} \; = \; 0 \, ,
\end{eqnarray}
where $\epsilon=\varphi/x$ and $\overline{\epsilon}=\overline{\varphi}/\y$. The same condition has already been encountered in \eqref{F1C0I1} and we refer to the discussion there regarding its solutions.}
\begin{eqnarray}
\left( \z - \x \z_{\x} \right) \left( \zb - \y \zb_{\y} \right)- \x \y \z_{\y}\zb_{\x}  =0
 \label{NoninvertibilityJack}
\end{eqnarray}
is non-invertible with one null eigenvalue, while if it satisfies the additional condition 
\begin{eqnarray}
\label{Invert2cond}
 \z - \x \z_{\x} + \zb -  \y \zb_{\y} =0 \, ,
\end{eqnarray}
it has two null directions. In that case, the disformal transformation is necessarily of the form $\z = \x / \x_0$ and $\zb = \y / \y_0$ where $\x_0$ and $\y_0$ are constants, as shown in appendix \ref{App:disformal} where we present an alternative way to compute the (non-)invertibility conditions of a disformal transformation.

\medskip

Let us now focus on singular conformal transformations. In this case the Jacobian has at most one null eigenvalue (given that its null eigentensor must be proportional to the metric) and the singular transformations are of the form
\begin{eqnarray}
  \z=\zb = \x c\left(\frac{ \y }{\x} \right) \,,\label{MimeticConf}
\end{eqnarray}
for any arbitrary function $c$ of the ratio of the EM invariants, in agreement with the result of \cite{Gumrukcuoglu:2019ebp} (see the equation (9) therein). This is to be contrasted with the ``mimetic" transformation in the context of scalar-tensor theories 
\begin{eqnarray}
\tilde{g}_{\mu\nu} = (g^{\rho\sigma} \partial_\rho \phi  \partial_\sigma \phi) \, g_{\mu\nu} \, ,
\end{eqnarray}
which is the unique non-invertible conformal transformation. 

In both cases $\tilde{g}_{\mu\nu}$ is invariant under conformal transformations of $g_{\mu\nu}$, i.e. under the transformation $g_{\mu\nu} \to e^{\sigma} g_{\mu\nu}$ for any scalar field $\sigma$. Indeed, this can be easily seen when one express \eqref{MimeticConf} in terms of the field strength, 
\begin{eqnarray}
\tilde{g}_{\mu\nu} = \left(\mathcal{F}^2+\mathcal{G}^2 \right)^{\frac{1}{4}} C\left(\frac{\mathcal{G}}{\mathcal{F}}\right)  g_{\mu\nu} \, ,
\end{eqnarray}
where $C$ is related to $c$ in a straightforward way. Recalling that 
\begin{eqnarray}
\mathcal{F}= -\frac{1}{4}g^{\mu\rho}g^{\nu\sigma} F_{\rho\sigma}F_{\mu\nu} \,,\;\;\;\;\; \mathcal{G}^2=-\frac{1}{4^3}  g^{\mu\alpha} g^{\nu\beta}  g^{\rho\gamma}   g^{\sigma\delta}  F_{[\rho\sigma} F_{\mu\nu]}F_{\alpha\beta} F_{\gamma\delta}\,,
\end{eqnarray}
 the invariance under conformal transformation becomes obvious. 

\subsection{Disformal transformation of ghost-free theories}\label{Sec.GhostFree}

In this section, we study disformal transformations of the Einstein-Hilbert Lagrangian $L_{\text{GR}}$ as well as that of Horndeski's unique U(1) non-minimal coupling $L_{\text{NMC}}$. This yields new DHOME which are by construction free of Ostrogradski ghosts and incorporate, as far as we know, all the previously-studied HOME theories with this property \cite{Gumrukcuoglu:2019ebp,Goulart:2013laa, Goulart:2020wkq, Minamitsuji:2020jvf, Minamitsuji:2021rtw, Bittencourt:2023ikm}. Interestingly, we have managed to obtain rather simple expressions for these disformed theories. Given the very complicated nature of U(1) disformal transformations, this shows how powerful the ``projector formalism" used in this paper is compared to the use of the field strength. Moreover, if a more complete classification of DHOME theories than ours should be carried out, this would enable to identify which of these is actually GR+NMC in disguised.

\subsubsection{A general discussion on the action}

Let us consider the following action
\begin{eqnarray}
I_{\text{disf}}= I_{\text{grav}} + \int d^4 x \sqrt{-\mathfrak{g}}\; \mathcal{L}_{\text{m}}\left[\Psi , \mathfrak{g}_{\mu\nu} \right]+  \int d^4 x \sqrt{-\mathfrak{g}} \,  \lambda^{\mu\nu\sigma} \nabla_{[\mu} F_{\nu\sigma]} \, ,
\end{eqnarray}
where $\mathcal{L}_{\text{m}}$ contains matter fields $\Psi$ minimally coupled to $\mathfrak{g}_{\mu\nu} \in \{ g_{\mu\nu},\tilde{g}_{\mu\nu}\}$, in particular it may contain a non-linear electrodynamics term $V\left(\mathcal{F},\mathcal{G}\right)$, while the gravitational action is given by 
\begin{eqnarray}
I_{\text{grav}}= \int d^4 x \Big( L_{\text{GR}}\left[\tilde{g}_{\mu\nu}\left( g_{\mu\nu} , F_{\rho\sigma}\right) \right] + \gamma  L_{\text{NMC}}\left[\tilde{g}_{\mu\nu}\left( g_{\mu\nu} , F_{\rho\sigma}\right),F_{\rho\sigma} \right] \Big) \, ,
\end{eqnarray}
where $\gamma$ is a coupling constant.

There are several possibly interesting routes to follow from this gravitational action. The first one is to consider an invertible disformal transformation with $\mathfrak{g}_{\mu\nu} = g_{\mu\nu}$, so that the theory differs from the usual Maxwell-Einstein-Horndeski theory by its non-standard matter coupling, but remains free of Ostrogradski ghost. Notice that the U(1) theory obtained from conformal transformation with $\gamma=0$ was studied from this point of view in \cite{Gumrukcuoglu:2019ebp}.

It is also possible to consider singular transformations \eqref{NoninvertibilityJack} with $\mathfrak{g}_{\mu\nu} = \tilde{g}_{\mu\nu}$, in which case the theory is analogous to mimetic gravity in the context of scalar-tensor theories. In this case, the resulting action should have more degrees of freedom than the two of gravity (and the two of the electromagnetic field if either $\gamma\neq0$, or if $\gamma=0$ and $V$ is non-trivial) and it would be interesting to understand under which conditions (if any) they are not ghosts-like. 
\medskip 

Finally, notice that unlike what might be expected from a naive counting of degrees of freedom, the flat space limit ($g_{\mu\nu}=\eta_{\mu\nu}$) of these theories do propagate additional ghost-like degrees of freedom \cite{Colleaux:2024ndy, Gumrukcuoglu:2019ebp}. However, if both the metric and gauge (or two-form) field equations are restricted to flat space, the naive counting is correct\footnote{A question following from this simple observation is the following : given a higher-order field theory in flat space, can it always be made ghost-free in curved space introducing suitable gravitational non-minimal couplings ? Are there equivalent mechanisms for other fields than gravity ?}. This is illustrated in \cite{Gumrukcuoglu:2019ebp}, from the study of the U(1) conformal transformation of GR.

\subsubsection{U(1) frame of General Relativity \& ``gauge-mimetic" gravity}

Under a generic U(1) preserving disformal transformation, General Relativity is mapped into the following higher-order Maxwell-Einstein theory,
\begin{eqnarray}
\boxed{L_{\text{GR}}=\sqrt{-\tilde{g}} \tilde{R} =s \bar{s}\sqrt{-g} \left( \varphi R - \Delta  R_{\mu\nu} \p^{\mu\nu}+ \mathscr{L} +\bar{\mathscr{L}}  + \nabla_{[\mu} \left[  \left(\tilde{g}^{-1}\right)^{\rho\sigma} \delta\Gamma^\mu_{\rho]\sigma} \right] \right)}  \label{DisformeGR}
\end{eqnarray}
where we recall that $\Delta = \varphi - \bar{\varphi}$ and
\begin{eqnarray}
\boxed{\begin{split}
\mathscr{L} =& - \frac{\Delta^2}{2\varphi} \left( \mathcal{Q}_2^{\bar{0}} - \mathcal{Q}_3^0 \right) - \frac{\Delta}{\bar{\varphi}} \left( \bar{\varphi}_{\x} \mathcal{Q}^+_{1\bar{0}} + \bar{\varphi}_{\y} \mathcal{Q}^-_{1\bar{0}} \right) + \frac{1}{\varphi\bar{\varphi}} \left( \bar{\varphi} \varphi_{\x}\varphi_{\y}+\bar{\varphi}_{\x} \varphi \varphi_{\y}+\bar{\varphi}_{\y} \varphi \varphi_{\x}\right) \mathcal{Q}_0 \\
&+ \frac{1}{2\varphi\bar{\varphi}} \left[  \varphi_{\x} \left(\varphi_{\x}   \bar{\varphi} +2 \varphi \bar{\varphi}_{\x} \right) \mathcal{Q}_+ + \varphi_{\y} \left(\varphi_{\y}   \bar{\varphi} +2 \varphi \bar{\varphi}_{\y} \right) \mathcal{Q}_- \right]
\end{split}}
\end{eqnarray}
Thus, it involves only $6\times2$ independent invariants quadratic in the derivative of the field strength. Remark that no (other) boundary term and no dimensional or Bianchi identities have been used to obtain this result, other than the DDIs encoded into the projectors, i.e. the Cayley-Hamilton identity.  It is noticeable that $\mathscr{L}$ (and its conjugate) can be decomposed into three pieces, two of these being related by a transformation acting on the first \eqref{FirstClassInv} and second class of invariants \eqref{SecondClassInv}, i.e.
\begin{eqnarray}
\mathscr{L}=- \frac{\Delta^2}{2\varphi} \left( \mathcal{Q}_2^{\bar{0}} - \mathcal{Q}_3^0 \right) + \ell + \hat{\ell} \, ,
\end{eqnarray}
where
\begin{eqnarray}
\ell =  -  \frac{\Delta \bar{\varphi}_{\x}}{\bar{\varphi}}   \mathcal{Q}^+_{1\bar{0}}  + \left(\frac{\varphi_{\x}}{2\varphi}+\frac{\bar{\varphi}_{\x}  }{\bar{\varphi}}\right) \left( \varphi_{\y} \mathcal{Q}_0 +\varphi_{\x}  \mathcal{Q}_+ \right) \,,
\end{eqnarray}
where the ``hat" transformation partially interchange $\x$ and $\y$ as follows
\begin{eqnarray}
\mathcal{Q}^+ \leftrightarrow \mathcal{Q}^- \,,\;\;\;\; \varphi^a_{\x} \leftrightarrow \varphi^a_{\y} \,, \label{hatTransf}
\end{eqnarray}
with $\varphi^a= \{\varphi,\bar{\varphi}\}$. This additional transformation is quite useful to simplify the expressions of more complicated disformal theories in terms of the invariants classified in \ref{Classification}. For conformal transformations, $\Delta=0$, and up to the previous boundary term, it reduces to the expected expression
\begin{eqnarray}
\begin{split}
\sqrt{-\tilde{g}} \tilde{R}  &= \sqrt{-g} \left( \varphi R +  \frac{3}{2\varphi} \left(  \varphi_{\x}^2  \left( \mathcal{Q}_+ + \bar{\mathcal{Q}}_- \right) + \varphi_{\y}^2 \left(  \bar{\mathcal{Q}}_+ + \mathcal{Q}_-   \right)+ 2 \varphi_{\x} \varphi_{\y} \left( \mathcal{Q}_0 + \bar{ \mathcal{Q}}_0 \right) \right) \right) \\
&= \sqrt{-g} \Big( X^2 R  +6 \,g^{\mu\nu} \partial_\mu X \partial_\nu X \Big)\,,
\end{split}
\end{eqnarray}
where $X\left(\x,\y\right) = \sqrt{\varphi\left(\x,\y\right)}$. The Hamiltonian formalism and degrees of freedom of this theory restricted to flat space-time has been studied in \cite{Colleaux:2024ndy}.
 Using the non-invertible conformal transformation \eqref{MimeticConf}, we obtain the analogue of mimetic gravity when the scalar is replaced by a U(1) gauge field (or a two-form field) setting $\varphi=\x c\left(\frac{ \y }{\x} \right)$ in the previous expression, or in terms of the field strength
\begin{eqnarray}
\boxed{X\left(\mathcal{F},\mathcal{G}\right) = \left(\mathcal{F}^2+\mathcal{G}^2 \right)^{\frac{1}{8}} C\left(\frac{\mathcal{G}}{\mathcal{F}}\right)} \,.
\end{eqnarray}

\subsubsection{U(1) frame of Horndeski's non-minimal coupling}

We now turn to the unique U(1) non-minimal coupling yielding second order field equations in four dimensions, written in terms of the disformal metric $\tilde{g}$, and consider the following Lagrangian
\begin{eqnarray}
L_{\text{NMC}}= -\frac{1}{16} \delta^{\mu}_{[\alpha} \delta^{\nu}_{\beta}\delta^{\rho}_{\gamma}\delta^{\sigma}_{\delta]}  \sqrt{-\tilde{g}}  \left(\tilde{g}^{-1}\right)^{\kappa\alpha} \tilde{R}_{\mu\nu\kappa}{}^{\beta} F_{\rho\sigma}\tilde{F}^{\gamma\delta}   + \mathcal{B}_{\text{NMC}} \,,
\end{eqnarray}
where we simply added the following boundary term
\begin{eqnarray}
\mathcal{B}_{\text{NMC}} = -\frac{1}{8} \delta^{\mu}_{[\alpha} \delta^{\nu}_{\beta}\delta^{\rho}_{\gamma}\delta^{\sigma}_{\delta]}  \sqrt{-g} \,  \nabla_\mu \left[ \z \zb F_{\rho\sigma} \tilde{F}^{\gamma\delta}  \left(\tilde{g}^{-1}\right)^{\alpha\kappa}  \delta \Gamma^\beta_{\nu\kappa} \right]  
\end{eqnarray}
and used the following notation
\begin{eqnarray}
\tilde{F}^{\gamma\delta}= \left(\tilde{g}^{-1}\right)^{\gamma\mu}\left(\tilde{g}^{-1}\right)^{\delta\nu}  F_{\mu\nu} \,.
\end{eqnarray}
After a long calculation and making use of Bianchi and dimensional identities, we obtain 
\begin{eqnarray}
\boxed{L_{\text{NMC}} =-\frac{s \bar{s}}{2 \z\zb} \sqrt{-g}  \left(  \x \y \left( \z + \zb \right) \mathcal{R}_1 +  \y^2 \z \mathcal{R}_2 -\x^2 \zb  \mathcal{R}_3+ \mathscr{L}_{\text{NMC}}-\bar{\mathscr{L}}_{\text{NMC}}  \right)} \label{DisfNMC}
\end{eqnarray}
where 
\begin{eqnarray}
\boxed{\mathscr{L}_{\text{NMC}}=\Delta \left( \mathscr{Q}_0- \frac{\z}{\zb} \mathscr{Q}_1 + \mathscr{Q}_2 -\mathscr{Q}_3 \right) -2 \x \left( \zb_{\x} \mathcal{Q}_+ + \zb_{\y}\mathcal{Q}_0  \right) + \ell_{\text{NMC}} + \hat{\ell}_{\text{NMC}} } \label{DisfHorn}
\end{eqnarray}
involves the following combinations of quadratic invariants
\begin{eqnarray}
\begin{split}
\mathscr{Q}_0 &= \frac{3 \y^2}{2} \left( \mathcal{Q}_2^{\bar{0}} - \mathcal{Q}_3^0 \right) \,, \;\;\;\;\; \mathscr{Q}_1 = \frac{\y^2}{2} \left( 2 \mathcal{Q}_1^{\bar{0}} + \mathcal{Q}_2^{\bar{0}} +\mathcal{Q}_3^{0}  \right)\,,\\
\mathscr{Q}_2 &= \x \y \left(\mathcal{Q}_1^{\bar{1}} +\mathcal{Q}_2^{\bar{1}}  \right) \,, \;\;\;\;\;\;\;  \mathscr{Q}_3 = \y \mathcal{Q}^{+}_{0\bar{1}}  + \x  \mathcal{Q}^{-}_{0\bar{1}} \,,
\end{split}
\end{eqnarray}
 and 
 \begin{eqnarray}
\boxed{ \begin{split}
 \ell_{\text{NMC}}& = \frac{\x^2 \z_{\x}}{\z} \left( \zb_{\y} \mathcal{Q}_0 +\zb_{\x} \mathcal{Q}_+ \right)- \frac{\y^2 \z_{\x}}{2\zb} \left( \z_{\y} \mathcal{Q}_0 +\z_{\x} \mathcal{Q}_+ \right) 
\\&+ \frac{\x \y}{\z \zb} \left( \z^2 \zb_{\x}+\zb^2 \z_{\x} - 2 \z \zb \zb_{\x} \right)  \mathcal{Q}^{+}_{0\bar{1}}+ \left( \frac{\x^2 \zb \zb_{\x} + \y^2 \z \z_{\x}}{\zb} \right) \mathcal{Q}^{+}_{1\bar{0}}
\end{split}}
 \end{eqnarray}
 while $\hat{\ell}_{\text{NMC}}$ is obtained from the previous expression via a ``hat" transformation \eqref{hatTransf}.  For conformal transformations, $\Delta=0$, it reduces to 
 \begin{eqnarray}
L_{\text{NMC}} =\frac{\sqrt{-g}}{\z}  \left(\frac{1}{4} \,\hodge F^{\mu\nu} \hodge F^{\rho\sigma}  R_{\mu\nu\rho\sigma} -\frac{1}{2 \z} \left(\mathscr{L}_{\text{NMC}}-\bar{\mathscr{L}}_{\text{NMC}}  \right) \right)\,,
\end{eqnarray}
where
\begin{eqnarray}
\begin{split}
\mathscr{L}_{\text{NMC}} &= \left(\x^2 + \y^2 \right) \left( \z_{\x} \mathcal{Q}^{+}_{1\bar{0}} + \z_{\y} \mathcal{Q}^{-}_{1\bar{0}} \right) + \frac{\z_{\y}^2}{2 \z}\left( 2 \x^2 - \y^2 \right) \mathcal{Q}_- \\
&-\frac{\z_{\x}}{2\z} \left( 4 \x \z - 2 \x^2 \z_\x +\y^2 \z_{\x} \right) \mathcal{Q}_+ - \frac{\z_{\y}}{\z}\left(  \x \z - 2 \x^2 \z_{\x} + \y^2 \z_{\x}\right) \mathcal{Q}_0 \,.
\end{split}
\end{eqnarray}
Although more complicated than its GR counterpart, it is quite remarkable that a relatively simple expression can be obtained for this theory as well. This provides another concrete example of a degenerate higher-order Maxwell-Einstein theory without Ostrogradski ghost beyond the U(1) conformal transformation of GR.
 
\section{Discussion}

In this paper, we have laid the foundation for a comprehensive and systematic study of higher-order Maxwell-Einstein (HOME) theories. 
We have restricted to the quadratic case and classified as much as possible theories with a degenerate kinetic matrix. 

\medskip
Our main result has been to obtain the generalisation of quadratic Horndeski scalar-tensor theory for a U(1) gauge field, given by \eqref{L*}, whose field equations involve respectively second and third order derivatives of the metric and gauge field. This theory is parametrised by two free functions of one EM invariant each. We showed that GR and Horndeski's unique non-minimal coupling \cite{Horndeski:1976gi} can be obtained when these functions are respectively assumed to be constant and quadratic \eqref{GRNMC}, while linear functions yield the unique pair of (even) totally degenerate theories \eqref{Ltde}, which turn out to be both conformal invariant and U(1)-preserving disformal invariant. A Galileon-like expression of this theory for cubic functions is given by \eqref{Galileonlike}.

Indeed, assuming field equations containing second order derivatives of the metric and at most third order derivatives of the scalar field uniquely yields Horndeski's theory in scalar-tensor theories, meaning that the usual assumption of second order field equations can be weakened. Furthermore, the no-go result of \cite{Deffayet:2013tca} regarding the non-existence of U(1) Galileons in flat space stands in stark contrast with the usual theorems classifying theories with second order field equations, because contrary to the latter, it is dimensionally independent, and thus much stronger.  This is another hint that the assumption of second order field equations for U(1) theories is in some sense too strong. Therefore, it would be quite interesting to obtain a generalisation of these theorems for a generic four-dimensional higher-order Maxwell-Einstein theory $L\left(\partial ... \partial g, \partial ... \partial A \right)$, restricting to our weaker assumptions.

\medskip

Among the other degenerate theories we have obtained and classified, we would like to highlight the partially degenerate theories \eqref{RC0}, which also generalise GR and Horndeski's NMC, as well as the fully degenerate theories in the electromagnetic sector given by \eqref{R1theory}, \eqref{Squareroot} and \eqref{BeyondNMC}. Interestingly, we have also found two classes of degenerate theories, given by \eqref{R1theoryConf} and \eqref{ConfTh2}, which are conformally invariant. Furthermore, we applied U(1) disformal transformations on GR \eqref{DisformeGR} and Horndeski's NMC \eqref{DisfNMC} to obtain new ghost-free DHOME theories. The analogue of Mimetic theories for a gauge field have also been derived. Following the analysis of \cite{Domenech:2023ryc} in the context of HOST theories and our result \eqref{MimeticTDE} which shows that the totally degenerate even DHOME are mimetic, it would be interesting to understand if all conformally and disformally invariant HOME theories are mimetic as well.

Concerning (quadratic) two-form field theories, we have obtained all those admitting second order field equations for the metric and first order for the two-form field, see \eqref{Ltdo} and \eqref{Ltf}.

\medskip

In a future work, we intend to carry out a detailed Hamiltonian analysis of these gauge theories, what will enable to characterise their degrees of freedom and assess if there exist ghost-free DHOME theories beyond those generated by disformal transformation, which we obtained here. Given the close similarity between quadratic Horndeski and the theory \eqref{L*}, if we could show that it is ghost-free, we would have one more argument (if needed) in favour of the more general criterion of degeneracy, compared to that of second order field equations.

Following the existing knowledge on scalar-tensor theories, it would be interesting to construct an analogue of the unitary gauge and generalise our action to include quartic interactions, meaning schematically
\begin{eqnarray}
\int \left( \mathscr{H}\left( g,F \right) \mathcal{R}^2 + \mathscr{O}\left( g,F \right)  \mathcal{R} \left( \nabla F\right)^2 + \mathscr{M}\left( g,F \right) \left( \nabla F\right)^4 \right)\,.
\end{eqnarray}
This would be analogous to cubic higher-order scalar-tensor theories, which famously include both ``quintic" Horndeski gravity and DHOST theories \cite{BenAchour:2016fzp}, for a gauge field.

As emphasised throughout this paper, our formalism naturally extends to non-integrable two-form fields, what is analogous to generalised Proca theories, and a systematic study of these theories, in particular the possible construction of ghost-free degenerate two-form theories would be desirable. In terms of more general field content, a natural generalisation of our work would be to consider non-Abelian Yang-Mills gauge fields and, in the non-integrable case, the corresponding multi-two-form field theories.

Finally, it would be interesting to investigate the possible physical implications of these theories. In particular, given the substantial amount of exact solutions existing in the Maxwell-Einstein system \cite{Stephani:2003tm} or more generally in GR with non-linear electrodynamics (see e.g. the recent solutions in ModMax \cite{BallonBordo:2020jtw, Barrientos:2024umq, Bokulic:2025usc}), but also when Horndeski's NMC theory is included \cite{Horndeski:1978dw, Gurses:1978zs,Feng:2015sbw,ACNMC}, the ghost-free DHOME \eqref{DisformeGR} and \eqref{DisfNMC} could be used as an interesting laboratory to study black hole and cosmological solutions, as well as exact non-linear gravitational waves, as it has been done recently using scalar-tensor theories in
\cite{BenAchour:2024hbg,BenAchour:2024tqt, BenAchour:2024zzk}.

\acknowledgments
The work of K.N. was supported by the French National Research Agency (ANR) via Grant No. ANR-22-CE31-0015-01 associated with the project Strong.
We would like to warmly thank David Langlois for his collaboration and participation in the early stages of this project and for suggesting the name DHOME. We would also like to thank Mokhtar Hassaine for very useful discussions. We acknowledge the use of the xAct package for Mathematica  \cite{martin2002xact}.

\appendix

\section{Classification of higher-order Lagrangians} \label{Classification of higher-order Lagrangian}
In this appendix, we provide a new  classification of higher-order Maxwell-Einstein Lagrangians which is equivalent to the one obtained recently in  \cite{Colleaux:2023cqu}. This classification is based on the parametrisation of the curvature tensor $F_{\mu\nu}$ in terms of $\po_{\mu\nu}$ and $\poo_{\mu\nu}$, which drastically simplifies the treatment of dimensionally dependent identities (DDIs).

\subsection{Independent electromagnetic matrices}\label{PropProj}

 One can easily obtain an upper bound on the number of independent matrices built from the curvature $F$ and metric field $g$ using the Cayley-Hamilton identity in four dimensions\footnote{For more details about this point, see  \cite{Colleaux:2023cqu} for example.},
\begin{eqnarray}
F_4^{\mu\nu} = (\x \y)^2 g^{\mu\nu} + \left( \x^2- \y^2 \right) F_2^{\mu\nu} \,,  \label{CH}
\end{eqnarray}
where we introduced the notation $F_I$ for the different powers  ($I\geq 1$) of $F$, i.e.
\begin{eqnarray}
F_0= g \, , \qquad F_I^{\mu\nu} = F^{\mu}{}_{\alpha}  F_{I-1}^{\alpha\nu} \, . 
\end{eqnarray}
As we also have the relation,
\begin{eqnarray}
F_3 = \x \y \,\hodge F +\left(\x^2 -\y^2\right) F \,,
\end{eqnarray}
it is clear that the vector space of independent matrices built from powers of $F$ and $g$ in four dimensions can be spanned by the family $\{g, F, \hodge F,F_2\}$.

\subsection{Derivatives of projectors}\label{Dproj}

A priori, any derivatives of $F$ can be trivially decomposed as a sum of the  derivatives $\{ \nabla \x , \nabla \y , \nabla q_I \}$ where  the matrices $q_I \in \{\po,\poo,p,\bar{p} \}$ have been defined in \eqref{quartet}. 

However, these derivatives are not independent.
Indeed, from \eqref{asprojdef}, it is clear that the derivatives of symmetric projectors ($p$ and $\bar{p}$) can be transformed into derivatives of the antisymmetric ones $(\po, \poo)$. Moreover, from the definitions and the orthogonality relations satisfied by the projectors, it is straightforward to show that 
\begin{eqnarray}
q_I^{\mu\nu}  \nabla_\alpha q_J{}_{\mu\nu}=0  \,,\;\;\;\;\;  q_i^{\mu \rho} q_j^{\nu \sigma}  \nabla_\alpha q_I{}_{\mu\nu}=0   \,,\;\;\;\;\;  q_{\bar{i}}^{\mu \rho} q_{\bar{j}}^{\nu \sigma}  \nabla_\alpha q_I{}_{\mu\nu}=0  \,, \label{DprojEq}
\end{eqnarray}
for any  $I,J \in \{0,1,\bar{0} , \bar{1}\}$, $i,j \in \{0,1\}$ and $\bar{i},\bar{j}=\in \{\bar{0} , \bar{1}\}$. Therefore, only terms of the form $q_i^{\mu\rho} q_{\bar{i}}^{\nu\sigma} \nabla_\alpha q_{I\mu\nu}$ are non vanishing.

In addition to that, using the orthogonality relations and  \eqref{asprojdef}  once again, it is clear that we can reduce further these terms to $q_i^{\mu\rho} q_{\bar{i}}^{\nu\sigma} \nabla_\alpha o_{\mu\nu}$. As a consequence, $\nabla F$ can always be decomposed as a sum of the terms $\{ \nabla \x , \nabla \y , \nabla \po \}$. 

\subsection{Dimensionally dependent identities}
As we said above, the parametrisation of the tensors $F$ in terms of $\po$ and $\poo$, and the decomposition of these antisymmetric tensors in terms of a tetrad (as shown in appendix \ref{App:tetrads}) drastically simplifies the treatment of DDIs. Indeed,  it becomes straightforward to show that the left-over dimensionally dependent identities involving quadratic invariants are 
\begin{eqnarray}
\begin{split}
& \mathcal{Q}_1^1 - \mathcal{Q}_2^1 =0\,, \\
&\mathcal{Q}_1^{\bar{0}} + \mathcal{Q}_2^{\bar{0}} - \mathcal{Q}_3^{\bar{0}} =0\,, \;\;\;\; \left( \bar{0} \to \bar{1} \right)=0\,, \\
&\mathcal{Q}_1^0 - \mathcal{Q}_2^0 + \mathcal{Q}_3^{\bar{0}} =0\,, \;\;\;\; \left( 0 \to \bar{1}, \bar{0}\to 1 \right)=0\,, \\
&\mathcal{Q}_1^{\bar{0}} - \mathcal{Q}_2^0 + \mathcal{Q}_3^0 =0\,,   \label{6DDIs}
\end{split}
\end{eqnarray}
together with their conjugate. It is interesting to notice that in this formalism, given that each $q_I$ is effectively two-dimensional, the non-trivial DDIs (i.e. those which are not given by \eqref{DprojEq}) are antisymmetrisations over three indices, just like Bianchi identities, while all DDIs deriving from the Cayley-Hamilton identity are automatically satisfied by definition of the projectors.

\subsection{Bianchi identity}
So far, the analysis of the previous subsections apply to any two-form field $F$ in four dimensions. Considering  the classification of higher order electromagnetic invariants, imposing that $F$ is the curvature of a U(1) connection, reduces further the dimensionality of the basis of invariants.  As shown in \cite{Colleaux:2023cqu}, there are $7\times2$ Bianchi identities  to take into account in that case to classify the kinetic terms. Using the projectors $q_I$, it is clear that these identities can be obtained from the relations
\begin{eqnarray}
\left( q_I q_J \right)^{\mu\nu\rho\sigma} \nabla_\mu z_a \nabla_{[\nu} F_{\rho\sigma]} =0 \,,\;\;\;\; \left( q_I q_J q_K \right)^{\mu\nu\rho\sigma\alpha\beta} \nabla_\mu o_{\nu\rho} \nabla_{[\sigma} F_{\alpha\beta]}=0\,.
\end{eqnarray}
When written in terms of the invariants defined previously, they can be shown to be given by
\begin{eqnarray}
\begin{split}
 \x \mathcal{Q}^{1}_1 - \mathcal{Q}^+_{0\bar{0}} &= 0\,,\\
 \y \mathcal{Q}^{1}_2 +  \mathcal{Q}^+_{1\bar{1}} &= 0\,,\\
 \x \mathcal{Q}_1^{\bar{1}} - \y \mathcal{Q}_1^{0} - \mathcal{Q}^+_{0\bar{1}} &= 0\,,\\
 \y \mathcal{Q}_1^{\bar{1}} - \x \mathcal{Q}_1^{\bar{0}} + \mathcal{Q}^+_{1\bar{0}} &= 0\,,\\
\mathcal{Q}_{\bar{0}} - \x \mathcal{Q}^-_{1\bar{0}} + \y \mathcal{Q}^-_{0\bar{1}}   &=0 \,,\\
\mathcal{Q}_{\bar{1}} - \x \mathcal{Q}^-_{1\bar{1}}  - \y \mathcal{Q}^-_{0\bar{0}}   &=0 \,, \label{Bianchi}
\end{split}
\end{eqnarray}
and 
\begin{eqnarray}
\mathcal{Q}_+ =  \x^2 \mathcal{Q}_1^{\bar{0}} - 2 \x \y \mathcal{Q}_1^{\bar{1}} + \y^2 \mathcal{Q}_1^{0}  \,,
\end{eqnarray}
together with their $7$ conjugated identities. 

Let us remark that only the first two and the last identities in \eqref{Bianchi} involve odd invariants. As a consequence, considering in addition the odd DDIs, there are $3\times 2$ independent odd invariants remaining  and $8\times 2$ even ones at this stage.

\subsection{Boundary terms}
Let us now classify the possible boundary terms relating the previous invariants together. In order to remain within the class of theories linear in the Riemann tensor or quadratic in the derivative of the field strength, the boundary terms must be of the form 
\begin{eqnarray}
 \nabla_\sigma \left( \Phi\left(\x,\y\right) Z^{\mu\nu[\sigma\rho]} \nabla_\rho F_{\mu\nu} \right)\,,
\end{eqnarray}
for some function $\Phi$ of the EM invariants and some tensor $Z$, polynomial in the metric and field strength.

We start considering the odd boundary terms which can be divided into 2 classes. The first class of $(2\times 2)$ boundary terms can be written as
\begin{eqnarray}
\mathcal{B}_a= \nabla_\sigma \left( \Phi\left(\x,\y\right) V_a^\sigma \right) \;\;\;\; \text{with} \;\;\; V_a^\sigma = \po^{\rho\sigma} \nabla_\rho z_a \,,
\end{eqnarray}
together with the conjugate vectors and associated divergences. The second class is given by 
\begin{eqnarray}
V^\sigma =  \pp^{\rho[\sigma} \p^{\mu]\nu} \nabla_\mu \po_{\rho\nu} \,,
\end{eqnarray}
and its conjugate $\bar{V}$. Thus, there are $3\times 2$ odd boundary terms whose divergences decompose in terms of the algebraic basis as follows,
\begin{eqnarray}
\begin{split}
\mathcal{B}_+ \left[ \Phi \right]&= - \Phi \mathscr{Q}_+ + i \Phi_{\y} \bar{\mathcal{Q}}_1 \,, \\
\mathcal{B}_- \left[ \Phi \right]&= - \Phi \mathscr{Q}_- + i \Phi_{\x} \bar{\mathcal{Q}}_1  \,, \\
\mathcal{B}_{\phantom{a}} \left[ \Phi \right]&=- \Phi_{\x} \mathscr{Q}_+ - \Phi_{\y} \mathscr{Q}_-   \,, \label{BTodd}
\end{split}
\end{eqnarray}
where 
\begin{eqnarray}
\mathscr{Q}_+=\mathcal{Q}^-_{0\bar{0}} +i \bar{\mathcal{Q}}^+_{1\bar{1}} \,,\;\;\;\; \mathscr{Q}_- = \mathcal{Q}^+_{0\bar{0}} -i \bar{\mathcal{Q}}^-_{1\bar{1}}\,.
\end{eqnarray}
Their conjugates $\bar{\mathcal{B}}_a\left[\bar{\Phi} \right]$ and $\bar{\mathcal{B}}\left[\bar{\Phi} \right]$ are obtained by assuming that the conjugation transformation extends to the scalar field according to
\begin{eqnarray}
\Phi  \leftrightarrow \bar{\Phi}  \,.
\end{eqnarray}
Furthermore, the set of these $6$ odd boundary terms $\{ \mathcal{B}_a,\bar{\mathcal{B}}_a,\mathcal{B},\bar{\mathcal{B}}\}$ can be reduced to only $3$ independent ones when one takes into account the following Bianchi identity,
\begin{eqnarray}
\mathscr{V}= \left( V_- -\y V \right)  + i\left( \bar{V}_- -  \x \bar{V} \right)  =0 \,,
\end{eqnarray}
and also the fact that there exists a couple of divergenceless vectors $\mathcal{V}$ and $\bar{\mathcal{V}}$ given by 
\begin{eqnarray}
\mathcal{V}^\sigma \left[ \Phi \right]= \nabla_\rho \left( \Phi\left(\x,\y\right) \po^{\rho\sigma} \right) = \Phi V^\sigma - \sum_a \Phi_{a} V^{\sigma}_a  \,,
\end{eqnarray}
where $\Phi_a = \partial_a \Phi$, for $a \in \{\x, \y\}$. Remark that these three vectors are independent, except for the following choice of scalars
\begin{eqnarray}
\mathscr{V} = \bar{\mathcal{V}} \left[\x  \right] -\mathcal{V} \left[   \y \right]  \,.
\end{eqnarray}

Finally, there are 2 independent even vectors, which are proportional to their own conjugates, and given by
\begin{eqnarray}
\begin{split}
V_0^\sigma &=  \pp^{\rho[\sigma} \po^{\mu]\nu} \nabla_\mu \po_{\rho\nu} \,, \\
V_1^\sigma &=  \p^{\rho[\sigma} \poo^{\mu]\nu} \nabla_\mu \po_{\rho\nu} \,.
\end{split}
\end{eqnarray}
Their divergences yield the following even boundary terms involving the curvature,
\begin{eqnarray}
\begin{split}
\mathcal{B}_0 \left[ \Phi \right]&= \Phi \left(- \mathcal{R}_0 + \bar{\mathcal{Q}}_1^{\bar{0}} + \bar{\mathcal{Q}}_2^{\bar{0}} + \mathcal{Q}_1^{\bar{0}}+ \mathcal{Q}_2^{\bar{0}}  \right) -\Phi_{\x} \left( \bar{\mathcal{Q}}^+_{1 \bar{0}}+\mathcal{Q}^-_{1 \bar{0}} \right) -\Phi_{\y} \left(  \mathcal{Q}^+_{ 1\bar{0}}+\bar{\mathcal{Q}}^-_{1 \bar{0}} \right)  \,, \\
\mathcal{B}_1 \left[ \Phi \right]&=   \Phi \left(- \mathcal{R}_1 + \bar{\mathcal{Q}}_1^{\bar{1}} - \bar{\mathcal{Q}}_2^{\bar{1}} - \mathcal{Q}_1^{\bar{1}}+ \mathcal{Q}_2^{\bar{1}}  \right) -\Phi_{\x} \left( \bar{\mathcal{Q}}^+_{0 \bar{1}}-\mathcal{Q}^-_{0 \bar{1}} \right) -\Phi_{\y} \left(  \mathcal{Q}^+_{0 \bar{1}}-\bar{\mathcal{Q}}^-_{0 \bar{1}} \right)  \,. 
\end{split} \label{BTEven}
\end{eqnarray}

In summary, we end up with only $5$ independent boundary terms in the U(1) case, which implies that the basis for quadratic-higher-order Maxwell-Einstein theories contains 21 $\left(=4 + 7\times 2 + 3 \right)$ terms:  4 of them are linear in the Riemann, $7\times 2$ are even and $3$ are odd. 

\section{Electromagnetic tetrad }
\label{App:tetrads}
It has been shown that any non-null curvature $F_{\mu\nu}$ (i.e. $\x\neq 0$ and/or $\y\neq 0$)  decomposes according to\footnote{A null two-form admits a different decomposition which can be found in section 5.2. of \cite{Stephani:2003tm} for instance.}
\begin{eqnarray}
F_{\mu\nu} = \x \po_{\mu\nu} + \y \poo_{\mu\nu} \,,
\end{eqnarray}
in terms of the normalised antisymmetric rank 2-matrices $\po_{\mu\nu}$ and $\poo_{\mu\nu}$. More concretely, $\po$ and $\poo$ can be viewed as  simple bi-vectors which means 
that there exist a tetrad $e_\mu^I$, where $I \in \{0,1,2,3\}$, such that
\begin{eqnarray}
\label{bivectors}
\po = e^0 \wedge e^1 \, , \qquad
\poo =- \hodge (e^0 \wedge e^1 ) = e^2 \wedge e^3 \, .
\end{eqnarray}
We choose  $e^0$ to be a time-like vector field while the three other vectors are space-like, i.e. 
\begin{eqnarray}
g^{\mu\nu} \, e^I_\mu \, e^J_\nu \; = \; \eta^{IJ} \, ,
\end{eqnarray}
where $\eta$ is the flat Minkowski metric. The orientation of the tetrad is chosen such that
\begin{eqnarray}
\varepsilon_{\mu\nu\rho\sigma} \, e_0^\mu \, e_1^\nu \, e_2^\rho \, e_3^\sigma \; = \; 1 \, ,
\end{eqnarray}
where $\varepsilon_{\mu\nu\rho\sigma} $ is the 4-dimensional Levi-Civita tensor. 

When written in terms of the tetrad, the electromagnetic projectors obviously read 
\begin{eqnarray}
\begin{split}
  \p_{\mu\nu} = -e^0_\mu e^0_\nu + e^1_\mu e^1_\nu \,, \qquad \pp_{\mu\nu} =  e^2_\mu e^2_\nu + e^3_\mu e^3_\nu \,, \label{ProjTetrad}
\end{split}
\end{eqnarray}
which makes clear that $\p$ and $\pp$ project onto a space-time 2-surface and a purely space-like 2-surface respectively.
These expressions, together with the previous ones \eqref{bivectors}, are particularly useful to easily detect and take into account dimensionally dependent identities involved in tensor fields constructed out of the two-form $F$ and its derivatives.

Notice that the conditions \eqref{bivectors} together with the previous normalisation conditions do not uniquely define the tetrad $e_\mu^I$. 
  Indeed, the pair $(e^0,e^1)$ is defined up to a special Lorentz transformation characterised by the rapidity $w \in \mathbb R$
\begin{eqnarray}
\label{Ltransfo}
\begin{pmatrix}e^0 \\
e^1
\end{pmatrix} \; \longmapsto \; \begin{pmatrix}f^0 \\
f^1
\end{pmatrix} = 
\begin{pmatrix}
\cosh w & \sinh w \\
\sinh w & \cosh w
\end{pmatrix}
\begin{pmatrix}e^0 \\
e^1
\end{pmatrix} \, ,
\end{eqnarray}
while $(e^2,e^3)$ is defined up to a 2-dimensional rotation characterised by the angle $\theta \in [0,2\pi[$,
\begin{eqnarray}
\label{Rtransfo}
\begin{pmatrix}e^2 \\
e^3
\end{pmatrix} \; \longmapsto \; \begin{pmatrix}
f^2 \\
f^3
\end{pmatrix} = 
\begin{pmatrix}
\cos \theta & \sin \theta \\
-\sin \theta & \cos \theta 
\end{pmatrix}
\begin{pmatrix}
e^2 \\
e^3
\end{pmatrix} \, .
\end{eqnarray}

In order to define the components of the electromagnetic field, one need to introduce a time-like unit vector $n_\mu$. Indeed, this enables us to construct the electric field $E_\mu$,
the magnetic field $B_\mu$ and also the Pontying vector $P_\mu$ as follows,
\begin{eqnarray}
\label{EBP}
E_\mu = n^\rho h^\sigma_\mu F_{\rho\sigma} \,,\;\;\;\;\; B_\mu = n^\nu \hodge F_{\nu\mu} = \frac{1}{2} \varepsilon_{\rho\sigma\mu} F^{\rho\sigma}  \,,\;\;\;\;\; P_\mu=  \varepsilon_{\mu\rho\sigma} B^\rho E^\sigma \,,
\end{eqnarray}
where $\varepsilon_{\rho\sigma\mu} = n^\nu \varepsilon_{\nu\rho\sigma\mu}$. Of course, the vector $n$ can be decomposed into the basis defined by the tetrad $\{e^I\}$ with $I=0,...,3$.
As we can make use of the freedom to choose $(e^0,e^1)$ up to a Lorentz transformation \eqref{Ltransfo} and $(e^2,e^3)$ up to a rotation $\eqref{Rtransfo}$, we can set without loss of generality,
\begin{eqnarray}
n \cdot e^1 = 0 \, , \qquad
n \cdot e^3 = 0 \, ,
\end{eqnarray}
and we parametrise $n$ in term of the real number $\chi$ as follows\footnote{A more general parametrisation would be 
$n_\mu  = \cosh \chi e^0_\mu + \sinh \chi (\cos \alpha e^2_\mu + \sin\alpha \cos\beta   e^3_\mu+  \sin\alpha \sin\beta e^1_\mu)$ for instance.}
\begin{eqnarray}
n_\mu \; = \; \cosh \chi \, e^0_\mu \, + \, \sinh \chi \, e^2_\mu \, . 
\end{eqnarray}
We see that $\chi$ measures the deviation between the two time-like vector $n$ and $e^0$ and it is a gauge dependent quantity. For simplicity and without loss of generality, we
fix $\chi \geq 0$.

As a consequence, the expressions of electromagnetic fields  \eqref{EBP}  reduce to those given in  section \ref{SEC3+1},
\begin{eqnarray}
\label{EeBb}
B_\mu = \x \psi b_\mu + \y \bar{\psi} e_\mu \,, \;\;\;\; E_\mu= \x \bar{\psi} e_\mu - \y \psi b_\mu \,, 
\end{eqnarray}
where 
\begin{eqnarray}
\bar{\psi}= \cosh\chi \, , \qquad
\psi=\sinh\chi \, , \qquad
e_\mu = e^1_\mu \, , \qquad
b_\mu = e_\mu^3 \, .
\end{eqnarray}
Notice that the relations between $\psi$, $\bar\psi$ and $\chi$ is consistent with the condition \eqref{psi}. 
The Poynting vector can also be written in the form given in  section \ref{SEC3+1},
\begin{eqnarray}
 P_\mu = - \left(\x^2 + \y^2 \right) \psi \bar{\psi} r_\mu \,,
\end{eqnarray}
where the direction $r_\mu$ is given by
\begin{eqnarray}
 \text{sign}(\chi)  r_\mu = \sinh \chi \, e^0_\mu + \cosh \chi \, e^2_\mu \, .
\end{eqnarray}
It is useful to compute the norm of the Poynting vector which is given by
\begin{eqnarray}
\label{Psquare}
P^2 = (\x^2 + \y^2)^2 \cosh^2 \chi \, \sinh^2 \chi \, .
\end{eqnarray}

We can easily invert the previous equations in order to express the components of the electromagnetic tetrad in terms of $(n,E,B,P)$. After a short calculation, we
obtain,
\begin{eqnarray}
&& e^0_\mu  = \cosh \chi \, n_\mu +  \sinh \chi  \, r_\mu \, , \quad e^1_\mu  =  \frac{\sinh \chi}{\vert P \vert } (\x E_\mu + \y B_\mu) \, , \\
&& e^2_\mu  =  - \sinh \chi \, n_\mu - \cosh \chi \,  r_\mu ,\, \quad e^3_\mu  =  \frac{\cosh \chi}{\vert P \vert }  (\x B_\mu - \y E_\mu) \, .
\end{eqnarray}
We have used the relations
\begin{eqnarray}
\label{EBsquare}
\bar{\psi}^2=\cosh^2\chi = \frac{B^2 + \x^2}{\x^2 + \y^2} \, = \,  \frac{E^2 + \y^2}{\x^2 + \y^2} \, ,
 \end{eqnarray}
 or equivalently
\begin{eqnarray}
\psi^2=\sinh^2\chi  = \frac{B^2- \y^2 }{\x^2+\y^2 } = \,  \frac{E^2- \x^2 }{\x^2+\y^2 } \label{psiB}\, .
\end{eqnarray}

\section{Classification of degenerate non-minimally coupled theories and beyond}\label{AppNMC}

In this appendix, we gather various results concerning the classification of degenerate theories undertaken in Sec.\ref{NMCDeg}.

\subsection{Kinetic matrix of non-minimally coupled theories}
 
 Choosing a basis for EM velocities given by $\{ \LL_{+} , \LL_{-} , \LL \}$, the non-vanishing building blocks of the kinetic matrix of the non-minimally coupled theories \eqref{NMCAction} can be written as follows, starting from the cross terms,
\begin{eqnarray}
\mathscr{D}=\left(
\begin{array}{cccccc}
\zeta_1 &  \zeta_2 & 0 & \zeta_3 & \zeta_4 & 0 \\
\bar{\zeta}_3 & \bar{\zeta}_2 & 0 & \bar{\zeta}_1 & \bar{\zeta}_4 & 0 \\
0 & 0 & \bar{\psi} \xi_1 & 0 & 0 & \psi \bar{\xi_1}
\end{array}
\right)\,,
\end{eqnarray}
while the matrix of quadratic terms in the extrinsic curvature is given by
\begin{eqnarray}
\mathscr{C}=2\left(
\begin{array}{cccccc}
- \frac{ \x \psi^3 \bar{\xi_1}}{\bar{\psi}}  &\frac{\psi^2 \left( \y \bar{\xi_1} - \x \xi_1 \right)}{2} & 0 & \frac{\xi_2+\xi_3}{2} & \frac{ \xi_3}{2} & 0 \\
* & \kappa_1  & 0 & \frac{\bar{\psi}^2 \left(  \y \bar{\xi_1} - \x \xi_1\right)}{2} & \frac{\x \xi_1  + \y \bar{\xi}_1}{2}  & 0 \\
* & * & \kappa_2 & 0 & 0 &\frac{ \y \psi^2 \bar{\xi}_1 - \x \bar{\psi}^2 \xi_1}{2} \\
*&*&*&   \frac{\y \bar{\psi}^3\xi_1}{\psi}  & \frac{\xi_3 - \frac{ \xi_2}{\psi^{2}}}{2} & 0 \\
*&*&*&*&0&0\\
*&*&*&*&*&- \xi_3  - \frac{ \x \psi^3 \bar{\xi_1}}{\bar{\psi}}
\end{array}
\right)\,,
\end{eqnarray}
where
\begin{eqnarray}
\kappa_1 =  \psi \bar{\psi} \left( \y \xi_1 - \x \bar{\xi}_1 \right) - \xi_2- \xi_3 \,,\;\;\;\;\;\; \kappa_2 =  \frac{\psi^2}{\bar{\psi}^2} \left( \y \psi \bar{\psi} \xi_1 + \xi_2 \right) - \xi_2 - \xi_3 \,,
\end{eqnarray}
and
\begin{eqnarray}
\begin{split}
\zeta_1 &= \frac{\psi^2 \alpha_{0\x}}{2} - \bar{\psi}^2 \alpha_{2\x} - \frac{\psi^2}{\x^2 +\y^2} \left( \y \alpha_1 + \x \left( \alpha_0 - 2 \alpha_2 \right) \right)\,, \\
\zeta_2 &= - \psi \bar{\psi} \alpha_{1\x} + \frac{\psi}{\bar{\psi}\left(\x^2 + \y^2 \right)} \left( - \left( \psi^2 + \bar{\psi}^2 \right) \left( \y \alpha_0 - \x \alpha_1 \right) + 2 \y \left( \bar{\psi}^2 \alpha_2 + \psi^2 \alpha_3 \right)\right)\,, \\
\zeta_3 &=- \frac{\bar{\psi}^2 \alpha_{0\x}}{2}+\psi^2 \alpha_{3\x} + \frac{\psi^2}{\x^2 +\y^2} \left( \y \alpha_1 + \x \left( \alpha_0 - 2 \alpha_3 \right) \right) \,,\\
\zeta_4 &= - \frac{\alpha_{0\x}}{2} \,,\\
\xi_1 &=\frac{\psi}{\bar{\psi}\left( \x^2 +\y^2 \right)} \left( - \x \alpha_1 + \y \left( \alpha_0 - 2 \alpha_3 \right) \right)\,,\\
\xi_2 &= \frac{\psi^2}{\x^2 + \y^2} \left(2 \x \y \alpha_1 + \left( \x^2 - \y^2 \right) \left( \alpha_0 - \alpha_2 - \alpha_3 \right) \right)\,, \\ 
\xi_3 &= \frac{1}{\x^2 + \y^2} \left( \x \y \alpha_1 - \x^2 \alpha_2 + \y^2 \left( \alpha_2-\alpha_0 \right) \right)\,,
\end{split}
\end{eqnarray}
where we have assumed that $\alpha_I$ transforms under a conjugation as $\mathcal{R}_I$, meaning $\alpha_1 \to \alpha_1$, $\alpha_2 \to - \alpha_2$, $\alpha_3 \leftrightarrow \alpha_4$, in order to simplify the notation. This should not be viewed as a restriction on their specific dependence on $(\x,\y)$. It can be shown that the determinant of the kinetic matrix factorises as follows
\begin{eqnarray}
\text{det}\left( \mathbb{M} \right) = m_0 m_1 \,, 
\end{eqnarray}
where 
\begin{eqnarray}
m_0=  - \psi \bar{\psi} \xi_1 \bar{\xi}_1 \left( \x \xi_1 + \y \bar{\xi}_1 \right) - \bar{\xi}_1^2 \xi_2 + \left( \bar{\psi}^2 \xi_1^2 + \psi^2 \bar{\xi}_1^2 \right) \xi_3 \,, 
\end{eqnarray}
while $m_1$ is complicated and we will not give its full expression. 

\subsection{Full degeneracy in the electromagnetic sector}

We now focus on theories which are fully degenerate in the electromagnetic sector, meaning $\text{rank}(\mathbb{M})=\text{rank}(\mathscr{C})=6$. 

\subsubsection{$\mathcal{R}_{\text{C}_1\text{\RNum{2}}} $}
In order to prove that the theory \eqref{R1theory}, given by
\begin{eqnarray}
\mathcal{R}_{\text{C}_1\text{\RNum{2}}} = \alpha_1 \left(\x , \y\right) \mathcal{R}_1  = \alpha_1 \left(\x , \y\right)  \po^{\mu\nu} \poo^{\rho\sigma} R_{\mu\rho\nu\sigma}
\end{eqnarray}
admits a fully degenerate EM sector, we introduce the following effective extrinsic curvatures
\begin{eqnarray}
\begin{split}
\mathcal{K}_1 &= K_1  + \left( \frac{\alpha_1 - \x \alpha_{1\x}}{2 \x \alpha_1} \right)  \LL_{+}  + \left( \frac{\alpha_1 - \y \alpha_{1\y}}{2 \y \alpha_1} \right) \LL_{-}  \,,\\
 \mathcal{K}_2 &= K_2 \,,\\
 \mathcal{K}_3 &= K_3  -\left( \frac{\y}{\psi \left( \x^2 - \y^2 \right)} \right) \LL  \,,\\
 \mathcal{K}_5 &= K_5  +\left( \frac{- \Psi^2 \alpha_1 + \x \alpha_{1\x}}{2 \x \alpha_1} \right) \LL_{+}  + \left( \frac{\Psi^2 \alpha_1 + \y \alpha_{1\y}}{2 \y \alpha_1} \right) \LL_{-} 
\end{split}
\end{eqnarray}  
and their conjugates, where $\Psi^2 = \psi^2 + \bar{\psi}^2$, in terms of which the kinetic term of the theory reads 
\begin{eqnarray}
 \int d^4 x \sqrt{-g} \left( \mathscr{C}^{AB} K_{A} K_{B} +2 \mathscr{D}^{Ai} K_{A} \LL_i  \right)=  \int d^4 x \sqrt{-g}\,   C^{AB} \mathcal{K}_{A} \mathcal{K}_{B}\,,
\end{eqnarray}
where the kinetic matrix $C$ is given by
\begin{eqnarray}
C =\alpha_1 \left(
\begin{array}{cccccc}
 2 \mathcal{G} \psi ^3 \bar{\psi }^2 & x^2 \psi ^4 \bar{\psi }+ \bar{x}^2 \psi ^2 \bar{\psi }^3 & 0 &
   -\mathcal{G} \Psi \psi  \bar{\psi }^2   & -\mathcal{G} \psi  \bar{\psi }^2 & 0 \\
 * & 4 \mathcal{G}\Psi  \psi   \bar{\psi }^2  & 0 & x^2 \psi ^2 \bar{\psi }^3+\bar{x}^2 \bar{\psi
   }^5 & \bar{x}^2 \bar{\psi }^3-x^2 \psi ^2 \bar{\psi } & 0 \\
 * & * & 2 \mathcal{G} \psi ^3 \bar{\psi }^2 & 0 & 0 &X  \psi ^2  \bar{\psi }^3 \\
 * & * & * & 2 \mathcal{G} \psi  \bar{\psi }^4 & \mathcal{G} \psi  \bar{\psi
   }^2 & 0 \\
 * & * & * & * & 0 & 0 \\
 * & * & * & * & * & 2 \mathcal{G} \psi  \bar{\psi }^4 \\
\end{array}
\right)
\end{eqnarray}
and $X^2= \x^2 + \y^2$.

\subsubsection{$\mathcal{R}_{\text{C}_1\text{\RNum{3}}_\pm}$}

A similar calculation can be done for the second class of fully degenerate theories in the EM sector \eqref{Squareroot}, given by
\begin{eqnarray}
\begin{split}
\mathcal{R}_{\text{C}_1\text{\RNum{3}}_\pm} &= \gamma \left(\mathcal{R}_0+\mathcal{R}_2+\mathcal{R}_3 \right) - \frac{1}{2 \gamma \delta} \left( \x \y \pm \sqrt{\left( \x^2 - 2 \gamma^2 \delta \right) \left( \y^2 + 2 \gamma^2 \delta \right)}  \right)\mathcal{R}_1 \\
&= \gamma \left(\mathcal{R}_0+\mathcal{R}_2+\mathcal{R}_3 \right)+\alpha_\pm \left( \x, \y \right) \mathcal{R}_1 \,,
\end{split}
\end{eqnarray}
where $\alpha_\pm$ is a fixed function of the EM invariants defined by comparing the two previous lines. Recall that $\gamma$ and $\delta$ are coupling constants transforming under a rotation as $\gamma \to \gamma$ and $\delta \to - \delta$. Indeed, just as before, we introduce the effective extrinsic curvature scalars, given by 
\begin{eqnarray}
\begin{split}
\mathcal{K}_1^{\pm} &= K_1  +\left( \frac{\gamma}{2 \bar{\phi}^2 \bar{\psi}^2} \right) \mathcal{J}^{\pm}  \,,\\
 \mathcal{K}_2^{\pm} &= K_2  -\left( \frac{\gamma}{2 \bar{\phi} \phi \bar{\psi} \psi}  \right) \mathcal{J}^{\pm} \,, \\
 \mathcal{K}_3^{\pm} &= K_3  -\left(\frac{ \bar{\phi}}{\psi \left(\y \bar{\phi} + \x \phi \right)} \right) \LL  \,,\\
 \mathcal{K}_5^{\pm} &= K_5  -\left( \frac{\theta}{2 \bar{\phi}^2 \phi^2 \bar{\psi}^2 \psi^2}  \right) \mathcal{J}^{\pm} \,,
\end{split}
\end{eqnarray}  
together with their conjugates, where 
\begin{eqnarray}
\mathcal{J}^{\pm} = \bar{\phi} \psi^2 \LL_{+}  + \phi \bar{\psi}^2 \LL_{-}  \,,\;\;\;\;\; \phi =  \x \alpha_{\pm} + \y \gamma
\end{eqnarray}
and 
\begin{eqnarray}
\begin{split}
\theta &= \gamma \left( \alpha_{\pm}^2 - \left( \gamma^2 + \alpha_{\pm}^2 \right) \psi^4 \right) \y^2 + \gamma \left( \gamma^2 \bar{\psi}^4  + \frac{\alpha_{\pm}^2 \psi^2}{2}  \left( 3 + \Psi^2 \right) \right) \x^2\\
& - 2 \alpha_{\pm} \left(\alpha_{\pm}^2 \psi^2 \bar{\psi}^2 + \gamma^2 \left( 1 + \psi^2 \bar{\psi}^2 \right) \right) \x \y  \,.
\end{split}
\end{eqnarray}
This enables to show that the kinetic term of the theory can be written solely in terms of these effective extrinsic curvature scalars. Indeed, it factorises as follows,
\begin{eqnarray}
 \int d^4 x \sqrt{-g} \left( \mathscr{C}^{AB} K_{A} K_{B} +2 \mathscr{D}^{Ai} K_{A} \LL_i  \right) = \int d^4 x \sqrt{-g}\, C_{\pm}^{AB} \mathcal{K}^{\pm}_{A} \mathcal{K}^{\pm}_{B} \,,
\end{eqnarray}
with the kinetic matrix $C_{\pm}$ given by 
\begin{eqnarray}
C_{\pm} =\left(
\begin{array}{cccccc}
 2 x \bar{\phi}  \psi ^3 \bar{\psi }^2 & \kappa  \psi ^2 \bar{\psi } & 0 & \zeta  \psi  \bar{\psi }^2 & -x \bar{\phi} 
   \psi  \bar{\psi }^2 & 0 \\
 * & -4 \zeta  \psi  \bar{\psi }^2 & 0 & \kappa  \bar{\psi }^3 & -\xi  \bar{\psi } & 0 \\
 * & * & -2\bar{x} \phi  \psi ^3  \bar{\psi }^2 & 0 & 0 &  X  \psi ^2
   \bar{\psi }^3 \alpha_\pm \\
 * & * & * & -2 \bar{x} \phi  \psi   \bar{\psi }^4 & - \bar{x}  \phi  \psi \bar{\psi}^2 & 0 \\
 * & * & * & * & 0 & 0 \\
 * & * & * & * & * & 2 x \bar{\phi}  \psi  \bar{\psi }^4 \\
\end{array}
\right)\,,
\end{eqnarray}
where 
\begin{eqnarray}
     \zeta = \y \phi \psi^2 - \x \bar{\phi} \bar{\psi}^2  \,,\;\;\;\;\;\;  \xi = \x \phi \psi^2 + \y \bar{\phi} \bar{\psi}^2 \,,\;\;\;\;\;\;     \kappa = \x^2 \psi^2 \alpha_{\pm} - \x\y \gamma  +\y^2 \bar{\psi}^2 \alpha_{\pm}  \,.
\end{eqnarray}

\subsubsection{Quadratic ten-functions family}

Finally, we consider the theory \eqref{BeyondNMC} given by 
\begin{eqnarray}
\mathcal{L}_{\text{quadratic}}= \mathscr{L}_{\text{quadratic}} +\bar{\mathscr{L}}_{\text{quadratic}}\,,
\end{eqnarray}
where
\begin{eqnarray}
\mathscr{L}_{\text{quadratic}}=\beta \left( - \x \mathcal{Q}^{\bar{0}}_1 + \mathcal{Q}^+_{1\bar{0}}  \right) + \beta^0_1 \mathcal{Q}^{0}_1 + \beta^{\bar{1}}_2  \mathcal{Q}^{\bar{1}}_2   + \beta^+_{0\bar{0}}  \mathcal{Q}^+_{0\bar{0}} + \beta^-_{0\bar{1}}  \mathcal{Q}^-_{0\bar{1}} \,.
\end{eqnarray}
 The null eigenvectors associated with its kinetic matrix expressed in the basis $\{\LL_+,\LL_-,\LL, K_1, ... , K_6 \}$ are given by 
\begin{eqnarray}
\begin{split}
&\left(1,0,0,-\mu,0, -\x \sigma \bar{\psi}  , \nu , \psi^2 \left( \frac{1}{\x} - \mu \right) - \nu \bar{\psi}^2,   -\y \sigma \psi \right)  \,,\\
&\left(0,1,0,\bar{\nu},0, -\x \bar{\sigma} \bar{\psi}  , - \bar{\mu} , \bar{\psi}^2 \left( -\frac{1}{\y} + \bar{\mu} \right) + \bar{\nu} \psi^2,   -\y \bar{\sigma} \psi \right)  \,,\\
&\left(0,0,1,0,0, -\frac{\rho}{\y \psi}  , 0 , 0,  \frac{1-\rho}{\x \bar{\psi}} \right)  \,,
\end{split}
\end{eqnarray}
where we have introduced the following quantities
\begin{eqnarray}
\mu = \frac{1}{2} \left( \frac{1}{\x} + \frac{\bar{\beta}^-_{0\bar{1}} }{\beta^{\bar{1}}_2} \right) \,,\;\;\;\;\; \nu = \frac{\y \beta}{2 \x \bar{\beta}^{\bar{1}}_2}\,,\;\;\;\;\; \rho = \frac{\bar{\beta} \nu}{\bar{\beta} \nu + \beta \bar{\nu}} \,,
\end{eqnarray}
together with their conjugates, as well as 
\begin{eqnarray}
\sigma = \frac{\nu \bar{\nu} \beta^+_{0\bar{0}}}{\y \left(\bar{\beta}\nu + \beta \bar{\nu} \right)} \,,\;\;\;\;\; \bar{\sigma} =- \frac{i \nu \bar{\nu} \bar{\beta}^+_{0\bar{0}}}{\x \left(\bar{\beta}\nu + \beta \bar{\nu} \right)}  \,.
\end{eqnarray}
Notice that as long as
\begin{eqnarray}
\text{det}\left( \mathscr{C}_{\text{quadratic}} \right) = - \frac{4 \x^4 \y^4 \rho^2 \beta^6 \psi^4 \bar{\psi}^4}{\nu^6 \left( 1-\rho \right)^4 \left(\x^2 + \y^2 \right)^4} \neq 0 \,,
\end{eqnarray}
the gravitational sector is not degenerate.

\section{Miscelleaneous formulae on disformal transformations}\label{AppDisf}

In this section, we give miscellaneous technical results concerning the disformal transformations that we have discussed in section \ref{SecDisfTransf}.

\subsection{(Non-)invertibility conditions of disformal transformations}
\label{App:disformal}
Let us present  an alternative way to compute the (non-)invertibility conditions of disformal transformations, which gives a nice interpretation of 
the relations \eqref{NoninvertibilityJack} and
\eqref{Invert2cond}. Let us recall that these transformations read
\begin{eqnarray}
&&\tilde{p}_{\mu\nu} = \z(\x,\y) \, p_{\mu\nu} \, , \qquad \tilde{\bar{p}}_{\mu\nu} = \zb(\x,\y) \bar{p}_{\mu\nu} \, , \label{pptilde}\\
&&\tilde \x = \frac{s \x}{\z(\x,\y)} \, , \qquad \tilde \y = \frac{\bar{s} \y}{\zb(\x,\y)} \,, \label{xxtilde}
 \end{eqnarray} 
 where we have introduced the notations $\tilde{q}$ and $\tilde z$ for the disformal transformation of the projectors $q \in \{p,\bar{p}\}$ and $z \in \{\x,\y\}$ for clarity. The transformation is invertible if and only if
 one can invert the relations \eqref{xxtilde} in order to express $z$ in terms of $\tilde z$. Indeed, if this is the case, the relations \eqref{pptilde} can also be inverted, then one
 can express the two projectors $q \in \{p,\bar{p}\}$ in terms of $\tilde{q}$ and finally one immediately gets the metric $g_{\mu\nu}$ in terms of $\tilde{g}_{\mu\nu}$. 
 
 Hence, the transformation is non-invertible if the Jacobian matrix of the transformation \eqref{xxtilde} is itself non-invertible, which means
 \begin{eqnarray}
 \tilde \x_{\x} \, \tilde \y_{\y} - \tilde \x_{\y} \, \tilde \y_{\x} \; = \; \frac{s \bar{s}}{\z^2 \zb^2} \left[  ( \z - \x \z_{\x})( \zb - \y \zb_{\y} )-  \x \y \z_{\y}\zb_{\x} \right]\; = 0 \, ,
 \end{eqnarray} 
that is equivalent to \eqref{NoninvertibilityJack} when $\z \neq 0$ and $\zb \neq 0$.

In order to understand the second condition \eqref{Invert2cond}, it is useful the introduce the following new variables,
\begin{eqnarray}
y = \frac{1}{\x} \, , \qquad \tilde y = \frac{\tilde s}{\tilde \x} \, , \qquad \bar{y}=\frac{1}{\y} \, , \qquad \tilde {\bar{y}} = \frac{\tilde s}{\tilde \y} \, ,
\end{eqnarray} 
in terms of which the relations \eqref{xxtilde} becomes
\begin{eqnarray}
\tilde{y} = y \, \z \, , \qquad \tilde{\bar{y}} = \bar{y} \, \zb \, .
\end{eqnarray}
The Jacobian of these transformations  is now given by the matrix
\begin{eqnarray}
\begin{pmatrix}
\z - x \z_x & -\frac{\y^2}{\x} \, \z_{\y} \\
- \frac{\x^2}{\y} \, \zb_{\x} & \zb - \y \zb_{\y} 
\end{pmatrix} \, .
\end{eqnarray} 
We see immediately that the transformation is not invertible when the condition \eqref{NoninvertibilityJack} is satisfied. Furthermore,  the condition  \eqref{Invert2cond} means that
the Jacobian is traceless. Hence, if the two conditions  \eqref{NoninvertibilityJack}  and   \eqref{Invert2cond}  are satisfied, the Jacobian matrix vanishes identically which means that the transformation is fully degenerate. It is then straightforward to show that these conditions lead to  $\z = \x / \x_0$ and $\zb = \y / \y_0$ where $\x_0$ and $\y_0$ are constants. In that case, the transformations  \eqref{xxtilde} are clearly fully degenerate.

\subsection{``Non-conformal component" of U(1) disformal transformations}

Let us start with the transformation law of the triad $\{e_\mu, b_\mu, r_\mu \}$ introduced in section \ref{SEC3+1}. Assuming the positive branch in the transformation of the normal \eqref{DisfTransfNormal} and that one of the conditions \eqref{CdtNormalDisf} is satisfied, i.e.
\begin{eqnarray}
n_\mu \longrightarrow \,  \Theta \, n_\mu \,,\;\;\;\;\; \Theta = \sqrt{\frac{\z \zb}{\zb \bar{\psi}^2 - \z \psi^2}} >0 \, ,
\end{eqnarray}
we obtain
\begin{eqnarray}
\begin{split}
 b_\mu &\longrightarrow   \ssb \, \text{sign}\left(\psi\right) \sqrt{\zb} \, b_\mu \,,\;\;\;\; \psi \longrightarrow \frac{  \text{sign}\left(\psi\right) \, \psi \, \Theta}{\sqrt{\zb}} \,,\\
 e_\mu &\longrightarrow \s\, \text{sign}\left(\bar{\psi}\right) \sqrt{\z} \, e_\mu  \,,\;\;\;\; \bar{\psi} \longrightarrow \frac{  \text{sign}\left(\bar{\psi}\right) \, \bar{\psi} \, \Theta}{\sqrt{\z}} \,,
\end{split}
\end{eqnarray}
where $\s$ and $\ssb$ are signs appearing in the transformations of the projectors \eqref{DisfTransfProj} and
\begin{eqnarray}
 r_\mu &\longrightarrow  \frac{\text{sign}\left(\psi\right) \text{sign}\left(\bar{\psi}\right) \Theta}{\sqrt{\z \zb}}  \left( \psi \bar{\psi} \Delta n_\mu +\left( \zb \bar{\psi}^2 -\z \psi^2 \right)  r_\mu \right) \,.
\end{eqnarray}
 Therefore, for transformations belonging to the ``conformal component", i.e. $0 < \z  \leq \zb $, the transformed triad and scalar fields remain real, while for those belonging to the ``non-conformal component", i.e. $ \zb < 0 < \z$, the dyad $\{ \tilde{b}_\mu, \tilde{r}_\mu \}$ and the scalar field $\tilde{\psi}$ become imaginary, although they are still spacelike, orthogonal and unit for the disformal metric. However, $\tilde{b}_\mu$ is time-like for the metric $g$ and $\tilde{r}_\mu$ as well for some configurations of the electromagnetic field (for instance for $0<\psi \ll 1$).

An example of invertible transformation belonging to this ``non-conformal component" of disformal transformations has been studied in \cite{Goulart:2020wkq} and is given by
  \begin{eqnarray}
 \tilde{g}_{\mu\nu} & = & \Omega\left(\x,\y\right) T_{\mu\nu} =\Omega\left(\x,\y\right)  \left( F_{\mu}{}^\sigma F_{\nu\sigma} - \frac{1}{4} F_{\rho\sigma}F^{\rho\sigma} g_{\mu\nu} \right) \, ,
 \end{eqnarray}
where $T_{\mu\nu}$ is the stress-energy tensor of Maxwell's theory. Assuming the transformation to be volume-preserving, i.e. $\text{det}\left(\tilde{g}\right)=\text{det}\left(g\right)$, implies
\begin{eqnarray}
\Omega\left(\x,\y\right)= - \frac{2\epsilon}{\x^2 + \y^2} \;\;\;\; \implies\;\;\;\;  \tilde{g}_{\mu\nu} =\epsilon \left( \p_{\mu\nu}-\pp_{\mu\nu} \right) \, ,
\end{eqnarray}
where $\epsilon^2=1$. For $\epsilon=1$, the transformation does satisfy the second inequality $ \zb < 0 < \z$. The effect of this opposite sign between $p$ and $\pp$ has been discussed in \cite{Goulart:2020wkq}. 

In order to illustrate what happens in that case, let us consider as an example the exterior of a static spherically symmetric or a Taub-NUT system in Schwarzschild coordinates $\left( t,r,\theta,\phi \right)$ with signature $\left(-+++\right)$. The effect of the transformation is to flip the sign of the angular part
\begin{eqnarray}
 \pp_{\mu\nu} dx^\mu dx^\nu  =r^2 \left( d\theta^2 + \sin^2\theta d\phi^2 \right) \longrightarrow - r^2 \left( d\theta^2 + \sin^2\theta d\phi^2 \right)\,.
\end{eqnarray}
Let us recall that the projector $\pp$ projects on two-dimensional space-like surfaces, as we saw in section \ref{App:tetrads}, so that the signature of the disformal metric is $\left(-+--\right)$. 

\subsection{Christoffel symbols and Riemann curvature}
In this section, we give the transformation laws of the Christoffel symbols and the Riemann curvature under disformal transformations.  Christoffel symbols transform according to
\begin{eqnarray}
\delta \Gamma_{\mu\nu}^\sigma =  \tilde{\Gamma}_{\mu\nu}^\sigma  -  \Gamma_{\mu\nu}^\sigma  = \omega_{\mu\nu}^\sigma + \bar{\omega}_{\mu\nu}^\sigma \,,
\end{eqnarray}
where we have introduced the notation
\begin{eqnarray}
\omega_{\mu\nu}^\sigma=  \frac{1}{2} \left( \omega_{\mu\nu}^{\sigma\rho} \nabla_\rho \log \z  +\Delta  \z^{-1}  \omega_{\mu\nu}^{\sigma\rho\gamma\delta}  \nabla_\rho \po_{\gamma\delta} \right) ,
\end{eqnarray}
with $\Delta = \z - \zb$ and where 
\begin{eqnarray}
\begin{split}
\omega_{\mu\nu}^{\sigma\rho} &= p^\sigma_{(\mu} p^\rho_{\nu)} + p^\sigma_{(\mu} \pp^\rho_{\nu)} - \p_{\mu\nu} \left(  \p^{\sigma\rho}  + \z \zb^{-1} \pp^{\sigma\rho} \right) \, ,
 \\
 \omega_{\mu\nu}^{\sigma\rho\gamma\delta} &= \p^{\rho\sigma} \po^{\gamma}{}_{(\mu} \, \pp^\delta_{\nu)} - \po^{\gamma\sigma} \left( \p^\rho_{(\mu} \pp^\delta_{\nu)} +  \pp^\rho_{(\mu} \pp^\delta_{\nu)} \right) \, .
 \end{split}
\end{eqnarray}
These formulae are valid for arbitrary scalar fields $\z$ and $\zb$, i.e. whether they depend on the electromagnetic invariants or not.  The Riemann tensor transforms according to
\begin{eqnarray}
\delta R_{\mu\nu\rho}{}^\sigma = \tilde{R}_{\mu\nu\rho}{}^\sigma - R_{\mu\nu\rho}{}^\sigma = \nabla_{[\nu} \delta \Gamma_{\mu]\rho}^\sigma + \delta \Gamma_{\rho[\mu}^{\gamma}  \delta \Gamma_{\nu]\gamma}^{\sigma} \, .
\end{eqnarray}
This formula holds for  any affine connection $\tilde{\Gamma}= \Gamma + \delta \Gamma$.

\section{Dictionary between projectors and field strength}
\label{App:dictio}
In this appendix, we provide  relations which are useful to convert expressions involving the projectors $\po$, $\poo$, $p$, $\bar{p}$ and the scalars $\x$ and $\y$ into expressions involving the more familiar field strength itself $F_{\mu\nu}$ and  metric $g_{\mu\nu}$. Such relations are particularly useful to derive the field equations of HOME theories, as well as their transformation properties under U(1)-preserving disformal transformations.

Let us start with the two scalars $\x$ and $\y$ which have been defined in \eqref{EMinvxy} by
\begin{eqnarray}
\mathcal{F} =\frac{1}{2} \left(\x^2 - \y^2\right)  \,,\;\;\;\;\; \mathcal{G}= - \x \y\,,
\end{eqnarray}
with the electromagnetic invariants given by \eqref{EMinv}
\begin{eqnarray}
\mathcal{F}= -\frac{1}{4} F^{\mu\nu}F_{\mu\nu} \,,\;\;\;\;\; \mathcal{G}= \frac{1}{4}\hodge F^{\mu\nu} F_{\mu\nu}  \,.
\end{eqnarray}
A straightforward calculation leads to
\begin{eqnarray}
\frac{\partial \mathcal{F}}{\partial F_{\mu\nu}} = - \frac{1}{2} F^{\mu\nu} = \x \frac{\partial x}{\partial F_{\mu\nu}} - \y  \frac{\partial \y}{\partial F_{\mu\nu}} \, , \qquad
\frac{\partial \mathcal{G}}{\partial F_{\mu\nu}} =  \frac{1}{2} \hodge F^{\mu\nu} = -\y \frac{\partial x}{\partial F_{\mu\nu}} - \x  \frac{\partial \y}{\partial F_{\mu\nu}} \, .
\end{eqnarray}
Using the defining relation of $\po$ and $\poo$, 
\begin{eqnarray}
\label{FetstarF}
F_{\mu\nu} = \x \po_{\mu\nu} + \y \poo_{\mu\nu} \,,\;\;\;\;\; \hodge F_{\mu\nu} =\y \po_{\mu\nu} - \x \poo_{\mu\nu} \,,
\end{eqnarray}
one gets the relations
\begin{eqnarray}
 \frac{\partial \x }{\partial F_{\mu\nu}} =- \frac{1}{2} o^{\mu\nu} \,,\;\;\;\;\;  \frac{\partial \y }{\partial F_{\mu\nu}} =\frac{1}{2} \bar{o}^{\mu\nu} \, .
\end{eqnarray}
Notice that the sign difference between these two expressions is consistent with the transformation property of $F$ under conjugation \eqref{bartransfF}, which is multiplied by the imaginary unit.

In order to compute the derivative of $\x$ and $\y$ with respect to the metric, we proceed in a similar way and we use the relations\footnote{To compute the derivative of $\mathcal G$, we use the fact that $\varepsilon_{\mu\nu\rho\sigma}$ is a tensor with
$$
\frac{\partial \varepsilon_{\alpha\beta\rho\sigma}}{\partial g^{\mu\nu}} = - \frac{1}{2}\varepsilon_{\alpha\beta\rho\sigma} \, g_{\mu\nu} \, .
$$}
\begin{eqnarray}
\frac{\partial \mathcal{F}}{\partial g_{\mu\nu}} &=& -\frac{1}{2} (\x^2 p^{\mu\nu} - \y^2 \bar{p}^{\mu\nu}) = \x \frac{\partial \x}{\partial g_{\mu\nu}} - \y\frac{\partial \y}{\partial g_{\mu\nu}} \, , \\
\frac{\partial \mathcal{G}}{\partial g_{\mu\nu}} &=& \frac{1}{2} \x\y (p^{\mu\nu} + \bar{p}^{\mu\nu}) = -\y \frac{\partial \x}{\partial g_{\mu\nu}} - \x\frac{\partial \y}{\partial g_{\mu\nu}} \, .
\end{eqnarray}
As a consequence, we obtain
\begin{eqnarray}
 \frac{\partial \x }{\partial g_{\mu\nu}} = -\frac{1}{2} \x \, \p^{\mu\nu} \,,\;\;\;\;\;  \frac{\partial \y }{\partial g_{\mu\nu}} =-\frac{1}{2} \y \, \pp^{\mu\nu} \, .
\end{eqnarray}

Now, we can compute the derivative of the projectors $\po$ and $\poo$. The method is a straightforward generalisation of the previous one. Indeed, we derive the two relations above \eqref{FetstarF} with respect to the field strength and also to the metric and after some calculations, we obtain
 \begin{eqnarray}
 \frac{\partial o^{\rho\sigma}}{\partial F^{\mu\nu}} & = &\frac{1}{2 \left( \x^2 + \y^2 \right)} \left( \x \delta_{[\mu}^\rho \delta_{\nu]}^\sigma + \x \left( o_{\mu\nu} o^{\rho\sigma}- \bar{o}_{\mu\nu} \bar{o}^{\rho\sigma}\right) - \y o_{[\mu}{}^{[\sigma} \, \bar{o}_{\nu]}{}^{\rho]}    \right) \, , \\
 \frac{\partial o_{\rho\sigma}}{\partial g_{\mu\nu}} & = & \frac{1}{2}  \po_{\rho\sigma} p^{\mu\nu} +\frac{1}{2 \left( \x^2 + \y^2 \right)} \left( \x \y  \poo^{(\mu}{}_{[\rho}\, \p^{\nu)}_{\sigma]} - \y^2 \po^{(\mu}{}_{[\rho}\, \pp^{\nu)}_{\sigma]}   \right) \, .
\end{eqnarray}
The derivatives of $\poo$ can be easily recovered using the transformations of the fields under the conjugation. 
Finally, the derivatives of the projector $p$  are obtained from the defining relation $p=\po^2$ which leads to 
\begin{eqnarray}
 \frac{\partial p^{\rho\sigma}}{\partial F^{\mu\nu}} & = & \frac{1}{2 \left( \x^2 + \y^2 \right)} \left( z^{\rho\sigma}_{\mu\nu} +i \bar{z}^{\rho\sigma}_{\mu\nu} \right) \,,\;\;\;\;\; z^{\rho\sigma}_{\mu\nu}= \x \left( \delta^{(\rho}_{[\mu} o_{\nu]}{}^{\sigma)} +2 \po_{\mu\nu} \p^{\rho\sigma} \right)
 \, , \\
 \frac{\partial p_{\rho\sigma}}{\partial g_{\mu\nu}}  & =& \frac{1}{2 \left( \x^2 + \y^2 \right)} \Big(\x\y  \po^{(\mu}{}_{(\rho}\, \poo^{\nu)}{}_{\sigma)}  + \x^2 \left( \po^{\mu}{}_{(\rho}\, \po^{\nu}{}_{\sigma)}+2 \p^{\mu\nu} \pp_{\rho\sigma} \right) \nonumber \\
 &+& \y^2 \left( \poo^{\mu}{}_{(\rho}\, \poo^{\nu}{}_{\sigma)} + \p^\mu_{(\rho}\p^\nu_{\sigma)} +\p^{(\mu}_{(\rho}\pp^{\nu)}_{\sigma)} + \pp^\mu_{(\rho}\pp^\nu_{\sigma)}  -2 \pp^{\mu\nu} \pp_{\rho\sigma}\right) \Big) \, .
\end{eqnarray}
Once again, the derivatives of $\bar{p}$ can be  recovered using the transformations of the fields under the conjugation. 

\section{$\mathscr{L}_{\hodge}$ from the 21-dimensional basis \& 2D wave equation}\label{AppLstar}

In this section,  we present an expression of $\mathscr{L}_{\hodge}$ and its conjugate, defined by \eqref{L*}, obtained at early stages of this work in terms of the explicit 21-dimensional basis of Lagrangians considered in \cite{Colleaux:2023cqu}. We refer the reader to Sec.VI of this paper for the definitions of the elementary quadratic Lagrangians $\mathcal{F}_i$, for $i  \in \{1,..., 18\}$.

As we will see, this early result illustrates how inconvenient it is to fix a priori a basis of Lagrangians built from high-rank tensors with symmetries, and thus, how convenient it is to use the projectors \eqref{proj} and \eqref{asproj} with their built-in Cayley-Hamilton identity, even though more Lagrangians have to be taken into consideration in this case (see \eqref{Lag48}). In particular, the alternative expressions for $\mathscr{L}_{\hodge}$ and its conjugate which follow are non-polynomial in the field strength and involve second derivatives of the arbitrary coupling functions, which is due to fixing respectively Bianchi identities and boundary terms \textit{a priori} in order to obtain a basis. On the contrary, we have used all these identities and dimensional ones to simplify as much as possible the form of $\mathscr{L}_{\hodge}$, given by \eqref{L*}.

\medskip

Starting from the 21-dimensional basis of \cite{Colleaux:2023cqu}, deriving its kinetic matrix in terms of the time derivative of the electric field and imposing $\mathscr{D}=\mathscr{E}=0$, we obtain (excluding the theory \eqref{F17}, which corresponds to $\mathcal{F}_{17}$ in \cite{Colleaux:2023cqu}),
\begin{eqnarray}
\begin{split}
\mathscr{L}&= \alpha R + 4 \partial_0 \alpha F_2^{\mu\nu} R_{\mu\nu} +\left(  \left( 2\mathcal{S}_0 \mathcal{S}_1^{-1} \partial_1 + \partial_0 + 2 \mathcal{S}_1 \partial_0\partial_1 \right)\alpha - \frac{\mathcal{S}_1^2}{16} \beta \right)F^{\mu\nu} F^{\sigma\rho} R_{\mu\nu\sigma\rho} \\
&+ \beta \left( -\frac{\mathcal{S}_1^2}{4} \mathcal{F}_{1} -2\mathcal{S}_0 \mathcal{F}_{6} -4 \left( \mathcal{F}_{10} -\mathcal{F}_{11} + \mathcal{F}_{13} \right) \right) - \frac{16\partial_1\alpha}{\mathcal{S}_1} \left(\mathcal{F}_{2}-\mathcal{F}_{4}+\mathcal{F}_{6}+3 \mathcal{F}_{7}-2 \mathcal{F}_{8} \right) \\
&+ 16\partial^2_1\alpha \left(\mathcal{F}_{8}-\mathcal{F}_{7} \right) + \frac{4 \partial_0\partial_1 \alpha }{\mathcal{S}_1}\left(\mathcal{S}_1^2\mathcal{F}_{1} + 8 \mathcal{S}_0 \left(2 \mathcal{F}_{6}-\mathcal{F}_{7}+2 \mathcal{F}_{8} \right)  + 16 \left( 2\mathcal{F}_{10} + \mathcal{F}_{13} + \mathcal{F}_{14}-\mathcal{F}_{15} \right) \right)\,, \label{L*basis21}
\end{split}
\end{eqnarray}
where we have used the notations
\begin{eqnarray}
\mathcal{S}_0 = F^{\mu\nu}F_{\mu\nu}  = -4 \mathcal{F} \,,\;\;\;\;\;\; \mathcal{S}_1 = \hodge F^{\mu\nu} F_{\mu\nu} = 4 \mathcal{G} \,,\;\;\;\;\;\;\; \partial_0 = \frac{\partial}{\partial \mathcal{S}_0} \,,\;\;\;\;\;\; \partial_1 =   \frac{\partial}{\partial \mathcal{S}_1} \,,
\end{eqnarray}
while $\alpha$ and $\beta$ are functions of the two invariants $\mathcal{S}_0$ and $\mathcal{S}_1$.

Interestingly, $\alpha$ satisfies a two-dimensional wave equation. Indeed, introducing the variables 
\begin{eqnarray}
u = \mathcal{S}_0 + \sqrt{\mathcal{S}_0^2 + \mathcal{S}_1^2}  = 4 \y^2 \,,\;\;\;  v = - \mathcal{S}_0 + \sqrt{\mathcal{S}_0^2 +  \mathcal{S}_1^2} = 4 \x^2 \,,  
\end{eqnarray} 
 the left-over equation following from $\mathscr{D}=\mathscr{E}=0$ is
\begin{eqnarray}
 \frac{\partial^2}{\partial u \partial v} \left( (u+v)\alpha \right)=0  \;\;\;\; \implies \;\;\;\;\alpha (u,v)=\frac{f\left(u\right)+g\left(v\right)}{u+v},
\end{eqnarray}
for any functions $f$ and $g$, in terms of which the function $\beta$ is completely fixed and given by
\begin{eqnarray}
\beta(u,v) = \frac{32}{u^2v^2}\sum_{n=0}^2 \left( Q_n(u,v) \partial_u^n f(u) - Q_n(v,u)  \partial_v^n g(v) \right)\,,
\end{eqnarray}
where
\begin{eqnarray}
\begin{split}
Q_0 (u,v) &= \frac{v^2}{\left(u+v\right)^5} \left(27 u^3 - 17 u^2 v + 5 u v^2 + v^3 \right)\,,\\
Q_1 (u,v) &=- \frac{u v^2}{\left(u+v\right)^4} \left(17 u^2 - 6 u v + v^2 \right)\,,\\
Q_2 (u,v) &= \frac{4 u^3 v^2}{\left(u+v\right)^3}.
\end{split}
\end{eqnarray}
As emphasised in the main text, this theory can be seen as the generalisation of quadratic Horndeski scalar-tensor theory for a gauge field. From this perspective, it would be interesting to understand if the appearance of the 3D wave equation relating the functions of the three different ``kinetic scalars" in  bi-scalar-tensor theories with second order field equations \cite{Horndeski:2024hee} is related to the appearance of the 2D wave equation in our setting, which involves two ``kinetic scalars" $\mathcal{S}_0$ and $\mathcal{S}_1$.

\medskip

The relation between $\mathscr{L}$, $\mathscr{L}_{\hodge}$ and its conjugate, defined by \eqref{L*}, is given, up to boundary terms, Bianchi and dimensional identities, by
\begin{eqnarray}
\mathscr{L}\left[ f,g \right] \equiv \mathscr{L}_{\hodge}\left[ \upsilon \right]  + \bar{\mathscr{L}}_{\hodge}\left[ \bar{\upsilon} \right]  \,,
\end{eqnarray}
where
\begin{eqnarray}
\upsilon\left(\y\right) =\frac{f(4\y^2)}{4\y^2}\,,\;\;\;\;\; \bar{\upsilon}\left( \x \right) =\frac{g(4\x^2)}{4\x^2}\,.
\end{eqnarray}
Considering equal and linear functions, $f=g$ and $f(u)=u$, so that $\alpha=1$, reduces the theory to General Relativity, while opposite quadratic ones, $f=-g$ and $f(u)=u^2/2$, so that $\alpha=\mathcal{S}_0$, reduces the theory to Horndeski's NMC. A cubic term is obtained setting $f=g$ and $f(u)=u^3$, so that $\alpha=4 \mathcal{S}_0^2 + \mathcal{S}_1^2$, from which we get the following Lagrangian density
 \begin{eqnarray}
 \begin{split}
\mathscr{L}_\text{cubic}=\left(  \mathcal{S}_0^2 + \frac{\mathcal{S}_1^2}{4} \right) R +\mathcal{S}_0  \left( 8 F_2^{\mu\nu} R_{\mu\nu} +3 F^{\mu\nu} F^{\sigma\rho} R_{\mu\nu\sigma\rho}\right) - 8 \left(\mathcal{F}_{2} - \mathcal{F}_{4} +\mathcal{F}_{6}+4 \mathcal{F}_{7}-3 \mathcal{F}_{8}\right) .
\end{split}
 \end{eqnarray}
In four dimensions and up to boundary terms, it is equivalent to the following, ``Galileons-looking", expression
 \begin{eqnarray}
 \begin{split}
\mathscr{L}_\text{cubic}\equiv -\varepsilon^{\mu\nu\rho\sigma} \varepsilon_{\alpha\beta\gamma\delta}   \left( F_{2\rho}^\gamma F_{2\sigma}^\delta + \mathcal{S}_0 F_{\rho\sigma} F^{\gamma\delta} \right)  R^{\alpha\beta}_{\mu\nu} -8 \mathcal{S}_0 F_2^{\mu\nu} G_{\mu\nu} - 8 \left( \mathscr{L}_{1} -2 \mathscr{L}_{2} + \mathscr{L}_{3} \right) , \label{Galileonlike}
\end{split}
 \end{eqnarray}
where 
\begin{eqnarray}
\mathscr{L}_{1}  &= &  F_{2}^{\mu\nu}  \nabla_{[\gamma|} F_{\mu}{}^\gamma  \nabla_{|\lambda]} F_{\nu}{}^\lambda \,, \nonumber \\
\mathscr{L}_{2} & =   & F_{2}^{\mu\nu} \nabla_{[\mu|} F_{\nu\gamma}  \nabla_{|\lambda]} F^{\gamma\lambda} \,,\nonumber \\
\mathscr{L}_{3} &= & F^{\mu\nu}  \nabla_{[\mu} \mathcal{S}_0  \nabla_{\gamma]} F_{\nu}{}^\gamma \,.  \label{QLT2}
\end{eqnarray}

\bibliographystyle{utphys}

\bibliography{Bilbio_DHOEM}

\end{document}